\documentclass[showpacs,preprintnumbers,amsmath,amssymb,nofootinbib,floatfix]{revtex4}

\usepackage[dvips]{graphicx}
\usepackage{dcolumn}
\usepackage{bm}

\def\mapgeq{\mathbin{\lower.3ex\hbox{$\buildrel>\over{\smash{\scriptstyle\sim}\vphantom{_x}}$}}}
\def\mapleq{\mathbin{\lower.3ex\hbox{$\buildrel<\over{\smash{\scriptstyle\sim}\vphantom{_x}}$}}}
\def\mapgeqeq{\mathbi{\lower.3ex\hbox{$\buildrel>\over{\smash{\scriptstyle\approx}\vphantom{_2}}$}}}
\def\mapleqeq{\mathbin{\lower.3ex\hbox{$\buildrel<\over{\smash{\scriptstyle\approx}\vphantom{_2}}$}}}
\mathchardef\hanaO="724F
\def\Journal#1#2#3#4{{#1} {\bf #2} (#4) #3}

\def\NPB{Nucl. Phys. B}
\def\NPBOLD{Nucl. Phys.}
\def\NPSUPPL{Nucl. Phys. Proc. Suppl.}
\def\PLB{{Phys. Lett.} B}

\def\PLBOLD{Phys. Lett.}
\def\PRL{Phys. Rev. Lett.}
\def\RMP{Rev. Mod. Phys.}
\def\PRD{Phys. Rev. D}

\def\PTP{Prog. Theor. Phys.}
\def\JHEP{JHEP}

\def\EPJ{Euro. Phys. J. C}
\def\EPJA{Euro. Phys. J. A}
\def\JETPUSSR{Sov. Phys. JETP}

\def\ZETP{Zh. Eksp. Teor. Fiz.}

\def\IJMP{Int. J. Mod. Phys. A}

\def\JPG{J. Phys. G}

\def\SCI{Science}
\def\APJ{Astrophys. J.}

\def\NJP{New. J. Phys.}
\def\PPNP{Prog. Part. Nucl. Phys.}

\def\ACTAB{Acta. Phys. Pol. B}
\def\Erratum{Erratum-ibid}

\begin{document}

\preprint{TOKAI-HEP/TH-0602}

 \title{Two Categories of Approximately $\mu$-$\tau$ Symmetric Neutrino Mass Textures}

\author{Kenichi Fuki}
\email{fuki@phys.metro-u.ac.jp}
\affiliation{\vspace{3mm}%
\sl Department of Physics, Tokyo Metropolitan University,\\
1-1 Minami-Osawa, Hachioji, Tokyo 192-0397, Japan
}
\author{Masaki Yasu\`{e}}%
\email{yasue@keyaki.cc.u-tokai.ac.jp}
\affiliation{\vspace{3mm}%
\sl Department of Physics, Tokai University,\\
1117 Kitakaname, Hiratsuka, Kanagawa 259-1292, Japan
}

\date{August, 2006}

\begin{abstract}
Our approximately $\mu$-$\tau$ symmetric neutrino mass textures fall into two 
different categories, whose behaviors in the $\mu$-$\tau$ symmetric limit are 
characterized by either $\sin\theta_{13}\rightarrow 0$ (referred to as C1)), or 
$\sin\theta_{12}\rightarrow 0$ (referred to as C2)), where $\theta_{12}$ and $\theta_{13}$, 
respectively stand for the $\nu_e$-$\nu_\mu$ and $\nu_e$-$\nu_\tau$ mixing angles.  We present ten 
phenomenologically viable neutrino mass textures: two for the normal mass hierarchy,
 three for the inverted mass hierarchy, and five for the quasi degenerate mass pattern. 
 Tiny $\mu$-$\tau$ symmetry breaking ensures that 
$\sin^2\theta_{13}\ll 1$ for C1), and $\Delta m^2_\odot/\Delta m^2_{atm} (\equiv R) \ll 1$ 
for C2), where $\Delta m^2_\odot = m^2_2-m^2_1$ for solar neutrinos, 
and $\Delta m^2_{atm}=\vert m^2_3-(m^2_1+m^2_2)/2\vert$ for atmospheric neutrinos with $m_{1,2,3}$ being 
neutrino masses.  
A correlation among small quantities is provided by $\cos 2\theta_{23}\sim\sin\theta_{13}$ for 
C1), and by either $\cos 2\theta_{23}\sim R$, or 
$\cos 2\theta_{23}\sin\theta_{13}\sim R$ for C2), where $\theta_{23}$ is the $\nu_\mu$-$\nu_\tau$ mixing angle.  It is further shown that  
$\tan 2\theta_{12}\sim\cos 2\theta_{23}/\sin\theta_{13}$ is satisfied for C2).  We find the following 
properties for each mass ordering: 1) For the normal mass hierarchy, the other smallness 
of $R$ (or $\sin^2\theta_{13}$) in the case of C1) (or C2)) can be ascribed to an approximate electron number 
conservation; 2) For the normal and inverted mass hierarchies and for the 
quasi degenerate mass pattern II (with $m^2_{1,2,3}\gg \Delta m^2_{atm}$) exhibiting $m_1\sim m_2 \sim m_3$, all in the case of C2), either $R\sim\tan 2\theta_{12}\sin^2\theta_{13}$ or $R\sim\tan 2\theta_{12}\sin\theta_{13}$ is satisfied; 3) For the inverted mass hierarchy and the quasi degenerate mass pattern I (with $\vert m_{1,2,3}\vert\sim \sqrt{\Delta m^2_{atm}}$), both
 in the case of C1) with $m_1\sim -m_2$, the relation $R\sim \sin^2\theta_{13}$ arises from the 
contribution of ${\mathcal{O}}(\sin^2\theta_{13})$ in the estimation of neutrino masses; 
and 4) For the quasi degenerate mass 
pattern II, we observe that 
$m_{\beta\beta}$ can be as large as 0.5 eV, where $m_{\beta\beta}$ 
is the effective neutrino mass in $(\beta\beta)_{0\nu}$-decay, and 
that larger values of $m_{\beta\beta}$ favor larger values of $\sin \theta_{13}$ in the case of C1) for $m_1\sim m_2 \sim m_3$ and 
smaller values of $\sin \theta_{13}$ in the cases of C1) for $m_1\sim m_2 \sim -m_3$ and C2) for $m_1\sim m_2 \sim \pm m_3$.
\end{abstract}

\maketitle
\section{\label{sec:1}Introduction}
Neutrinos are massive particles \cite{PMNS}, which undergo neutrino oscillations that have been first 
confirmed to exist for atmospheric neutrinos by the Super-Kamiokande collaboration \cite{SK}.   
The similar neutrino oscillations \cite{Experiments} have also been observed in neutrinos from the Sun \cite{OldSolor,Sun}, 
from accelerators \cite{K2K}, and from reactors \cite{Reactor}.  Masses of such neutrinos, which are much lighter 
than the electron mass, can be  
created by theoretical mechanisms such as the seesaw mechanism \cite{Seesaw,type2seesaw}, and 
the radiative mechanism \cite{Zee,Babu}.
  The oscillations are interpreted as a result of the mixings among three flavor 
neutrinos $\nu_e$, $\nu_\mu$, and $\nu_\tau$, which are converted into three 
massive neutrinos $\nu_1$, $\nu_2$, and $\nu_3$ during their flight.  Their masses $m_{1,2,3}$ and mixing angles 
 $\theta_{12,23,13}$ are currently constrained to be \cite{NuData}:
\begin{eqnarray}
\Delta m^2_\odot = 7.92\left( 1\pm0.09\right) \times 10^{-5}~{\rm eV}^2,
\quad
\Delta m^2_{atm} = 2.4
{\footnotesize
\left( {1\begin{array}{*{20}c}
   { + 0.21}  \\
   { - 0.26}  \\
\end{array}} \right)
} \times 10^{-3}~{\rm eV}^2,
\label{Eq:NuDataMass}
\end{eqnarray}
where $\Delta m^2_\odot$, and $\Delta m^2_{atm}$ are 
neutrino mass  differences squared given by $\Delta m^2_\odot = m^2_2-m^2_1$ ($>0$ \cite{PositiveSolor}) for solar 
neutrinos, and $\Delta m^2_{atm} = \vert m^2_3-(m^2_1+m^2_2)/2\vert$ for atmospheric neutrinos, and
\begin{eqnarray}
\sin ^2 \theta _{12}  = 0.314
{\footnotesize
\left( {1\begin{array}{*{20}c}
   { + 0.18}  \\
   { - 0.15}  \\
\end{array}} \right)}
,
\quad
\sin ^2 \theta _{23}  = 0.44
{\footnotesize
\left( {1\begin{array}{*{20}c}
   { + 0.41}  \\
   { - 0.22}  \\
\end{array}} \right),
}
\quad
\sin ^2 \theta _{13}  = 
{\footnotesize
\left( 0.9 
{\begin{array}{*{20}c}
   { + 2.3}  \\
   { - 0.9}  \\
\end{array}} \right) \times 10^{-2}.
}
\label{Eq:NuDataAngle}
\end{eqnarray}
These mixing angles parameterize the PMNS unitary matrix $U_{PMNS}$ \cite{PMNS} to give
\begin{eqnarray}
U_{PMNS} &=&
    \left( 
    \begin{array}{ccc}
    c_{12}c_{13}                     &  s_{12}c_{13}   &  s_{13}\\
    -s_{12}c_{23}-c_{12}s_{23}s_{13} &  c_{12}c_{23}-s_{12}s_{23}s_{13}  &  s_{23}c_{13}\\
    s_{12}s_{23}-c_{12}c_{23}s_{13}  &  -c_{12}s_{23}-s_{12}c_{23}s_{13} & c_{23}c_{13}\\
    \end{array} 
    \right),
\label{Eq:PMNS-noCP}
\end{eqnarray}
which transforms $\nu_f$ ($f=e,\mu,\tau$) into $\nu_i$ ($i=1,2,3$): 
$\nu_f = (U_{PMNS})_{fi}\nu_i$, where $c_{ij} \equiv \cos\theta_{ij}$, and 
$s_{ij} \equiv \sin\theta_{ij}$.  Two distinct properties 
are present in these observed data \cite{NeutrinoSummary}.  One is that the mixing 
angle $\theta_{13}$ is suppressed to show $\sin^2\theta_{13}\ll 1$ but the 
atmospheric, and solar mixing angles $\theta_{12,23}$ are not suppressed, and 
satisfy $\sin^2 2\theta_{12,23} ={\mathcal{O}}(1)$.  The other is that 
$\Delta m^2_\odot$, and $\Delta m^2_{atm}$ show the hierarchy 
$\Delta m^2_\odot/\Delta m^2_{atm}\ll 1$.  These properties may provide a great 
hint on how the underlying neutrino physics looks like.

It has been argued that the $\mu$-$\tau$ symmetry, which is based on the invariance 
of flavor neutrino mass terms under the interchange of $\nu_\mu$, and $\nu_\tau$ 
\cite{Nishiura,mu-tau,mu-tau0,mu-tau1,mu-tau2}, well describes the suppression 
of $\sin^2\theta_{13}\ll 1$ as well as the large mixings of 
$\sin^2 2\theta_{12,23} ={\mathcal{O}}(1)$.  The suppression reflects the fact 
that this symmetry requires $\sin\theta_{13}= 0$ and its tiny breaking induces $\sin\theta_{13}\ll 1$.  
To obtain the hierarchy $\Delta m^2_\odot/\Delta m^2_{atm}\ll 1$ may need an additional assumption such 
as that of the approximate conservation of the electron number \cite{eNumber-example,eNumber}.  
As another consequence of the tiny breaking of the $\mu$-$\tau$ symmetry, 
we have discussed in Ref.\cite{Another-mu-tau} 
that the hierarchy $\Delta m^2_\odot/\Delta m^2_{atm}\ll 1$
 can appear instead of the suppression of $\sin^2\theta_{13}\ll 1$.  In this case,
 we obtain $\sin\theta_{12}\rightarrow 0$ (referred to as C2)) instead of 
$\sin\theta_{13}\rightarrow 0$ (referred to as C1)) in the $\mu$-$\tau$ symmetric 
limit.\footnote{The possible choice of $\sin\theta_{12}=0$ has also been mentioned in 
Ref.\cite{Another-mu-tau-pre}.}  In both cases, we have found that $\cos 2\theta_{23}$ vanishes 
in the $\mu$-$\tau$ symmetric limit.  Therefore, $\cos 2\theta_{23}$ is a good 
measure of the $\mu$-$\tau$ symmetry breaking.  To obtain $\sin^22\theta_{12}={\mathcal{O}}(1)$
 from $\sin\theta_{12}\sim \varepsilon$ in C2), where $\varepsilon$ represents 
the $\mu$-$\tau$ symmetry breaking parameter, we have to introduce another small parameter 
denoted by $\eta$ to cancel $\varepsilon$.  In fact, the mixing angle $\theta_{12}$
 is determined to be $\tan 2\theta_{12}\sim \varepsilon/\eta$, leading to 
$\sin^22\theta_{12}={\mathcal{O}}(1)$ if $\eta\sim \varepsilon$ while it vanishes 
as $\varepsilon\rightarrow 0$ for a fixed $\eta$. Namely, we find that 
$\tan 2\theta_{12}\sim \cos 2\theta_{23}/\sin\theta_{13}$.  Any textures 
in C2) must be constrained so as to appropriately contain this small factor $\eta$.  

Useful relations among $\sin^2\theta_{13}$, 
$\Delta m^2_\odot/\Delta m^2_{atm}$, and $\cos 2\theta_{23}$ \cite{AtmDeviation,Theta31AndMass} have 
been derived on the general ground \cite{Another-mu-tau}, where we rely upon the $\mu$-$\tau$ symmetry 
breaking but not on details of textures.  Namely, we have found that 
$\cos 2\theta_{23}\sim\sin\theta_{13}$ for C1), and $\cos 2\theta_{23}\sim R$ for 
C2). It is also discussed in Ref.\cite{Another-mu-tau} that the hierarchy 
$\Delta m^2_\odot/\Delta m^2_{atm}\ll 1$ in C1) can be accounted for by the 
relation $\Delta m^2_\odot/\Delta m^2_{atm} \sim \sin^2\theta_{13}$ induced by 
effects of ${\mathcal{O}}(\sin^2\theta_{13})$ appearing in the estimation of neutrino masses.  This relation is specific to 
textures with $m_1+m_2 \sim 0$ \cite{PlusMinusNu} and is found to be satisfied by the inverted mass hierarchy, 
and the quasi degenerate mass pattern with $\vert m_{1,2,3}\vert \sim \sqrt{\Delta m^2_{atm}}$. On the other hand,
in C2), because of $\tan 2\theta_{12}\sim \cos 2\theta_{23}/\sin\theta_{13}$, 
it turns out that $\Delta m^2_\odot/\Delta m^2_{atm}\sim\tan 2\theta_{12}\sin^2\theta_{13}$ and  
$\Delta m^2_\odot/\Delta m^2_{atm}\sim\tan 2\theta_{12}\sin\theta_{13}$ are, respectively, 
satisfied for the normal and inverted mass hierarchies.

It is known that the $\mu$-$\tau$ symmetry is badly broken by the charged leptons.  
Since the neutrinos and charged leptons form the $SU(2)_L$-doublets, the assumed 
$\mu$-$\tau$ symmetry for neutrinos may not be well preserved.  This fact apparently 
disfavors the requirement of the $\mu$-$\tau$ symmetry.
However, if neutrinos are Majorana particles, it is theoretically possible 
to have an approximate $\mu$-$\tau$ symmetry for neutrinos but not for charged leptons who 
are Dirac particles.  This difference can supply approximately $\mu$-$\tau$ symmetric flavor 
structure for neutrinos and the dominance of Type II seesaw mechanism \cite{type2seesaw} is thus 
preferred \cite{type2seesawinSO10}.  
There are several examples to have the $\mu$-$\tau$ symmetry 
for neutrinos \cite{mu-tau0,mu-tau2,leptonic-mu-tau,recent_mu-tau-breaking}. 
One way to reconcile with the difficulty is to introduce a few Higgs 
scalars with, say, different $Z_2$ parity, or other different quantum numbers associated with 
extra symmetries that can discriminate among various Higgs scalars, where their vevs can provide charged 
lepton masses in such a way that the charged 
leptons acquire almost diagonal masses.  In this case, there necessarily arise flavor-changing 
neutral current interactions due to the direct exchanges of these Higgs scalars.  Such 
effects become sizable for quarks when quarks and leptons are unified and 
should be suppressed. 

In this article, we construct phenomenologically viable neutrino mass textures 
in our categories C1), and C2) that account for the present observed 
properties of neutrino oscillations.  There are ten textures: two of them provide 
the normal mass hierarchy, other three of them provide the inverted mass hierarchy,
 and the remaining five of them provide the quasi degenerate mass pattern.  Among 
five textures for the quasi degenerate mass pattern, one textures for C1) describe 
neutrinos with $\vert m_{1,2,3}\vert \sim \sqrt{\Delta m^2_{atm}}$ \cite{PlusMinusNu,NewMassTexture,QDGone}.  
To estimate neutrino 
masses, and mixings, we use general formula shown in the Appendix \ref{sec:Appendix}, where 
terms of ${\mathcal{O}}(\sin^2\theta_{13})$ are property taken into account to 
see that $\Delta m^2_\odot/\Delta m^2_{atm} \sim \sin^2\theta_{13}$ is naturally realized in some of textures for C1).  In the 
next section, we review what was derived in Ref.\cite{Another-mu-tau}.  In Sec.\ref{sec:3}, 
we present ten textures, where the general relations obtained in 
Sec.\ref{sec:2} are found to be indeed satisfied. In the quasi degenerate mass 
pattern with $m^2_{1,2,3} \gg \Delta m^2_{atm}$, the effective 
neutrino mass in $(\beta\beta)_{0\nu}$-decay is shown to be as large as 0.5 eV
 \cite{TheoryMass-ee}. The final section Sec.\ref{sec:4} is 
devoted to summary, and discussions.

\section{\label{sec:2}Neutrino Masses and Mixing Angles up to ${\hanaO}(\sin^2\theta_{13})$
}
Our neutrino mass matrix $M_\nu$ is parameterized by the sum of the $\mu$-$\tau$
 symmetric part $M_{sym}$, and the symmetry breaking part $M_b$:\footnote{It is 
understood that the charged leptons, and neutrinos are rotated, if necessary, to 
give diagonal charged-current interactions, and to define the flavor neutrinos 
of $\nu_e$, $\nu_\mu$, and $\nu_\tau$.} $M_\nu = M_{sum}+M_b$ with
\begin{eqnarray}
M_{sym} = 
\left( {\begin{array}{*{20}c}
   a & b & { - \sigma b}  \\
   b & d & e  \\
   { - \sigma b} & e & d  \\
\end{array}} \right),
\quad
M_b = \varepsilon \left( {\begin{array}{*{20}c}
   0 & {b^\prime} & {\sigma b^\prime}  \\
   {b^\prime} & {d^\prime} & 0  \\
   {\sigma b^\prime} & 0 & { - d^\prime}  \\
\end{array}} \right),
\label{Eq:M_13-break}
\end{eqnarray}
where $\varepsilon$ stands for a tiny $\mu$-$\tau$ symmetry breaking parameter.
We focus on clarifying flavor structure of $M_\nu$ to see how large or small each flavor neutrino mass is, 
and calculate neutrino masses, and mixing angles directly from $M_\nu$ in the normal mass hierarchy, the inverted mass 
hierarchy, and the quasi degenerate mass pattern.\footnote{See Ref.\cite{AotherAnalysis} for indirect 
discussions on the approximately $\mu$-$\tau$ symmetric flavor neutrino masses, which 
 utilize the mass eigenstates instead of the flavor eigenstates.}  
Various flavor structures consistent with the observed data to be obtained in the next section are visually analyzed. 

There is a simple reason why Eq.(\ref{Eq:M_13-break}) leads to either $\sin\theta_{13}=0$ or 
$\sin\theta_{12}=0$.  One of the eigenvectors for $M_{sym}$ is given by $(0, \sigma/\sqrt{2}, 1/\sqrt{2})^T$.
 It supplies a column vector placed in $U_{PMNS}$, which can be either 
$(s_{13}, s_{23}c_{13}, c_{23}c_{13})^T$ or 
$(s_{12}c_{13}, c_{12}c_{23}-s_{12}s_{23}s_{13}, -c_{12}s_{23}-s_{12}c_{23}s_{13})^T$
 depending on the ordering of the eigenvalues. By comparing our eigenvector with 
one of these two column vectors, we obtain that
\begin{eqnarray}
&&
\tan 2\theta _{12}  = \frac{2\sqrt 2 b }{d  - \sigma e  - a},
\quad
\tan 2\theta _{23}  =\sigma,
\quad
\sin \theta _{13}  =0,
\label{Eq:mixing_13case}
\end{eqnarray}
for $(s_{13}, s_{23}c_{13}, c_{23}c_{13})^T$, and 
\begin{eqnarray}
&&
\sin\theta _{12}  = 0,
\quad
\tan 2\theta _{23}  = -\sigma,
\quad
\tan 2\theta _{13}  = -\frac{2\sqrt 2 \sigma b }{d  - \sigma e -a},
\label{Eq:mixing_12case}
\end{eqnarray}
for $(s_{12}c_{13}, c_{12}c_{23}-s_{12}s_{23}s_{13}, -c_{12}s_{23}-s_{12}c_{23}s_{13})^T$,
 where $(0, 1/\sqrt{2}, \sigma/\sqrt{2})^T$ is used instead of  $(0, \sigma/\sqrt{2}, 1/\sqrt{2})^T$.

Since the $\mu$-$\tau$ symmetry 
breaking generally induces the deviation of the atmospheric neutrino mixing from 
the maximal one, we parameterize this deviation by $\Delta$:
\begin{eqnarray}
c_{23} = \frac{{1 + \Delta }}{{\sqrt {2\left( {1 + \Delta ^2 } \right)} }},
\quad
s_{23} = \pm \sigma \frac{{1 - \Delta }}{{\sqrt {2\left( {1 + \Delta ^2 } \right)} }},
\label{Eq:AtmDeviation}
\end{eqnarray}
giving $\sin 2\theta _{23} = \pm\sigma(1-\Delta^2)/(1+\Delta^2)$, and 
$\cos 2\theta_{23} = 2\Delta/(1+\Delta^2)$.  The plus (minus) sign in $s_{23}$ 
corresponds to the sign for $\tan 2\theta_{23}$ in Eq.(\ref{Eq:mixing_13case}) 
(Eq.(\ref{Eq:mixing_12case})).  The estimation of the masses and the mixing 
angles is summarized in the Appendix \ref{sec:Appendix}, where 
corrections of ${\mathcal{O}}(\sin^2\theta_{13})$ are included. It should be noted that this 
estimation does not recourse to the conventional perturbative expansions which provide order by order 
construction of eigen values (for masses) and eigen vectors (for mixing angles) because 
we know the exact expressions of masses and mixing angles given by 
Eqs.(\ref{Eq:ExactMixingAngle12})-(\ref{Eq:ExactMasses}) in the Appendix \ref{sec:Appendix}.

The suppression of $\Delta m^2_\odot$ is obviously possible if either
\begin{eqnarray}
&&
m_1+m_2\sim 0, ~{\rm or}~ m_1-m_2 \sim 0,
\label{Eq:SolarCond}
\end{eqnarray}
is satisfied.  This trivial condition can give useful relations when $m_1$ and $m_2$ 
calculated in the Appendix \ref{sec:Appendix} are used.  Namely, the mass difference $m_1-m_2$ is
connected to the quantity $X$ defined in Eq.(\ref{Eq:Parameters}).  In C2), the smallness of $X$ is a direct consequence of the tiny 
$\mu$-$\tau$ symmetry breaking as can been seen form Eq.(\ref{Eq:X-Y-Delta-12}).  Since 
$\Delta m_ \odot ^2$ is expressed in terms of $X$ as
\begin{eqnarray}
&&
\Delta m_ \odot ^2  = \frac{2\sqrt 2 (m_1  + m_2 )X}{\sin 2\theta _{12}},
\label{Eq:SolarCond2}
\end{eqnarray}
the appearance of $\Delta m^2_\odot/\Delta m^2_{atm}\ll 1$ is a natural result in C2).  
On the other hand, in C1), since $X$ is not necessarily suppressed, 
the appearance of $\Delta m^2_\odot/\Delta m^2_{atm}\ll 1$ is not a natural result. 
We have to suppress $X$ unless the condition $m_1+m_2\sim 0$ is satisfied.  

The additional requirements are present in C1) and C2).  First, in C1), we may require that
\begin{eqnarray}
&&
d-\sigma e+a= 0,
\label{Eq:SolarAtmMassHierachy}
\end{eqnarray}
to realize $m_1+m_2\sim 0$ in Eq.(\ref{Eq:Masses-13}). More precisely, we demand that $\vert d-\sigma e+a\vert\mapleq\varepsilon^2$.  If 
this is the case, we find that
\begin{eqnarray}
&&
\frac{\Delta m^2_\odot}{\Delta m^2_{atm}}\sim \sin^2\theta_{13},
\label{Eq:Delta-13}
\end{eqnarray}
which arises from textures with $m_2 \approx -m_1 \approx \sqrt{2} X/\sin 2\theta_{12}$ 
and $m_3 \sim d+e\sigma$
\begin{itemize}
\item for the inverted mass hierarchy if 
$d+e\sigma\sim 0$, giving $\vert m_3\vert \sim\sin^2\theta_{13}\vert m_{1,2}\vert$;
\item for the quasi degenerate mass pattern if 
$\vert d+e\sigma\vert\sim \vert X/\sin 2\theta_{12}\vert$, giving 
$\vert m_{1,2,3}\vert \sim\sqrt{\Delta m^2_{atm}}$.
\end{itemize}
These two mass patterns in C1) allow $\Delta m^2_\odot/\Delta m^2_{atm}$ to correlate with $\sin^2\theta_{13}$.
The relation Eq.(\ref{Eq:Delta-13}) has been already derived in Ref.\cite{AtmDeviation,Theta31AndMass},
 but our reason for the appearance of this relation is a purely perturbative one 
focusing on the effects of ${\mathcal{O}}(\sin^2\theta_{13})$ \cite{Another-mu-tau,NewMassTexture}.  

Next, in C2), there are two constraints, which guarantee that 
$\sin^2\theta_{13}\ll 1$ and $\sin^22\theta_{12}={\mathcal{O}}(1)$.  For 
$\sin^2\theta_{13}\ll 1$, we need $Y\sim0$ because $\tan 2\theta_{13}\propto Y$ 
in Eq.(\ref{Eq:Mixing-12}).  Since $Y$ is mainly determined by $b$, the $\mu$-
$\tau$ symmetric mass, we must realize $b\sim 0$.  For $\sin^22\theta_{12}={\mathcal{O}}(1)$,
 we require that $\vert d + \sigma e - a\vert \propto \vert\varepsilon\vert$ in Eq.(\ref{Eq:Mixing-12}) 
leading to $\tan 2\theta_{12}\propto X/\varepsilon$ only if $X$ cancels the effect from $\varepsilon$ in
the denominator to yield $\tan 2\theta_{12}={\mathcal{O}}(1)$.  Therefore, we must realize that
\begin{eqnarray}
&&
b\sim 0,
\quad
\vert d + \sigma e - a\vert \propto \vert\varepsilon\vert,
\label{Eq:tan12-C2}
\end{eqnarray}
in C2).  Hereafter, to suppress the contribution from $b$, another small parameter denoted 
by $\eta$ presumably of ${\mathcal {O}}(\varepsilon)$ is introduced.

\section{\label{sec:3}Neutrino Mass Hierarchy}
In this section, neutrino mass textures, which describe the present observed 
patterns of neutrino masses and mixings, are constructed.  We present ten textures,
 two for the normal mass hierarchy, three for the inverted mass hierarchy, one for the 
quasi degenerate mass pattern I (with $\vert m_{1,2,3}\vert \sim \sqrt{\Delta m^2_{atm}}$) 
and four for the quasi degenerate mass pattern II 
(with $m^2_{1,2,3} \gg \Delta m^2_{atm}$).\footnote{See Ref.\cite{AtmDeviation}, 
where some of the results are shared.}  Since we would like to see the effects from $\sin^2\theta_{13}$
 in $\Delta m^2_\odot/\Delta m^2_{atm}$, we calculate, in the Appendix \ref{sec:Appendix},
 all the quantities up to ${\mathcal{O}}(\varepsilon^2)$, where the terms of ${\mathcal{O}}(\varepsilon^2)$ 
 are responsible for $\Delta m^2_\odot/\Delta m^2_{atm}\ll 1$ in some textures. 

To construct textures, we consider general constraints on the flavor 
neutrino masses found in Sec.\ref{sec:2}.  First, the constraint from Eq.(\ref{Eq:SolarCond}) to obtain 
$\Delta m^2_\odot/\Delta m^2_{atm}\ll 1$ is satisfied in various mass patterns.  Namely, we require
\begin{itemize}
\item for the normal mass hierarchy with $m^2_{1,2}\sim 0(\ll m^2_3)$, $X\sim 0$ 
 in Eq.(\ref{Eq:Normal-13-1}) for C1), and Eq.(\ref{Eq:Normal-12-1}) for C2);
\item for the inverted mass hierarchy with $m^2_{1,2}\gg m^2_3$,
 either $X\sim 0$ in Eq.(\ref{Eq:Inverted-13-1}) for C1), and Eq.(\ref{Eq:Inverted-12-1}) for C2), 
 or $m_2+m_1\sim 0$ in Eq.(\ref{Eq:Inverted-13-2}) for C1);
\item for the quasi degenerate mass pattern I with $\vert m_{1,2,3}\vert \sim \sqrt{\Delta m^2_{atm}}$,
 $m_1+m_2 \sim 0$ in Eq.(\ref{Eq:Normal-13-2}) for C1);
\item for the quasi degenerate mass pattern II with $m^2_{1,2,3} \gg \Delta m^2_{atm}$, $X\sim 0$ in 
Eqs.(\ref{Eq:Degenerate-13-1}) and (\ref{Eq:Degenerate-dash-13-1}) for C1), 
and Eqs.(\ref{Eq:Degenerate-12-1}) and (\ref{Eq:Degenerate-dash-12-1}) for C2).
\end{itemize}
Furthermore, we can realize $\Delta m^2_\odot/\Delta m^2_{atm}\sim\sin^2\theta_{13}$ for C1) if we choose
\begin{itemize}
\item for the inverted mass hierarchy with $m_1 + m_2 \sim 0$, $d + \sigma e\sim 0$ to give $m_3\sim 0$;
\item for the quasi degenerate mass pattern I with $m_1 + m_2 \sim 0$, $\vert d + \sigma e\vert\sim \vert \sqrt{2}X/\sin 2\theta_{12}\vert$ 
to give $m_{3}\sim \vert m_{1,2}\vert$.
\end{itemize}
Finally, for C2), we find that Eq.(\ref{Eq:tan12-C2}) to obtain $\tan 2\theta_{12}={\mathcal{O}}(1)$ 
is satisfied in such a way that
\begin{itemize}
\item for the normal mass hierarchy, $d + \sigma e\sim 0$ and $a\sim 0$ are imposed;
\item for the inverted mass hierarchy, $d + \sigma e\sim a (\neq 0)$ is imposed;
\item for the quasi degenerate mass pattern II with $m_{1,2}\sim m_3$, $d\sim a(\neq 0)$ and $e\sim 0$ 
\item for the quasi degenerate mass pattern II with $m_{1,2}\sim -m_3$, $d\sim 0$ and $e\sim \sigma a$ 
are imposed.
\end{itemize}

The predictions from these textures are plotted in FIG.\ref{Fig:normal-R}-FIG.\ref{Fig:degenerate-ee} 
for $\tan 2\theta_{12}>0$ and $\sigma=1$, where $\vert\varepsilon\vert \leq 1/3$ is taken to satisfy 
$\varepsilon^2\mapleq 0.1$.  The results from $\tan 2\theta_{12}<0$ are covered by the changes of the 
signs of other parameters, 
and all figures are depicted as functions of $\vert \sin\theta_{13}\vert$, 
where the difference in the sign of $\sigma$ becomes irrelevant.  
No constraint on the range of $\eta$ is applied but 
we find that $\eta$ is phenomenologically bounded to satisfy $\eta^2\ll 1$ 
because $\eta$ is linked to either $\Delta m^2_\odot/\Delta m^2_{atm}$, or 
$\sin\theta_{13}$.  All calculations to show our scattered plots 
are based on the formula shown in the Appendix \ref{sec:Appendix}, which are derived by 
the approximation due to $\varepsilon^2\ll 1 $ but not due to $\eta^2\ll 1$.
The approximation due to $\eta^2\ll 1$ is solely used to show compact forms of 
calculate masses and mixing angles in terms of the flavor neutrino masses 
in order to explain behaviors of the scattered plots. 
There are additional parameters of ${\mathcal O}(1)$, $p$ for $M_{ee}$, $q$ for 
$M_{e\mu,e\tau}$, $r$ and $r^\prime$ for $M_{\mu\mu,\tau\tau}$, and $s$ for $M_{\mu\tau}$, 
whose magnitudes run from 1/3 to 3 to show the scattered plots.

\subsection{\label{subsec:3-1}Normal Mass Hierarchy}
\subsubsection{\label{subsec:3-1-1}Category C1) with $s_{23}\sim\sigma /\sqrt{2}$}
The mass ordering is given by $m^2_1 < m^2_2 \ll m^2_3$, which is realized by 
\begin{eqnarray}
&&
M^{\rm C1)}_\nu
=
d_0 \left( {\begin{array}{*{20}c}
   {p\eta } & {\eta  + \varepsilon } & { - \sigma \left( {\eta  - \varepsilon } 
\right)}  \\
   {\eta  + \varepsilon } & {1 + r^\prime \varepsilon } & {\sigma \left( {1 - s
\eta } \right)}  \\
   { - \sigma \left( {\eta  - \varepsilon } \right)} & {\sigma \left( {1 - s\eta }
 \right)} & {1 - r^\prime \varepsilon }  \\
\end{array}} \right).
\label{Eq:Normal-13-1}
\end{eqnarray}
From the texture, neutrino masses are predicted to be:
\begin{eqnarray}
&&
m_1  \approx \left( {\frac{{2\left( {p + s} \right)\eta  - \left( {2 + r^{\prime 2} } \right)\varepsilon ^2 }}{4} - \frac{{2\eta  - r^\prime \varepsilon ^2 }}{{\sqrt 2 \sin 2\theta _{12} }}} \right)d_0,
\nonumber\\
&&
m_2  \approx \left( {\frac{{2\left( {p + s} \right)\eta  - \left( {2 + r^{\prime 2} } \right)\varepsilon ^2 }}{4} + \frac{{2\eta  - r^\prime \varepsilon ^2 }}{{\sqrt 2 \sin 2\theta _{12} }}} \right)d_0,
\nonumber\\
&&
m_3  \approx \frac{{2\left( {2 - s\eta } \right) + \left( {2 + r^{\prime 2} } \right)\varepsilon ^2 }}{2}d_0, 
\label{Eq:Normal-13-1-masses}
\end{eqnarray}
and
\begin{eqnarray}
&&
\Delta m_ \odot ^2  \approx 
\frac{{\left[ {2\left( {p + s} \right)\eta  - \left( {2 + r^{\prime 2} } \right)\varepsilon ^2 } \right]\left( {2\eta  + r^\prime \varepsilon ^2 } \right)}}{{\sqrt 2 \sin 2\theta _{12} }}d_0^2, 
\quad
\Delta m^2_{atm}  \approx 4d_0^2,
\label{Eq:Normal-13-1-squared}
\end{eqnarray}
and mixing angles are calculated to be
\begin{eqnarray}
&&
\tan 2\theta _{12}  \approx 
\frac{{2\sqrt 2 \left( {2\eta  - r'\varepsilon ^2 } \right)}}{{2\left( {s - p} \right)\eta  + \left( {2 - r'^2 } \right)\varepsilon ^2 }},
\quad
\tan 2\theta _{13}  \approx \sqrt 2 \sigma \varepsilon,
\quad
\cos 2\theta_{23} \approx - r^\prime\varepsilon.
\label{Eq:Normal-13-1-angles}
\end{eqnarray}

This texture utilizes 
both $X\approx 0$ and $m_1+m_2\approx 0$ to have $\Delta m^2_\odot/\Delta m^2_{atm} \ll 1$. As a result, 
$\vert\eta\vert\gg \varepsilon^2$ is preferred because, for $\vert\eta\vert\mapleq \varepsilon^2$, 
$\Delta m^2_\odot/\Delta m^2_{atm} \sim\varepsilon^4\sim\sin^4\theta_{13}$, which is experimentally too small 
to account for the observed size of $\Delta m^2_\odot/\Delta m^2_{atm}$.  It is thus expected that
\begin{eqnarray}
&&
\eta \sim \sqrt{\frac{\Delta m_ \odot ^2}{\Delta m^2_{atm}}},
\label{Eq:Normal-13-1-simplified-eta}
\end{eqnarray}
is satisfied.  There is a simple relation between $\theta_{13}$, and $\theta_{23}$, which is
\begin{eqnarray}
\cos 2\theta_{23}  \approx -\sqrt 2 \sigma r^\prime\sin\theta _{13}.
\label{Eq:Normal-13-1-simple}
\end{eqnarray}
It should be noted that $p$ can be neglected in these predictions and that the smallness 
of $p$ can be ascribed to the approximate electron number ($L_e$) 
conservation \cite{eNumber-example,eNumber}.  Suppose that the  
$L_e$-conservation is violated by tiny $\vert\Delta L_e\vert=1$ interactions 
\cite{eNumber-example}, which are characterized by the strength $\eta$, then, it is 
reasonable to expect the relation
\begin{eqnarray}
&&
M_{ee}:M_{ei}:M_{ij}\sim \eta^2: \eta:1,
\label{Eq:Normal-13-1-simplified-ee}
\end{eqnarray}
to arises, where $i,j=\mu,\tau$.  The smallness of $\Delta m^2_\odot/\Delta m^2_{atm}$
 is a result of tiny breaking of the electron number conservation.

\begin{figure}[!htbp]
\begin{flushleft}
\includegraphics*[20mm,203.6mm][300mm,264mm]{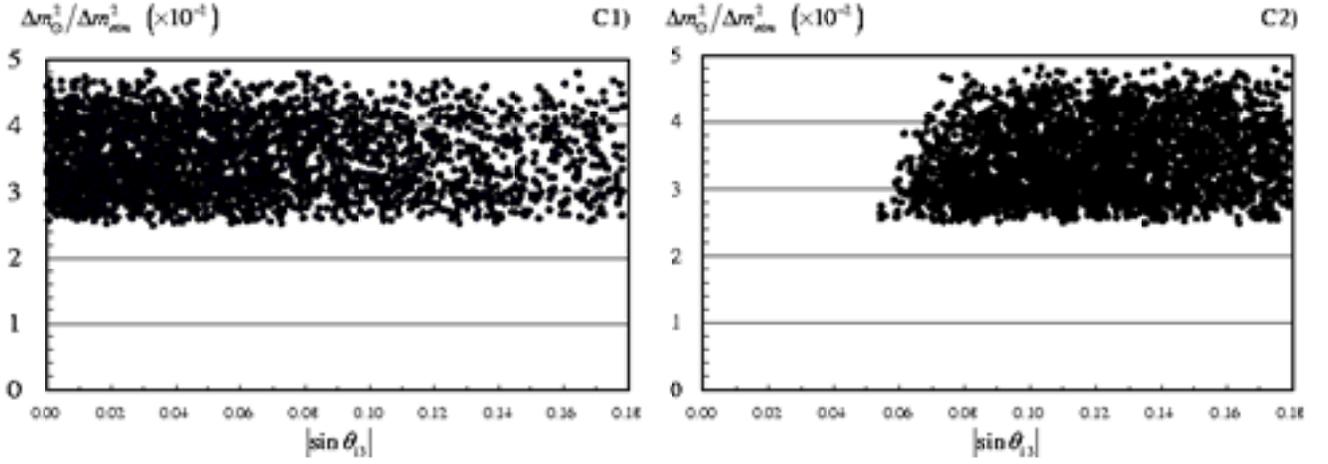}
\end{flushleft}
\vspace{-3mm}
\caption{The predictions of $\Delta m^2_\odot/\Delta m^2_{atm}$ as function of $\vert\sin\theta_{13}\vert$ for C1) and C2) in the normal mass hierarchy, where $m_{1,2}>0$ are taken and the experimental upper and lower bounds are applied.}
\label{Fig:normal-R}
\end{figure}

\begin{figure}[!htbp]
\begin{flushleft}
\includegraphics*[20mm,203.6mm][300mm,264mm]{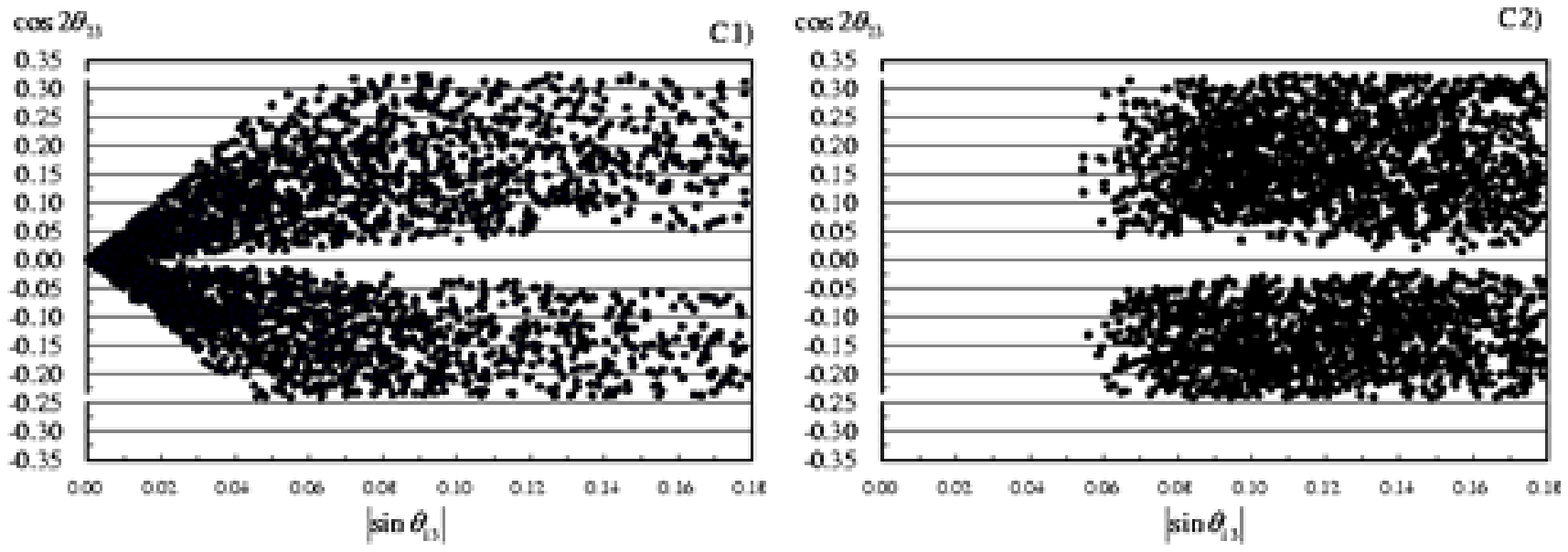}
\end{flushleft}
\vspace{-3mm}
\caption{The same as in FIG.\ref{Fig:normal-R} but for $\cos 2\theta_{23}$.}
\label{Fig:normal-cos23}
\end{figure}

The predictions $\Delta m^2_\odot/\Delta m^2_{atm}$
 and $\cos 2\theta_{23}$ are plotted in FIG.\ref{Fig:normal-R}-C1) and FIG.
\ref{Fig:normal-cos23}-C1) as functions of $\vert\sin\theta_{13}\vert$.  Since there is 
no efficient constraint on the region of $\eta$ arising from $\vert\eta\vert\mapgeq\varepsilon^2$, 
\begin{itemize}
\item $\Delta m^2_\odot/\Delta m^2_{atm}(\sim \eta^2)$ can have values for all ranges 
of $\sin\theta_{13}(\sim\varepsilon)$
\end{itemize}
as in FIG.\ref{Fig:normal-R}-C1). For $\cos 2\theta_{13}$, the proportionality of 
$\cos 2\theta_{23}$ to $\sin\theta_{13}$ in Eq.(\ref{Eq:Normal-13-1-simple}) indicates that
\begin{itemize}
\item  two straight boundaries for $\cos 2\theta_{23} > 0$ or $\cos 2\theta_{23} < 0$ correspond to 
 $\vert\cos 2\theta_{23}\vert\sim 3\sqrt{2}\vert\sin\theta_{13}\vert$
and $\vert\cos 2\theta_{23}\vert\sim \sqrt{2}\vert\sin\theta_{13}\vert /3$ from $1/3 \leq \vert r^\prime\vert\leq 3$, and 
\item the forbidden region near $\cos 2\theta_{23}=0$ shows up for $\sin\theta_{13}\neq 0$
\end{itemize}
as in FIG.\ref{Fig:normal-cos23}-C1).  
Since $p$ can vanish, the figures have been produced by adopting $\vert p\vert \leq 3$ 
instead of $1/3 \leq \vert p\vert \leq 3$. 

\subsubsection{\label{subsec:3-1-2}Category C2) with $s_{23}\sim -\sigma /\sqrt{2}$}
We also find the similar texture 
with $\sigma\rightarrow -\sigma$ in the $\mu$-$\tau$ entry of Eq.(\ref{Eq:Normal-13-1}). 
To obtain $\sin^2\theta_{13}\ll 1$, $b$ must be suppressed and the texture indeed provides 
extra suppression due to $\eta$.  The texture is given by
\begin{eqnarray}
&&
M^{\rm C2)}_\nu
=
d_0 \left( {\begin{array}{*{20}c}
   {p\eta } & {\eta  + \varepsilon } & { - \sigma \left( {\eta - \varepsilon } 
\right)}  \\
   {\eta  + \varepsilon } & {1 + r^\prime \varepsilon } & { -\sigma \left( {1 - 
s\eta } \right)}  \\
   { - \sigma \left( {\eta  - \varepsilon } \right)} & { -\sigma \left( {1 - s
\eta } \right)} & {1 - r^\prime \varepsilon }  \\
\end{array}} \right).
\label{Eq:Normal-12-1}
\end{eqnarray}
The masses are calculated to be:
\begin{eqnarray}
&&
m_1  \approx 
\left( {\frac{{2\left( {p + s} \right)\eta  - r^{\prime 2} \varepsilon ^2 }}{4} - \frac{{\sqrt 2 \varepsilon }}{{\sin 2\theta _{12} }}} \right)d_0,
\quad
m_1  \approx 
\left( {\frac{{2\left( {p + s} \right)\eta  - r^{\prime 2} \varepsilon ^2 }}{4} + \frac{{\sqrt 2 \varepsilon }}{{\sin 2\theta _{12} }}} \right)d_0,
\nonumber\\
&&
m_3  \approx 
\frac{{2\left( {2 - s\eta } \right) + r^{\prime 2} \varepsilon ^2 }}{2}d_0,
\label{Eq:Normal-12-1-masses}
\end{eqnarray}
and
\begin{eqnarray}
&&
\Delta m_ \odot ^2  \approx  
\frac{{\sqrt 2 \left[ {2\left( {p + s} \right)\eta  - r^{\prime 2} \varepsilon ^2 } \right]\varepsilon }}{{\sin 2\theta _{12} }},
\quad
\Delta m^2_{atm}  \approx 4d_0^2,
\label{Eq:Normal-12-1-squared}
\end{eqnarray}
and mixing angles are
\begin{eqnarray}
&&
\tan 2\theta _{12}  \approx 
\frac{{4\sqrt 2 \varepsilon }}{{2\left( {s - p} \right)\eta  - r^{\prime 2} \varepsilon ^2 }},
\quad
\tan 2\theta _{13}  \approx  - \sqrt 2 \sigma \eta,
\quad
\cos 2\theta_{23} \approx  -r^\prime\varepsilon.
\label{Eq:Normal-12-1-angles}
\end{eqnarray}
There is also the possibility to have the approximate electron number conservation 
that accounts for the smallness of $\eta$, which in turn ensures the smallness 
of $\sin\theta_{13}$. To obtain $\tan 2\theta_{12}={\mathcal{O}}(1)$, we require that 
$\vert\eta\vert \gg \varepsilon^2$ suggesting $\vert\eta\vert\sim\vert\varepsilon\vert$.

Since the terms proportional to $\varepsilon^2$ can be neglected, we find that
\begin{eqnarray}
\cos 2\theta _{23}\approx\frac{\sigma \left( {s-p} \right)r^\prime\tan 2\theta _{12}}{2}\sin \theta _{13}.
\label{Eq:Normal-12-1-simple-2}
\end{eqnarray}
This relation suggests that
\begin{eqnarray}
\tan 2\theta _{12} \sim \frac{\cos 2\theta _{23}}{\sin \theta _{13}},
\label{Eq:Normal-12-1-simple-3}
\end{eqnarray}
which is specific to the category C2). There is another correlation
\begin{eqnarray}
&&
\frac{{\Delta m_ \odot ^2 }}{{\Delta m_{atm}^2 }} \approx \frac{{\sigma \left( {s + p} \right)}}{{r'\sin 2\theta _{12} }}\cos 2\theta _{{\rm{23}}} \sin \theta _{13}, 
\label{Eq:Normal-12-1-simple}
\end{eqnarray}
where $\Delta m^2_\odot/\Delta m^2_{atm}\sim \eta\varepsilon$ from Eq.(\ref{Eq:Normal-12-1-squared}). 
These two relations give
\begin{eqnarray}
&&
\frac{{\Delta m_ \odot ^2 }}{{\Delta m_{atm}^2 }}
\approx
\frac{s^2-p^2}{2\cos 2\theta_{12}}\sin^2\theta_{13}.
\label{Eq:Normal-12-R13}
\end{eqnarray}
In this texture, 
\begin{eqnarray}
&&
\frac{\Delta m_ \odot ^2}{\Delta m^2_{atm}} \sim \sin^2\theta_{13},
\label{Eq:Normal-12-1-simplified}
\end{eqnarray}
is realized.

The predictions 
$\Delta m^2_\odot/\Delta m^2_{atm}$ and $\cos 2\theta_{23}$ are depicted in FIG.\ref{Fig:normal-R}-C2) 
and FIG.\ref{Fig:normal-cos23}-C2) as functions of $\vert\sin\theta_{13}\vert$, where $\vert p\vert \leq 3$ is also used to 
include the case of $p=0$.  In these figures, we find that
\begin{itemize}
\item $\vert\sin\theta_{13}\vert\mapgeq 0.06$ because of Eq.(\ref{Eq:Normal-12-1-simplified}),
\end{itemize}
which requires that $\vert\sin\theta_{13}\vert\sim 0.1$.   
 Because of $\varepsilon \neq 0$ for $\tan 2\theta_{12}\neq 0$, 
\begin{itemize}
\item $\cos 2\theta_{23}$ is not allowed to vanish
\end{itemize}
as shown in FIG.\ref{Fig:normal-cos23}-C2).

\subsection{\label{subsec:3-2}Inverted Mass Hierarchy}
\subsubsection{\label{subsec:3-2-1}Category C1) with $s_{23}\sim\sigma /\sqrt{2}$}
The mass ordering is given by $m^2_3\ll m^2_1 <m^2_2$.  
There are two types depending on the relative sign of $m_1$, and $m_2$. 
The first texture gives $m_1\sim m_2$, and takes the form of
\begin{eqnarray}
&&
M^{\rm C1)}_\nu
=
d_0 \left( {\begin{array}{*{20}c}
   2 - p\eta & {\eta  + \varepsilon } & { - \sigma \left( {\eta  - \varepsilon }
 \right)}  \\
   {\eta  + \varepsilon } & {1 + r^\prime \varepsilon } & { - \sigma \left( 1-s\eta\right) }  \\
   { - \sigma \left( {\eta  - \varepsilon } \right)} & { - \sigma \left( 1-s\eta\right)} & {1 - r^
\prime \varepsilon }  \\
\end{array}} \right).
\label{Eq:Inverted-13-1}
\end{eqnarray}
Neutrino masses are predicted to be:
\begin{eqnarray}
&&
m_1  \approx \left( {2 - \frac{{2\left( p+s\right)\eta  - \left( {2 + r^{\prime 2} } \right)\varepsilon ^2 }}{4} - \frac{{2\eta  + r^\prime \varepsilon ^2 }}{{\sqrt 2 \sin 2\theta _{12} }}} \right)d_0,
\nonumber\\
&&
m_2  \approx \left( {2 - \frac{{2\left( p+s\right)\eta  - \left( {2 + r^{\prime 2} } \right)\varepsilon ^2 }}{4} + \frac{{2\eta  + r^\prime \varepsilon ^2 }}{{\sqrt 2 \sin 2\theta _{12} }}} \right)d_0,
\nonumber\\
&&
m_3  \approx  \left(s\eta - \frac{2 + r^{\prime 2} }{2}\varepsilon ^2\right) d_0,
\label{Eq:Inverted-13-1-masses}
\end{eqnarray}
and
\begin{eqnarray}
&&
\Delta m_ \odot \approx  \frac{{4\sqrt 2 \left( {2\eta  + r^\prime \varepsilon ^2 } \right)}}{{\sin 2\theta _{12} }}d_0^2,
\quad
\Delta m^2_{atm}  \approx 4d_0^2,
\label{Eq:Inverted-13-1-squared}
\end{eqnarray}
and mixing angles are
\begin{eqnarray}
&&
\tan 2\theta _{12}  \approx \frac{{2\sqrt 2 \left( {2\eta  + r'\varepsilon ^2 } \right)}}{{2\left( p-s\right)\eta  - \left( {2 - r^{\prime 2} } \right)\varepsilon ^2 }},
\quad
\tan 2\theta _{13}  \approx  - \sqrt 2 \sigma \varepsilon,
\quad
\cos 2\theta_{23}   \approx r^\prime\varepsilon.
\label{Eq:Inverted-13-1-angles}
\end{eqnarray}

\begin{figure}[!htbp]
\begin{flushleft}
\includegraphics*[20mm,203.6mm][300mm,264mm]{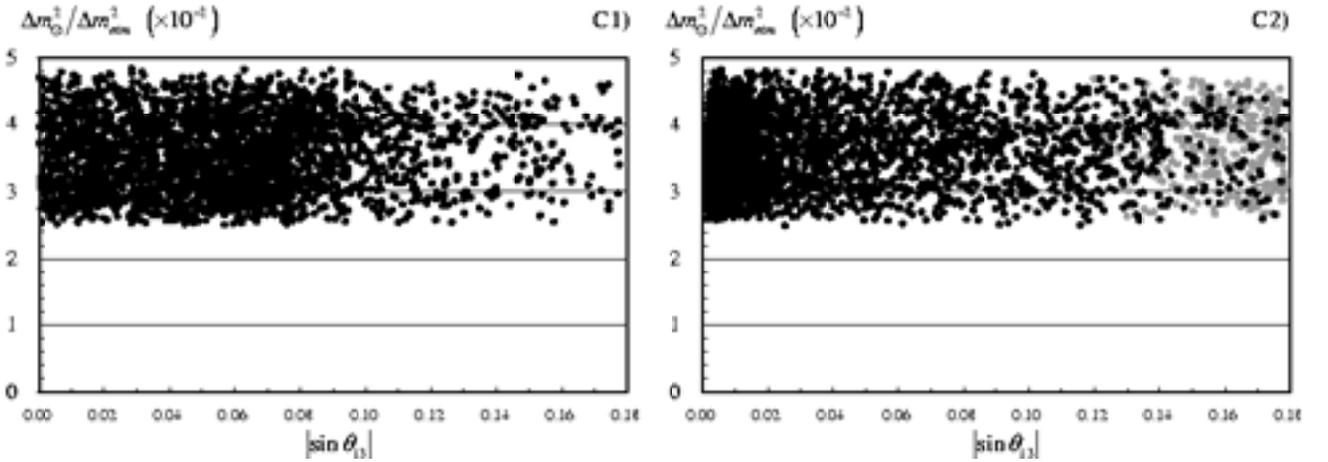}
\end{flushleft}
\vspace{-3mm}
\caption{The same as in FIG.\ref{Fig:normal-R} but in the inverted mass hierarchy with $m_1\approx m_2$, where the grey dots in 
the right figure represents the region of $\vert\eta\vert >1/3$, which lies beyond the approximation due to $\eta^2\ll 1$.}
\label{Fig:inverted-R}
\end{figure}

\begin{figure}[!htbp]
\begin{flushleft}
\includegraphics*[20mm,203.6mm][300mm,264mm]{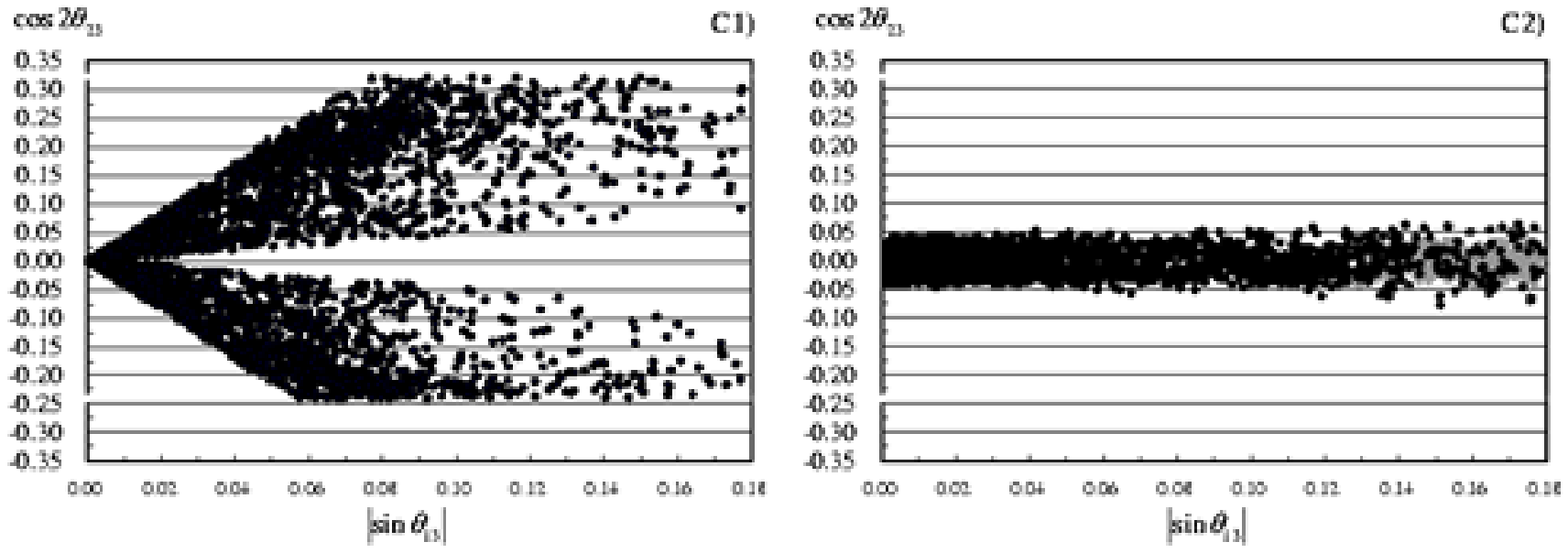}
\end{flushleft}
\vspace{-3mm}
\caption{The same as in FIG.\ref{Fig:inverted-R} but for $\cos 2\theta_{23}$}
\label{Fig:inverted-cos23}
\end{figure}

In this texture, we also have 
\begin{eqnarray}
&&
\cos 2\theta _{23}  \approx  - \sqrt 2 \sigma r^\prime \sin\theta_{13},
\label{Eq:Inverted-13-1-simple}
\end{eqnarray}
as in C1) for the normal mass hierarchy.  From the expression of $\tan 2\theta_{12}$, no a priori constraint on 
 $\eta$ arises because both $\eta$ and $\varepsilon^2$ are present in the denominator and numerator.\footnote{Unlike in the normal 
 mass hierarchy case, we obtain $\Delta m^2_\odot/\Delta m^2_{atm}\sim \varepsilon^2\sim\sin^2\theta_{13}$, which is the 
 right order of $\Delta m^2_\odot/\Delta m^2_{atm}$.  This possibility will be discussed elsewhere.} To realize 
 $\Delta m_ \odot ^2/\Delta m^2_{atm}={\mathcal O}(0.01)$, $\vert 2\eta  + r^\prime \varepsilon ^2\vert$ should be 
 ${\mathcal O}(0.01)$.  
The predictions $\Delta m^2_\odot/\Delta m^2_{atm}$ and $\cos 2\theta_{23}$ are 
depicted in FIG.\ref{Fig:inverted-R}-C1) and FIG.\ref{Fig:inverted-cos23}-C1) as 
functions of $\vert\sin\theta_{13}\vert$.  These predictions give similar dependences of $\sin\theta_{13}$ to those in  C1) 
for the normal mass hierarchy.  The proportionality of $\cos 2\theta_{23}$ to $\sin\theta_{13}$ also yields the straight boundaries 
in FIG.\ref{Fig:inverted-cos23}-C1).

The next texture is characterized by $m_1\sim -m_2$, and is given by
\begin{eqnarray}
&&
M^{\rm C1)}_\nu
=
d_0 \left( {\begin{array}{*{20}c}
   { - (2 - \eta)} & {q + \varepsilon } & { - \sigma \left( {q - \varepsilon } 
\right) }  \\
   {q + \varepsilon } & {1 + r^\prime \varepsilon } & { - \sigma }  \\
   { - \sigma \left( {q - \varepsilon } \right) } & { - \sigma } & {1 - r^
\prime \varepsilon }  \\
\end{array}} \right).
\label{Eq:Inverted-13-2}
\end{eqnarray}
Neutrino masses are predicted to be
\begin{eqnarray}
&&
m_1  \approx 
 - \left( {\frac{{\sqrt 2 q}}{{\sin 2\theta _{12} }} - \frac{{\eta  - 2\left[ {t_{13}^2  + \left( {\Delta  - \varepsilon r^\prime } \right)\Delta } \right]}}{2}} \right)d_0,
\nonumber\\
&&
m_2  \approx 
 \left( {\frac{{\sqrt 2 q}}{{\sin 2\theta _{12} }} + \frac{{\eta  - 2\left[ {t_{13}^2  + \left( {\Delta  - \varepsilon r^\prime } \right)\Delta } \right]}}{2}} \right)d_0,
\nonumber\\
&&
m_3  \approx 2\left( {t_{13}^2  + \left( {\Delta  - \varepsilon r^\prime } \right)\Delta } \right)d_0,
\label{Eq:Inverted-13-2-masses}
\end{eqnarray}
and 
\begin{eqnarray}
&&
\Delta m_ \odot ^2  \approx 
\frac{{2\sqrt 2 q\left[ {\eta  - 2\left( {t_{13}^2  + \left( {\Delta  - \varepsilon r^\prime } \right)\Delta } \right)} \right]}}{{\sin 2\theta _{12} }}d_0^2,
\quad
\Delta m_{atm}^2  \approx \frac{{2q^2 d_0^2 }}{{\sin ^2 2\theta _{12} }},
\label{Eq:Inverted-13-2-squared}
\end{eqnarray}
and mixing angles are
\begin{eqnarray}
&&
\tan 2\theta _{12}  \approx \frac{q}{{\sqrt 2 }},
\quad
\tan 2\theta _{13}  \approx \sqrt 2 \sigma \frac{{2 - qr^\prime }}{{2 + q^2 }}\varepsilon,
\quad
\cos 2\theta_{23}\approx 2\Delta,
\label{Eq:Inverted-13-2-angles}
\end{eqnarray}
where
\begin{eqnarray}
&&
\Delta  = \frac{{r^\prime  + q}}{{2 + q^2 }}\varepsilon.
\label{Eq:Inverted-13-2-Delta}
\end{eqnarray}
Since we have $\tan 2\theta_{12} > 0$, we know that $q>0$ and $q \sim 3$ to yield $\sin^22\theta_{12} \sim 0.8$.

We find that
\begin{eqnarray}
&&
\cos 2\theta_{23}\approx \frac{\sqrt{2}\sigma\left(r^\prime  + q\right)}{2 - 
qr^\prime}\sin\theta_{13}.
\label{Eq:Inverted-13-2-simple}
\end{eqnarray}
For $r^\prime > 0$, it just gives the proportionality of $\cos 2\theta_{23}$ to $\sin \theta_{13}$ 
for the fixed values of $q$ and $r^\prime$ because $r^\prime  + q$ in the coefficient of $\Delta$ 
does not vanish.  On the other hand, for $r^\prime < 0$, $\cos 2\theta_{23}$ can vanish.  
For $\vert\eta\vert\gg\varepsilon^2$, we find that 
\begin{eqnarray}
&&
\Delta m_ \odot ^2  \approx \frac{4\eta}{\cos 2\theta _{12} }d_0^2,
\quad
\eta \sim \frac{\Delta m^2_ \odot}{\Delta m^2_{atm}}.
\label{Eq:Inverted-13-2-eta}
\end{eqnarray}
For the opposite case, we may choose $\eta=0$. The mass hierarchy is taken care of by $\varepsilon^2\sim\sin^2\theta_{13}$ 
and $\Delta m_ \odot ^2$ becomes
\begin{eqnarray}
&&
\Delta m_ \odot ^2  \approx 
-\frac{4\sqrt 2 q \left[ t_{13}^2  + \left( \Delta  - \varepsilon r^\prime  \right)\Delta \right]}{{\sin 2\theta _{12} }}d_0^2,
\label{Eq:Inverted-13-2-DeltaSun}
\end{eqnarray}
which is converted into
\begin{eqnarray}
&&
\Delta m_ \odot ^2  \approx 
\frac{2\sqrt 2q}{\sin 2\theta _{12}}\left[ {\frac{{\left( {r^\prime  + q} \right)^2 }}{{2 + q^2 }} - 1} \right]\varepsilon ^2.
\label{Eq:Inverted-13-2-DeltaSun-2}
\end{eqnarray}
The constraint to obtain $\Delta m^2_\odot > 0$ can be satisfied if 
$r^\prime  <  - \sqrt {2 + q^2 }- q$ or $r^\prime  > \sqrt {2 + q^2 }  - q$, which gives $r^\prime \geq 1/3$ because 
$1/3\leq \vert r^\prime\vert \leq 3$ and $q\sim 3$. For $r^\prime \geq 1/3$, 
\begin{eqnarray}
&&
\frac{\Delta m^2_\odot}{\Delta m^2_{atm}}\approx \sin^2\theta_{13}
\label{Eq:Inverted-13-2-sin13-only}
\end{eqnarray}
is satisfied and $\vert \cos 2\theta_{23}\vert\sim \vert\sin\theta_{13}\vert$ is expected.

\begin{figure}[!htbp]
\begin{flushleft}
\includegraphics*[20mm,203.6mm][300mm,264mm]{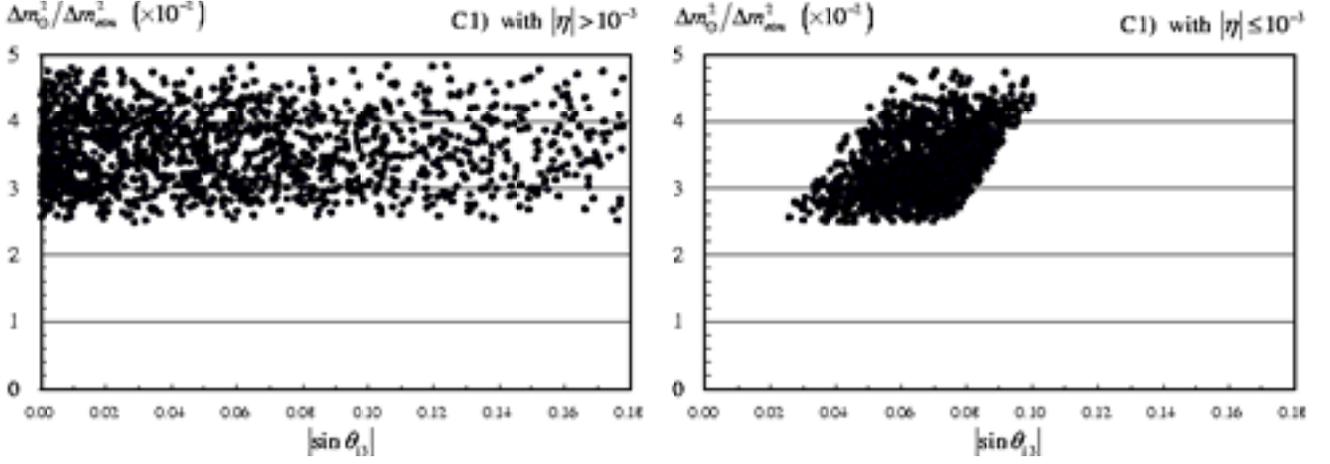}
\end{flushleft}
\vspace{-3mm}
\caption{The same as in FIG.\ref{Fig:normal-R} but in the inverted mass hierarchy for C1) with $m_1\approx -m_2$, 
where two regions of $\vert\eta\vert>0.001$ and $\vert\eta\vert\leq 0.001$ are separately shown.}
\label{Fig:inverted13-2-R}
\end{figure}

\begin{figure}[!htbp]
\begin{flushleft}
\includegraphics*[20mm,203.6mm][300mm,264mm]{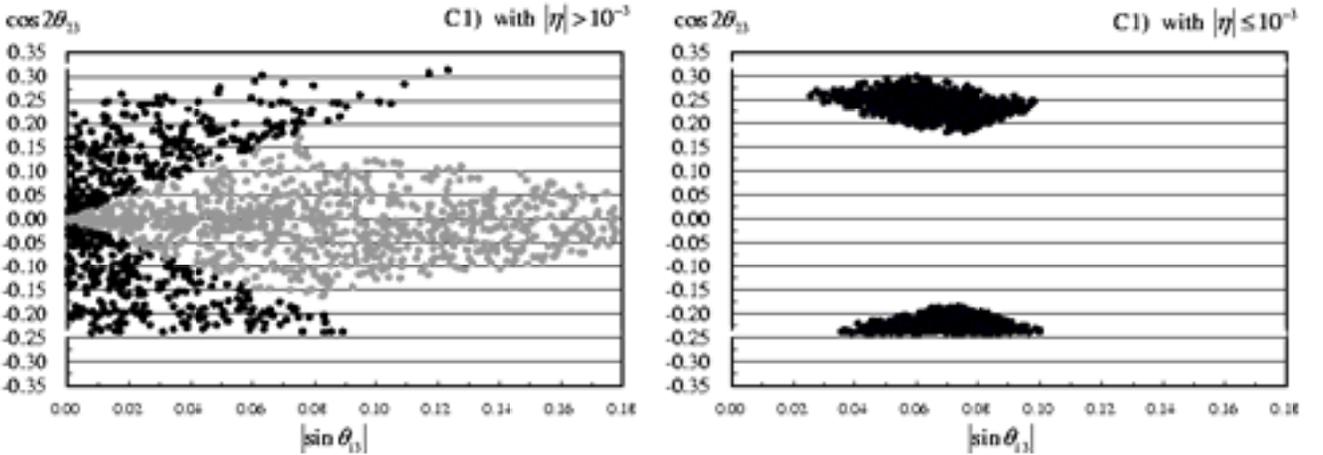}
\end{flushleft}
\vspace{-3mm}
\caption{The same as in FIG.\ref{Fig:inverted13-2-R} but for $\cos 2\theta_{23}$, where the black (grey) dots in the left figure correspond to
$r^\prime q > 0$ ($r^\prime q < 0$).}
\label{Fig:inverted13-2-cos23}
\end{figure}

Shown in FIG.\ref{Fig:inverted13-2-R}, and FIG.\ref{Fig:inverted13-2-cos23} are the predictions of $\Delta m^2_\odot/\Delta m^2_{atm}$,
 and $\cos 2\theta_{23}$ as functions of $\vert\sin\theta_{13}\vert$.   To enhance 
 the relationship of Eq.(\ref{Eq:Inverted-13-2-sin13-only}), 
we use two regions: one with $\vert\eta\vert > 0.001$ and the other $\vert\eta\vert \leq 0.001$ in the figures, 
In the region with $\vert\eta\vert \leq 0.001$, $\eta$ is hard to saturate the 
size of $\Delta m^2_\odot$/$\Delta m^2_{atm}$, which is ${\mathcal O}(0.01)$; therefore,
$\Delta m^2_\odot$/$\Delta m^2_{atm}$ should be determined by $\sin^2\theta_{13}$.
It is understood that FIG.\ref{Fig:inverted13-2-cos23} for $\vert\eta\vert > 0.001$ is 
produced by two contributions: one from $r^\prime >0$ marked by the black dots and the other from $r^\prime <0$ 
marked by the grey dots. Namely, 
\begin{itemize}
\item for $r^\prime >0$ giving $r^\prime  + q \neq 0$, the straight boundaries appear because of 
$\cos 2\theta_{23}\propto \sin\theta_{13}$, and
\item for $r^\prime <0$ giving $r^\prime  + q \sim 0$, $\cos 2\theta_{23}\sim 0$ is allowed,
\end{itemize}
as suggested by Eq.(\ref{Eq:Inverted-13-2-simple}).
In these figures, 
\begin{itemize}
\item  the contribution that gives $\Delta m^2_\odot/\Delta m^2_{atm} \approx\sin^2\theta_{13}$
 does manifest themselves for $\vert\eta\vert \leq 0.001$, 
\end{itemize}
which constrains $\vert\sin\theta_{13}\vert$ to be ${\mathcal O}(0.1)$.
  From FIG.\ref{Fig:inverted13-2-cos23} for $\vert\eta\vert \leq 0.001$, we find that this texture with 
  $\eta\sim 0$ can describes $\Delta m^2_\odot/\Delta m^2_{atm}\approx \sin^2\theta_{13}$
with
\begin{eqnarray}
&&
0.02\mapleq\vert\sin\theta_{13}\vert\mapleq 0.10,~{\rm and}~
-0.25\mapleq \cos 2\theta_{23}\mapleq -0.18,~{\rm or}~
0.18\mapleq \cos 2\theta_{23}\mapleq 0.30,
\label{Eq:Inverted-13-2-angle13-output}
\end{eqnarray}
where the range of $\vert\cos 2\theta_{23}\vert$ is in accord with Eq.(\ref{Eq:Inverted-13-2-simple}). 

For $\eta=0$, if we choose $q=3$ leading to $\sin^22\theta_{12}=9/11$, and $r^\prime = 2$, we obtain
\begin{eqnarray}
&&
\sin\theta_{13}\approx  -\frac{2\sqrt{2}\sigma\varepsilon}{11},
\quad
\cos 2\theta_{23} \approx \frac{10\varepsilon}{11},
\quad
\Delta m^2_\odot \approx 12d^2_0 \varepsilon^2,    
\quad
\Delta m^2_{atm} \approx 22d^2_0,
\label{Eq:Inverted-13-2-masses-angle1s}
\end{eqnarray}
which give
\begin{eqnarray}
&&
\frac{\Delta m^2_\odot}{\Delta m^2_{atm}}\approx 8\sin^2\theta_{13}.
\label{Eq:Inverted-13-2-R-angle13-2}
\end{eqnarray}
To be consistent with the observation of ${\Delta m^2_\odot}/{\Delta m^2_{atm}}$, we find that $0.06\mapleq \vert\sin\theta_{13} \vert\mapleq 0.07$.

\subsubsection{\label{subsec:3-2-2}Category C2) with $s_{23}\sim -\sigma /\sqrt{2}$}
There is only one texture that can explain the observed results.  The texture 
 yields $m_1\sim m_2$ and takes the following form:
\begin{eqnarray}
&&
M^{\rm C2)}_\nu=
d_0 \left( {\begin{array}{*{20}c}
   {2 - p\eta } & {\eta  + \varepsilon } & { - \sigma \left( {\eta  - \varepsilon }
 \right)}  \\
   {\eta  + \varepsilon } & {1 + r^\prime \varepsilon } & \sigma\left( 1-s\eta\right)   \\
   { - \sigma \left( {\eta  - \varepsilon } \right)} & \sigma\left( 1-s\eta\right)  & {1 - r^\prime 
\varepsilon }  \\
\end{array}} \right).
\label{Eq:Inverted-12-1}
\end{eqnarray}
The masses are calculated to be:
\begin{eqnarray}
&&
m_1  \approx 
\left( {2 - \frac{{2\left( p+s\right)\eta  - r^{\prime 2} \varepsilon ^2 }}{4} - \frac{{\sqrt 2 \varepsilon }}{{\sin 2\theta _{12} }}} \right)d_0,
\quad
m_2  \approx 
\left( {2 - \frac{{2\left( p+s\right)\eta  - r^{\prime 2} \varepsilon ^2 }}{4} + \frac{{\sqrt 2 \varepsilon }}{{\sin 2\theta _{12} }}} \right)d_0,
\nonumber\\
&&
m_3  \approx  \left( s\eta - \frac{r^{\prime 2}}{2} \varepsilon ^2 \right)d_0,
\label{Eq:Inverted-12-1-masses}
\end{eqnarray}
and
\begin{eqnarray}
&&
\Delta m_ \odot ^2  \approx \frac{{8\sqrt 2 \varepsilon d_0^2 }}{{\sin 2\theta _{12} }},
\quad
\Delta m^2_{atm}  \approx 4d_0^2,
\label{Eq:Inverted-12-1-squared}
\end{eqnarray}
and mixing angles are
\begin{eqnarray}
&&
\tan 2\theta _{12}  \approx \frac{{4\sqrt 2 \varepsilon }}{{2\left( p-s\right)\eta  + r^{\prime 2} \varepsilon ^2 }},
\quad
\tan 2\theta _{13}  \approx \sqrt 2 \sigma \eta,
\quad
\cos 2\theta_{23}  \approx r^\prime \varepsilon.
\label{Eq:Inverted-12-1-angles}
\end{eqnarray}

To obtain $\tan 2\theta_{12}={\mathcal{O}}(1)$, we require that 
$\vert\eta\vert \gg \varepsilon^2$, suggesting $\vert\eta\vert \sim \vert\varepsilon\vert$.
  This texture exhibits 
\begin{eqnarray}
&&
\cos 2\theta _{23} \approx \frac{\sigma\left(p-s\right) r^\prime\tan 2\theta_{12}}{2}\sin\theta _{13},
\label{Eq:Inverted-12-1-atm-angle}
\end{eqnarray}
which suggests $\tan 2\theta_{12} \sim \sin\theta _{13}/\cos 2\theta _{23}$ similar to 
Eq.(\ref{Eq:Normal-12-1-simple-2}) for C2) of the normal mass hierarchy. 
There is also another correlation
\begin{eqnarray}
\frac{{\Delta m_ \odot ^2 }}{{\Delta m_{atm}^2 }} \approx \frac{{2\sqrt 2 }}{{r'\sin 2\theta _{12} }}\cos 2\theta _{23}. 
\label{Eq:Inverted-12-1-simple-2}
\end{eqnarray}
These two relations show
\begin{eqnarray}
&&
\frac{{\Delta m_ \odot ^2 }}{{\Delta m_{atm}^2 }}
\approx
\frac{{\sqrt 2 \sigma \left( {p - s} \right)}}{{\cos 2\theta _{12} }}\sin \theta _{13}. 
\label{Eq:Inverted-12-R13}
\end{eqnarray}

The predictions $\Delta m^2_\odot/\Delta m^2_{atm}$ and $\cos 2\theta_{23}$ are 
depicted in FIG.\ref{Fig:inverted-R}-C2) and FIG.\ref{Fig:inverted-cos23}-C2) as 
functions of $\vert\sin\theta_{13}\vert$. In these figures,  $\sin\theta_{13}$ and $\cos 2\theta_{23}$ are constrained to satisfy 
\begin{eqnarray}
&&
0.002\mapleq \vert\sin\theta_{13}\vert\left( \mapleq 0.18\right),
\quad
0.005 \mapleq  \vert \cos 2\theta_{23} \vert \mapleq 0.05.
\label{Eq:Inverted-12-1-sin13}
\end{eqnarray}
The result is expected because
\begin{itemize}
\item  $\cos 2\theta_{23}$ determined by $\varepsilon$, whose magnitude is about $\sim 0.01-0.02$ from Eq.(\ref{Eq:Inverted-12-1-squared}), 
yields $\vert\cos 2\theta_{23}\vert\sim 0.03-0.06$ for $r^\prime = 3$,
\item  $\sin\theta_{13}$ determined by $\eta$, whose magnitude can be as large as 0.1 if $\vert p-s\vert\sim 0.1$
from Eq.(\ref{Eq:Inverted-12-1-angles}) to balance $\varepsilon\sim 0.01(\sim \Delta m^2_\odot/\Delta m^2_{atm})$, can have a broad range up to ${\mathcal O}(0.1)$.
\end{itemize}
This texture allows $\vert\eta\vert ={\mathcal O}(1)$, where the expressions of the masses and mixing angles based on 
$\eta^2 \ll 1$ are not appropriate.  We have to use the correct ones listed in the Appendix \ref{sec:Appendix}.  
The large $\vert\eta\vert$ region is marked by the grey dots in the figures.

\subsection{\label{subsec:3-3}Quasi Degenerate Mass Pattern I}
\subsubsection{\label{subsec:3-3-1}Category C1) with $s_{23}\sim\sigma /\sqrt{2}$}
This texture is only possible for C1). 
The mass ordering is given by $\vert m_1\vert\sim \vert m_2\vert\sim\vert m_3\vert \sim {\mathcal{O}}(\sqrt{\Delta m^2_{atm}})$.  
To describe $\Delta m^2_\odot \ll \Delta m^2_{atm}$, the relation $m_1\sim -m_2$ is imposed \cite{PlusMinusNu} 
 instead of suppressing both $m_1$ and $m_2$.  In this texture, both mass patterns $\vert m_1\vert<\vert m_2\vert<\vert m_3\vert$
 and $\vert m_3\vert<\vert m_1\vert<\vert m_2\vert$ are described consistently and are treated separately in our figures.  The 
texture is specified by 
\begin{eqnarray}
&&
M^{\rm C1)}_\nu
=
d_0 \left( {\begin{array}{*{20}c}
   { - (2 - \eta)} & {q + \varepsilon } & { - \sigma \left( {q - \varepsilon } 
\right)}  \\
   {q + \varepsilon } & 1 - r + r^\prime\varepsilon & { - \sigma \left( {1 + r} 
\right)}  \\
   { - \sigma \left( {q - \varepsilon } \right)} & { - \sigma \left( {1 + r} 
\right)} & 1 - r - r^\prime\varepsilon   \\
\end{array}} \right).
\label{Eq:Normal-13-2}
\end{eqnarray}
It should be noted that the texture with $r=0$ coincides with the texture Eq.(\ref{Eq:Inverted-13-2}) 
for the inverted mass hierarchy. The 
corresponding texture for C2) is obtained by replacing $\sigma$ by $-\sigma$ in 
$M_{\mu\tau}$ of Eq.(\ref{Eq:Normal-13-2}).  However, this replacement yields 
$\tan 2\theta_{13}={\mathcal{O}}(1)$.  Therefore, the texture for C2) with $m_1\sim -m_2$ is not phenomenologically 
allowed. 

The masses are calculated to be:
\begin{eqnarray}
&&
 m_1  \approx  - \left( {\frac{{\sqrt 2 q}}{{\sin 2\theta _{12} }} - \frac{{\eta  + 2\left( {r - 1} \right)t_{13}^2 
 - 2\left( {\left( {1 + r} \right)\Delta  - \varepsilon r'} \right)\Delta }}{2}} \right)d_0, 
\nonumber\\
&&
 m_2  \approx  \left( {\frac{{\sqrt 2 q}}{{\sin 2\theta _{12} }} + \frac{{\eta  + 2\left( {r - 1} \right)t_{13}^2 
 - 2\left( {\left( {1 + r} \right)\Delta  - \varepsilon r'} \right)\Delta }}{2}} \right)d_0, 
\nonumber\\
&&
 m_3  \approx \left[ { - 2r + 2\left( {1 - r} \right)t_{13}^2  
 + 2\left( {\left( {1 + r} \right)\Delta  - \varepsilon r'} \right)\Delta } \right]d_0, 
\label{Eq:Normal-13-2-masses}
\end{eqnarray}
and
\begin{eqnarray}
&&
\Delta m_ \odot ^2  \approx
\frac{{2\sqrt 2 q\left[ {\eta  + 2\left( {r - 1} \right)t_{13}^2  - 2\left( {\left( {1 + r} \right)\Delta  - \varepsilon r^\prime } \right)\Delta } \right]}}{{\sin 2\theta _{12} }}d_0^2,
\quad
\Delta m_{atm}^2 \approx 4\left| r^2  - \frac{2q^2 }{\sin ^2 2\theta _{12} } \right|d_0^2.
\label{Eq:Normal-13-2-squared}
\end{eqnarray}
The mixing angles are given by
\begin{eqnarray}
&&
\tan 2\theta _{12}  \approx \frac{q}{{\sqrt 2 }},
\quad
\tan 2\theta _{13}  \approx  \frac{{\sqrt 2 \sigma \left( {2 + 2r - qr^\prime} 
\right)}}{{q^2  - 2\left( {r^2  - 1} \right)}}\varepsilon,
\quad
\cos 2\theta_{23} \approx 2\Delta,
\label{Eq:Normal-13-2-angles}
\end{eqnarray}
where
\begin{eqnarray}
&&
\Delta  = \frac{{q + r^\prime \left( {1 - r} \right)}}{{q^2  - 2\left( {r^2  - 1}
 \right)}}\varepsilon,
\label{Eq:Normal-13-2-Delta}
\end{eqnarray}
from which $q (\approx\sqrt 2\tan 2\theta_{12})\sim 3$ is expected.

There are two distinct realizations of $\Delta m^2_\odot/\Delta m^2_{atm}\ll 1$.  
In the estimation of $\Delta m_ \odot ^2$, if future experiments observe that 
$\sin^2\theta_{13}\approx 0$, we obtain that
\begin{eqnarray}
&&
\Delta m_ \odot ^2  \approx \frac{{4\eta }}{{\cos 2\theta _{12} }}d_0^2, 
\quad
\eta \sim \frac{\Delta m^2_ \odot}{\Delta m^2_{atm}},
\label{Eq:Normal-13-2-eta}
\end{eqnarray}
leading to $\eta = {\mathcal{O}}(0.01)$, where we have used $q \approx\sqrt 2\tan 2\theta_{12}$.  
However, if $\sin^2\theta_{13}={\mathcal{O}}(10^{-2})$,
 we find an interesting possibility that $\sin^2\theta_{13}$ takes care of the 
mass hierarchy.  Suppose that $\eta=0$ in the texture, then, we reach
\begin{eqnarray}
&&
\Delta m_ \odot ^2  \approx \frac{4\left[ 2\left( {r - 1} \right)s^2_{13}  - \left( {
\left( {3 + r} \right)\Delta  - 2\varepsilon r^\prime} \right)\Delta \right]}{{\cos 2
\theta _{12} }}d_0^2.
\label{Eq:Normal-13-2-DeltaSun}
\end{eqnarray}
Because $\tan 2\theta_{13}$, and $\Delta$ are proportional to $\varepsilon$, 
$\Delta m_\odot$ is proportional to $\sin^2\theta_{13}$, leading to 
\begin{eqnarray}
&&
\frac{\Delta m^2_ \odot}{\Delta m^2_{atm}}\sim \sin^2\theta_{13}.
\label{Eq:Normal-13-2-DeltaSun-final}
\end{eqnarray}
There is also the proportionality of $\cos 2\theta_{23}$ to $\sin \theta _{13}$. which is given by
\begin{eqnarray}
&&
\cos 2\theta_{23}\approx \frac{{2\sqrt 2 \sigma \left[ {q + r^\prime \left( {1 - 
r} \right)} \right]}}{{2 + 2r - qr^\prime}}\sin \theta _{13}.
\label{Eq:Normal-13-2-simple}
\end{eqnarray}
Since the neutrino masses are roughly controlled by 
$m_2 \sim m_1 \sim 2d_0/\cos 2\theta_{12}$ for $q \approx\sqrt 2\tan 2\theta_{12}$ and $m_3\sim -2r d_0$, 
we have the normal mass ordering: $\vert m_1\vert < \vert m_2\vert < \vert m_3\vert$ 
if $\vert r\cos 2\theta_{12}\vert > 1$, and the inverted mass ordering: 
$\vert m_3\vert < \vert m_1\vert < \vert m_2\vert$ if $\vert r\cos 2\theta_{12}\vert < 1$.

\begin{figure}[!htbp]
\begin{flushleft}
\includegraphics*[20mm,203.6mm][300mm,264mm]{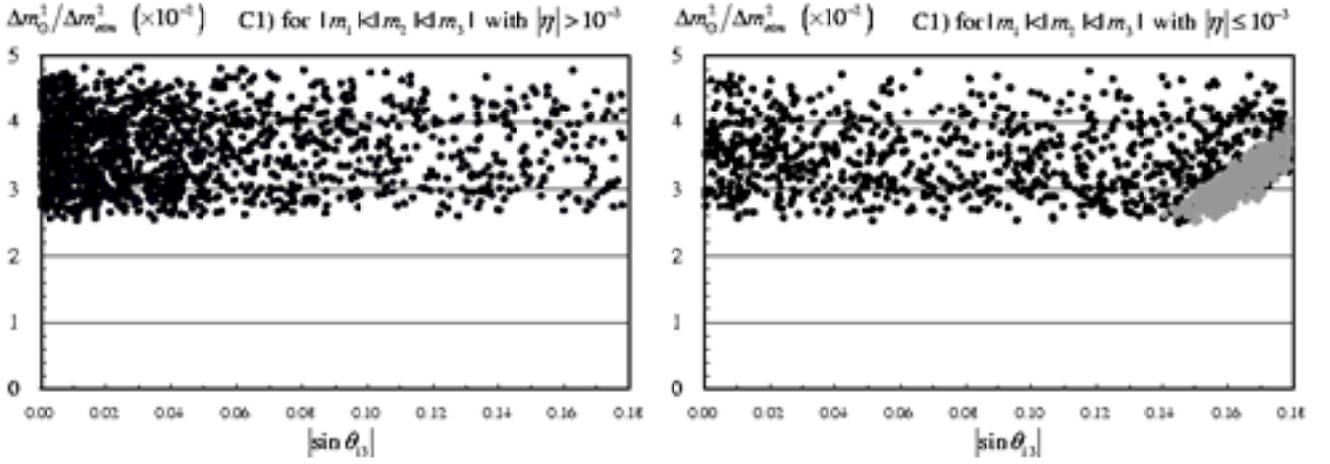}
\end{flushleft}
\vspace{-3mm}
\caption{The same as in FIG.\ref{Fig:normal-R} but in the quasi degenerate mass pattern I for C1) with the normal mass ordering $\vert m_1\vert < \vert m_2\vert < \vert m_3\vert$,  
where two regions of $\vert\eta\vert>0.001$ and $\vert\eta\vert\leq 0.001$ are separately shown. The black (grey) dots in the right figure correspond to $r^\prime > 0$ ($r^\prime < 0$).}
\label{Fig:normal13-2-R-plus}
\end{figure}

\begin{figure}[!htbp]
\begin{flushleft}
\includegraphics*[20mm,203.6mm][300mm,264mm]{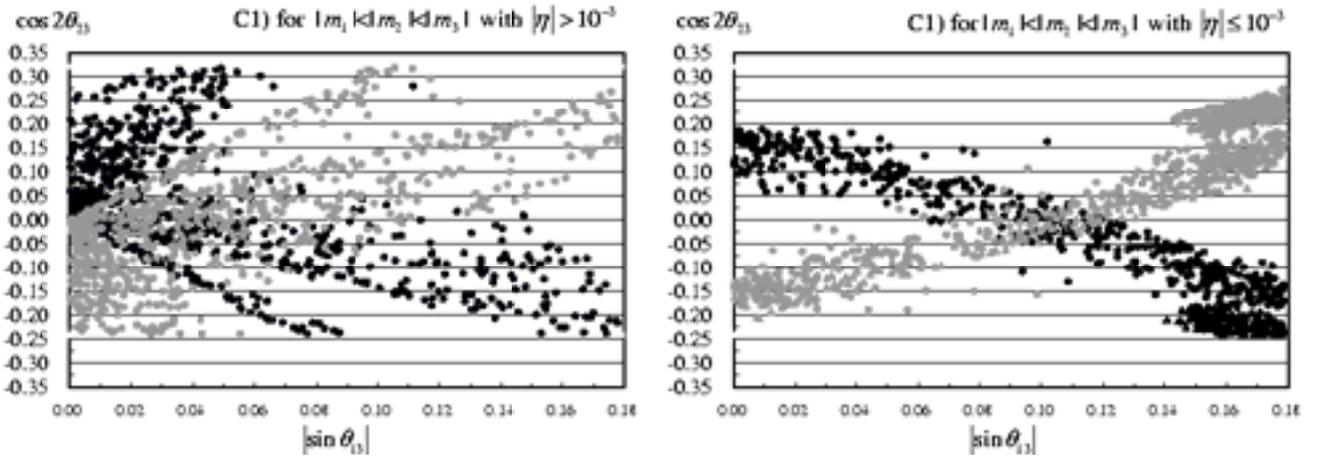}
\end{flushleft}
\vspace{-3mm}
\caption{The same as in FIG.\ref{Fig:normal13-2-R-plus} but for $\cos 2\theta_{23}$, where the black (grey) marks correspond to
$\epsilon > 0$ ($\epsilon < 0$) and, in the right figure, the dots (triangle marks
 for $\vert\sin\theta_{13}\vert \geq 0.14$ and $0.17\leq\vert\cos 2\theta_{23}\vert\leq 0.28$) correspond to $r^\prime > 0$ ($r^\prime < 0$).}
\label{Fig:normal13-2-cos23-plus}
\end{figure}

\begin{figure}[!htbp]
\begin{flushleft}
\includegraphics*[20mm,203.6mm][300mm,264mm]{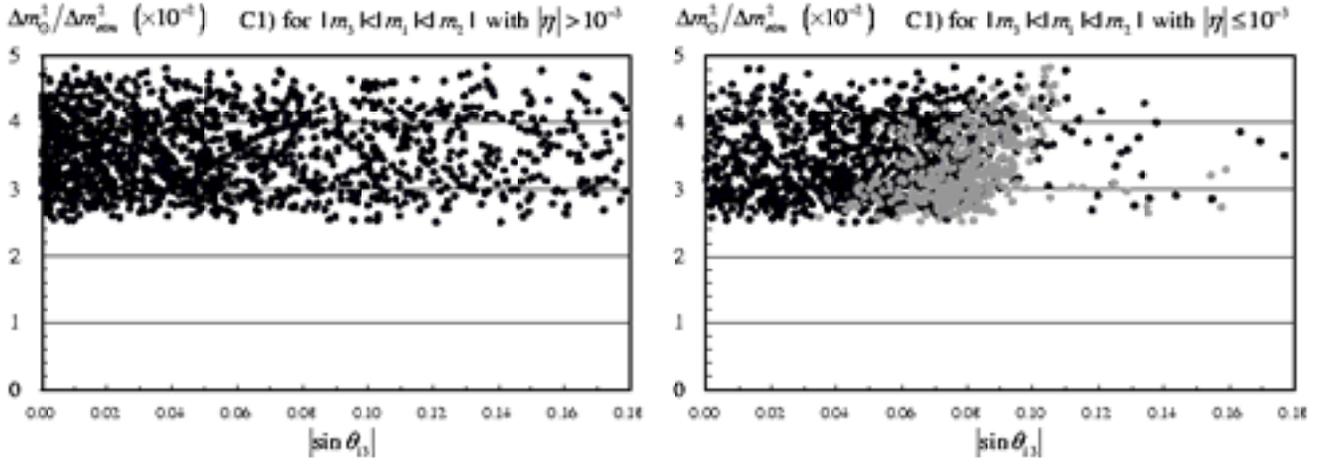}
\end{flushleft}
\vspace{-3mm}
\caption{The same as in FIG.\ref{Fig:normal13-2-R-plus} but with the inverted mass ordering $\vert m_3\vert < \vert m_1\vert < \vert m_2\vert$
, where the black (grey) dots in the right figure correspond to $r > 0$ ($r <0$)\label{Fig:normal13-2-R-minus}.}
\end{figure}

\begin{figure}[!htbp]
\begin{flushleft}
\includegraphics*[20mm,203.6mm][300mm,264mm]{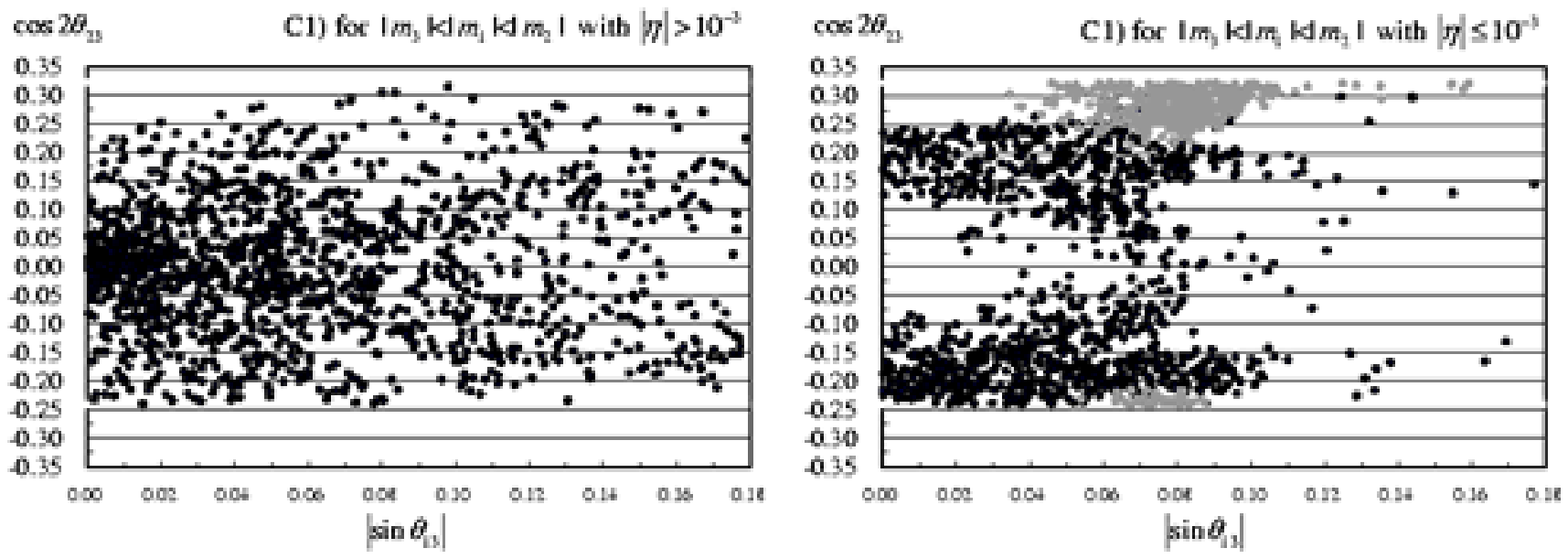}
\end{flushleft}
\vspace{-3mm}
\caption{The same as in FIG.\ref{Fig:normal13-2-R-minus} but for $\cos 2\theta_{23}$.}
\label{Fig:normal13-2-cos23-minus}
\end{figure}

These features are reflected in the predictions $\Delta m^2_\odot/\Delta m^2_{atm}$ and $\cos 2\theta_{23}$ 
depicted in FIG.\ref{Fig:normal13-2-R-plus} and FIG.\ref{Fig:normal13-2-cos23-plus} for 
the normal mass ordering $\vert m_1\vert < \vert m_2\vert < \vert m_3\vert$ and 
in FIG.\ref{Fig:normal13-2-R-minus} and FIG.\ref{Fig:normal13-2-cos23-minus} for 
the inverted mass ordering $\vert m_3\vert < \vert m_1\vert < \vert m_2\vert$. 
We again use two regions: one with $\vert\eta\vert > 0.001$ and the other $\vert\eta\vert \leq 0.001$ 
to enhance the relationship of Eq.(\ref{Eq:Normal-13-2-DeltaSun-final}). 
In the normal mass ordering, there are two clusters around $\sin\theta_{13}\sim 0$ and $\sin\theta_{13}\sim 0.18$. 
In the first cluster for $\eta\neq 0$, 
\begin{itemize}
\item the contribution from $\eta$ in $\Delta m^2_\odot$ is enhanced because $\varepsilon \sim 0$ for $\sin\theta_{13}\sim 0$
\end{itemize}
while in the second cluster for $\vert\eta\vert \leq 0.001$, 
\begin{itemize}
\item the contribution from $\varepsilon^2$ in $\Delta m^2_\odot$ is enhanced because $\eta \sim 0$. 
\end{itemize}
Furthermore, if the texture is restricted to satisfy $r^\prime <0$, the second cluster consisting of 
the grey dots in FIG.\ref{Fig:normal13-2-R-plus} 
becomes manifest itself.  Because $1-r < 0$ for the normal mass ordering with $r\cos 2\theta_{12} > 1$, 
$r^\prime (1-r) > 0$ for $r^\prime <0$ is satisfied in the numerator in Eq.(\ref{Eq:Normal-13-2-Delta}) for $\cos 2\theta_{23}$.  
Therefore, 
this negative $r^\prime$ can enhance the size of $\vert\cos 2\theta_{23}\vert$.  
This region is shown by the triangle marks in FIG.\ref{Fig:normal13-2-cos23-plus} with $\vert\eta\vert \leq 0.001$ 
and arises from the relation $\Delta m^2_ \odot/\Delta m^2_{atm}\sim \sin^2\theta_{13}$, leading to the cluster around
\begin{eqnarray}
&&
0.14\mapleq\vert\sin\theta_{13}\vert\mapleq 0.18, 
\quad
0.20\mapleq\vert\cos 2\theta_{23}\vert\mapleq 0.25, 
\label{Eq:Normal-13-2-sin13-simulation}
\end{eqnarray}
for the normal mass ordering. In the same figure for $r^\prime > 0$, there are points giving 
1) $\vert\cos 2\theta_{23}\vert\sim 0$ 
around $\vert\sin\theta_{13}\vert \sim 0.1$ and 2) $\vert\cos 2\theta_{23}\vert\sim 0.15$ 
around $\vert\sin\theta_{13}\vert \sim 0$.  These points arise because
\begin{itemize}
\item for 1), $\cos 2\theta_{23}$ in Eq.(\ref{Eq:Normal-13-2-simple}) vanishes at $q \approx  - r^\prime (1 - r)$ for $r > 1$, 
leading to $\tan 2\theta _{13}  \approx \sqrt 2 \sigma\varepsilon/(1 - r)$,
which is about $\sin\theta _{13}  \approx \sigma\varepsilon/[\sqrt 2 (1 - r)]$ numerically found to be 
around 0.1, namely around ${\mathcal O}(\sqrt{{\Delta m^2_ \odot}/{\Delta m^2_{atm}}})$,  
\item for 2), $\Delta m^2_\odot$ in Eq.(\ref{Eq:Normal-13-2-DeltaSun}) roughly requires that 
$\vert\Delta\vert\sim\vert\varepsilon\vert\sim 0.1$, namely, $\vert\cos 2\theta_{23}\vert\sim 0.1$,  
if $\sin\theta_{13}\sim 0$, which can be given by $2 + 2r - qr^\prime \sim 0$ even 
if $\varepsilon \neq 0$ in Eq.(\ref{Eq:Normal-13-2-angles}).
\end{itemize}
On the other hand, in the inverted mass ordering, there 
is one cluster consisting of the grey dots in FIG.\ref{Fig:normal13-2-cos23-minus} 
with $\vert\eta\vert \leq 0.001$, which gives the cluster around
\begin{eqnarray}
&&
0.05\mapleq\vert\sin\theta_{13}\vert\mapleq 0.10, 
\quad
0.20\mapleq\cos 2\theta_{23}\mapleq 0.32,
\label{Eq:Normal-13-2-sin13-simulation-2}
\end{eqnarray}
if $r$ is restricted to satisfy $r < 0$.  The reason to find this cluster is the same as that for the normal mass ordering. 
The positive $r^\prime (1-r) > 0$ is obtained for $r < 0$ because $r^\prime$ is found to be positive in this region.

\begin{figure}[!htbp]
\begin{flushleft}
\includegraphics*[20mm,203.6mm][300mm,264mm]{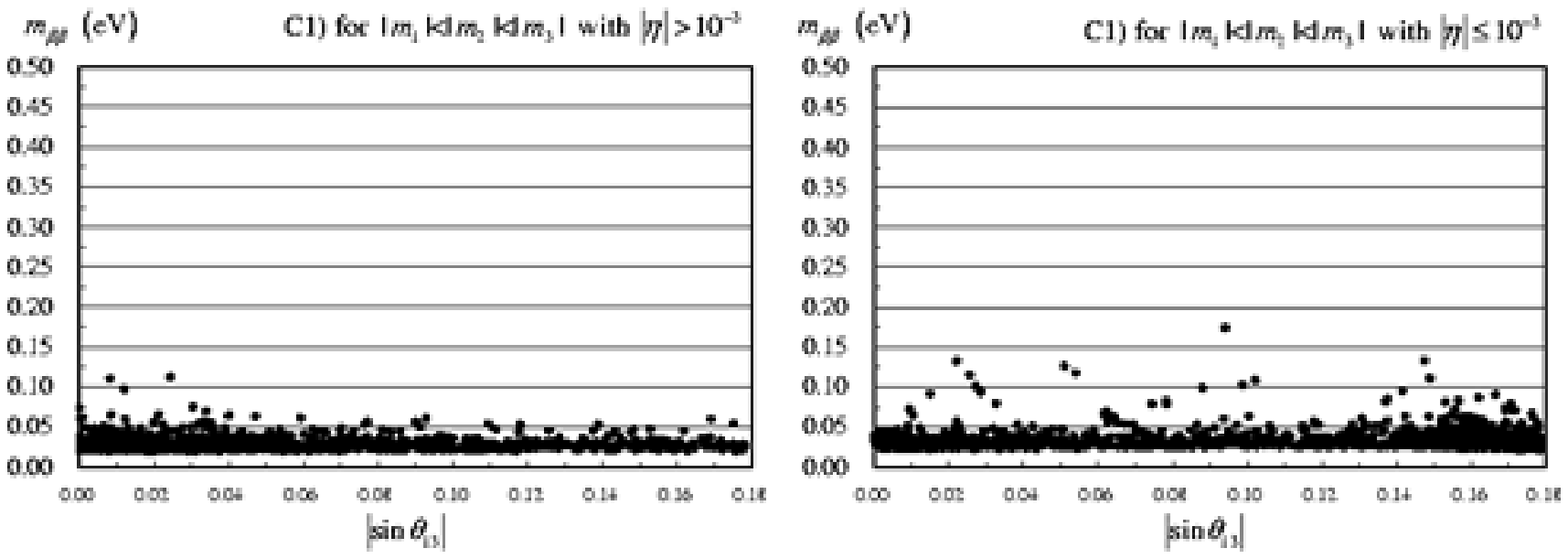}
\end{flushleft}
\vspace{-3mm}
\caption{The same as in FIG.\ref{Fig:normal13-2-R-plus} but for $m_{\beta\beta}$.}
\label{Fig:normal13-2-ee-plus}
\end{figure}

\begin{figure}[!htbp]
\begin{flushleft}
\includegraphics*[20mm,203.6mm][300mm,264mm]{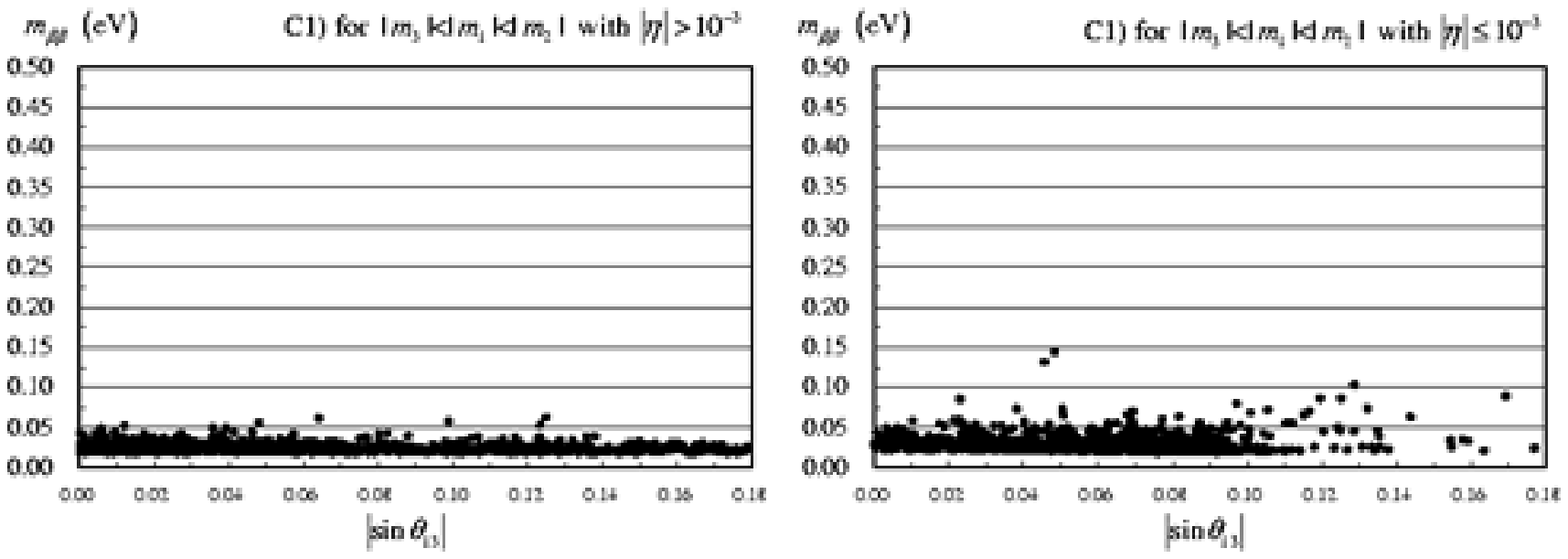}
\end{flushleft}
\vspace{-3mm}
\caption{The same as in FIG.\ref{Fig:normal13-2-R-minus} but for $m_{\beta\beta}$.}
\label{Fig:normal13-2-ee-minus}
\end{figure}

In our texture, the flavor mass of $M_{ee}$ estimated to be $-(2-\eta )d_0$ 
corresponds to the effective neutrino mass $m_{\beta\beta}$ \cite{TheoryMass-ee}
 used in the detection of the absolute neutrino mass \cite{AbsoluteMass}.  Since $d_0 \sim \sqrt{\Delta m^2_{atm}}$,
 the size of $m_{\beta\beta}$ is roughly 0.1 eV.
 Two figures shown in FIG.\ref{Fig:normal13-2-ee-plus} and FIG.\ref{Fig:normal13-2-ee-minus} are the predictions 
of $m_{\beta\beta}$, respectively, 
for the normal mass ordering $\vert m_1\vert < \vert m_2\vert < \vert m_3\vert$, 
and the inverted mass ordering $\vert m_3\vert < \vert m_1\vert < \vert m_2\vert$.
  From these figures, $m_{\beta\beta}$ is estimated to give $m_{\beta\beta}\mapleq$
0.05 eV as expected, and at most $m_{\beta\beta}\sim$0.15 eV for $\vert\eta\vert \leq 0.001$, which 
arises for $r\sim 1/\cos\theta_{12}$ that yields the larger $d_0$ from Eq.(\ref{Eq:Normal-13-2-squared}).

For the typical prediction with $\eta=0$, let us choose $q$ = $r$ = 3, thus 
giving the normal mass ordering, and $r^\prime=-1$, where $\sin^22\theta_{12}=9/11$ is obtained, 
and we obtain that
\begin{eqnarray}
&&
\sin\theta_{13}\approx - \frac{11\sigma\varepsilon}{7\sqrt{2}},
\quad
\cos 2\theta_{23} = -\frac{10\varepsilon}{7},
\quad
\Delta m^2_\odot \approx 21d^2_0 \varepsilon^2,
\quad
\Delta m^2_{atm} = 14d^2_0,
\nonumber\\
&&
m_2\approx -m_1\approx \sqrt{\frac{11\Delta m^2_{atm}}{7}},
\quad
m_3\approx -\sqrt{\frac{18\Delta m^2_{atm}}{7}},
\quad
m_{\beta\beta}\approx \sqrt{\frac{2\Delta m^2_{atm}}{7}},
\label{Eq:Normal-13-2-masses-angle1s}
\end{eqnarray}
leading to
\begin{eqnarray}
&&
\frac{\Delta m^2_\odot}{\Delta m^2_{atm}}\approx 1.2\sin^2\theta_{13},
\label{Eq:Normal-13-2-R-angle13-output}
\end{eqnarray}
which gives $0.15\mapleq \vert\sin\theta_{13}\vert\mapleq 0.18$ giving $0.12\mapleq \vert\cos 2\theta_{23}\vert\mapleq 0.15$ 
to recover the observed values of $\Delta m^2_\odot/\Delta m^2_{atm}$, which is consistent with the region of the cluster 
near $\vert\sin\theta_{13}\vert\sim 0.18$.

\subsection{\label{subsec:3-4}Quasi Degenerate Mass Pattern II}
The mass ordering is given by $\vert m_1\vert\sim \vert m_2\vert\sim\vert m_3\vert$
 with the hierarchy of $\vert m^2_i-m^2_j\vert \ll m^2_{1,2,3}$ ($i,j=1,2,3$).  For our numerical calculation, 
 we use $\vert m_{1,2,3}\vert \geq \sqrt{\Delta m^2_{atm}}$ and 
 we further require $4\vert\eta\vert \leq 1/3$ to favor  
 $\vert m^2_i-m^2_j\vert \ll m^2_{1,2,3}$ ($i,j=1,2,3$) because $m^2_{1,2,3}\sim\Delta m^2_{atm}/4\vert \eta\vert$ is 
 to be obtained.
Since there should be three distinct mass scales, textures will involve terms proportional to 
$\eta^2$, and $\eta\varepsilon$ in addition to the ordinary terms proportional to $\eta$, and 
$\varepsilon$ to produce the hierarchy. 
Textures are characterized by the relative sign of $m_3$ as $m_1 \sim m_2 \sim m_3$ or $m_1 \sim m_2 \sim -m_3$.  
In textures with $m_1\sim m_2\sim -m_3$, the flavor neutrino masses should satisfy that 
$\vert d - \sigma e - a\vert \mapleq {\mathcal O}(\eta^2)$ as well as 
$b + \varepsilon\Delta b^\prime= {\mathcal O}(\eta^2,\varepsilon^2)$ for C1) 
and $\vert d + \sigma e - a\vert \mapleq {\mathcal O}(\eta^2)$ as well as 
$\varepsilon b^\prime +\Delta b = {\mathcal O}(\eta\varepsilon)$ for C2).
The basic structure for each mass matrix is given by
\begin{eqnarray}
&&
\left( {\begin{array}{*{20}c}
   1 & 0 & 0  \\
   0 & 1 & 0  \\
   0 & 0 & 1  \\
\end{array}} \right), \quad
\left( {\begin{array}{*{20}c}
   1 & 0 & 0  \\
   0 & 0 & 1  \\
   0 & 1 & 0  \\
\end{array}} \right),
\label{Eq:Degenerate-13-12}
\end{eqnarray}
for the textures with $m_1 \sim m_2 \sim m_3$ and with $m_1 \sim m_2 \sim -m_3$, respectively.  The latter texture has 
an approximate $L_\mu -L_\tau$ conservation, whose influence on neutrino oscillations has been lately discussed in Ref.\cite{LeLmuPlot,LmueMinusLtau}. 

\subsubsection{\label{subsubsec:3-4-1}$m_1 \sim m_2 \sim m_3$}
\centerline{\label{subsec:3-4-1-1}{\small  \it 1-1. Category C1) with $s_{23}\sim\sigma /\sqrt{2}$}}
The texture is given by
\begin{eqnarray}
&&
M^{\rm C1)}_\nu
=
d_0 \left( {\begin{array}{*{20}c}
   {1 - \eta } & {\left( {q\eta  + \varepsilon } \right)\eta } & { - \sigma \left
( {q\eta  - \varepsilon } \right)\eta }  \\
   {\left( {q\eta  + \varepsilon } \right)\eta } & {1 + r^\prime \varepsilon \eta }
 & {\sigma \left( {1 - s\eta } \right)\eta }  \\
   { - \sigma \left( {q\eta  - \varepsilon } \right)\eta } & {\sigma \left( {1 - 
s\eta } \right)\eta } & {1 - r^\prime \varepsilon \eta }  \\
\end{array}} \right).
\label{Eq:Degenerate-13-1}
\end{eqnarray}
Neutrino masses are predicted to be:
\begin{eqnarray}
&&
m_1  \approx \left( 1 - \eta - \frac{\sqrt 2 \left(2q\eta  - r^\prime \varepsilon ^2 \right)\eta }{2\sin 2\theta _{12} }\right)d_0,
\quad
m_2  \approx \left( 1 - \eta + \frac{\sqrt 2 \left(2q\eta  - r^\prime \varepsilon ^2 \right)\eta }{2\sin 2\theta _{12} }\right)d_0,
\nonumber \\
&&
m_3  \approx \left( {1 + \eta} \right)d_0,
\label{Eq:Degenerate-13-1-masses}
\end{eqnarray}
and
\begin{eqnarray}
&&
\Delta m_ \odot ^2  \approx \frac{2\sqrt 2 \left(2q\eta  - r^\prime \varepsilon ^2 \right)\eta d_0^2 }{\sin 2\theta _{12} },
\quad
\Delta m_{atm}^2  \approx 4 \vert\eta\vert d_0^2,
\label{Eq:Degenerate-13-1-squared}
\end{eqnarray}
and mixing angles are
\begin{eqnarray}
&&
\tan 2\theta _{12}  \approx \frac{2\sqrt 2 \left( 2q\eta  - r^\prime \varepsilon ^2  \right)}{2s\eta  + \left( {2 - r^{\prime 2} } \right)\varepsilon ^2 },
\quad
\tan 2\theta _{13}  \approx \sqrt 2 \sigma \varepsilon, 
\quad
\cos 2\theta_{23}  \approx  -r^\prime \varepsilon.
\label{Eq:Degenerate-13-1-angles}
\end{eqnarray}

\begin{figure}[!htbp]
\begin{flushleft}
\includegraphics*[20mm,131.5mm][300mm,315mm]{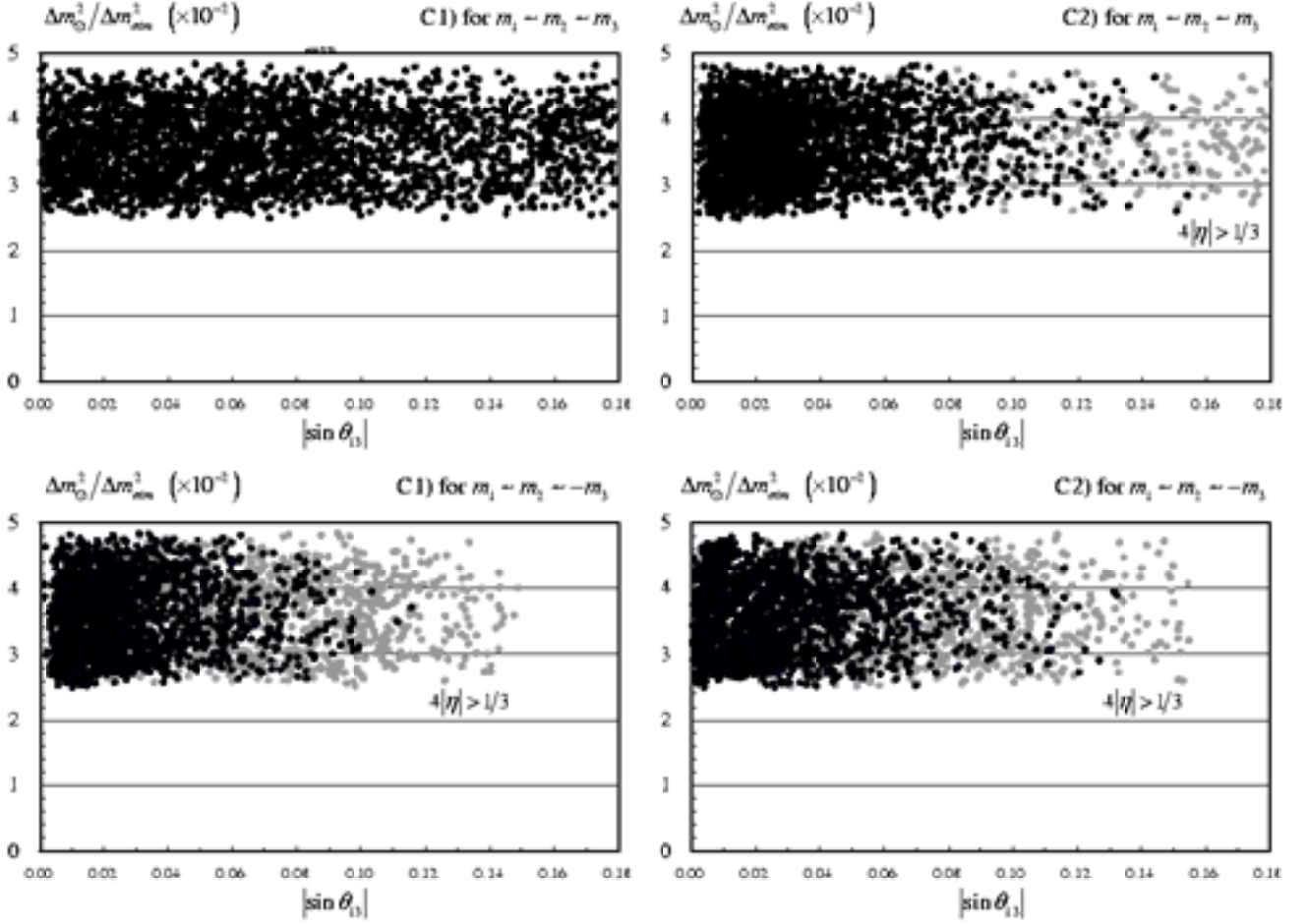}
\end{flushleft}
\vspace{-3mm}
\caption{The same as in FIG.\ref{Fig:normal-R} but for the quasi degenerate mass pattern II either with $m_1\sim m_2\sim m_3$ 
or $m_1\sim m_2\sim -m_3$, where the grey dots represent the region with $4\vert \eta\vert> 1/3$, which disturbs the condition 
$m^2_{1,2,3}\gg\Delta m^2_{atm}$.}
\label{Fig:degenerate-R}
\end{figure}

\begin{figure}[!htbp]
\begin{flushleft}
\includegraphics*[20mm,131.5mm][300mm,315mm]{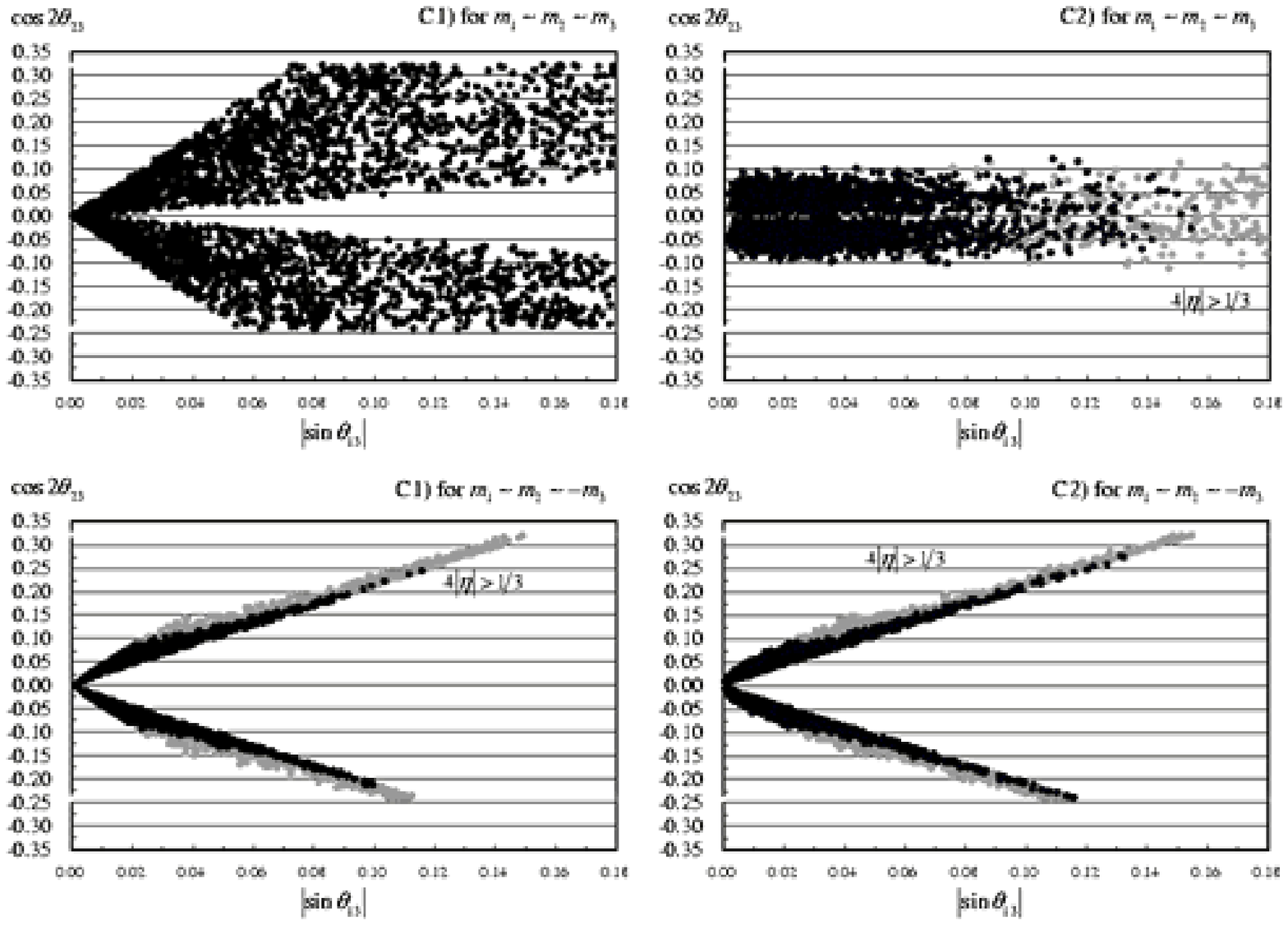}
\end{flushleft}
\vspace{-3mm}
\caption{The same as in FIG.\ref{Fig:degenerate-R} but for $\cos 2\theta_{23}$.}
\label{Fig:degenerate-cos23}
\end{figure}

The parameter $\varepsilon$ relates $\cos 2\theta_{23}$ to $\sin\theta_{13}$:
\begin{eqnarray}
\cos 2\theta_{23}  \approx -\sqrt 2\sigma r^\prime\sin\theta _{13},
\label{Eq:Degenerate-13-1-simple}
\end{eqnarray}
which indicates $\vert\cos 2\theta_{23}\vert\sim\vert\sin\theta _{13}\vert$.
It should be noted that the predictions of $\Delta m^2_{atm}/\Delta m^2_\odot$ and $\cos 2\theta_{23}$ are 
very similar to those for the inverted mass hierarchy in C1). 
Therefore, the figures FIG.\ref{Fig:degenerate-R}-C1) and FIG.\ref{Fig:degenerate-cos23}
-C1) for $m_1\sim m_2\sim m_3$ are almost identical to FIG.\ref{Fig:inverted-R}-C1) and FIG.\ref{Fig:inverted-cos23}
-C1).

\begin{figure}[!htbp]
\begin{flushleft}
\includegraphics*[20mm,131.5mm][300mm,315mm]{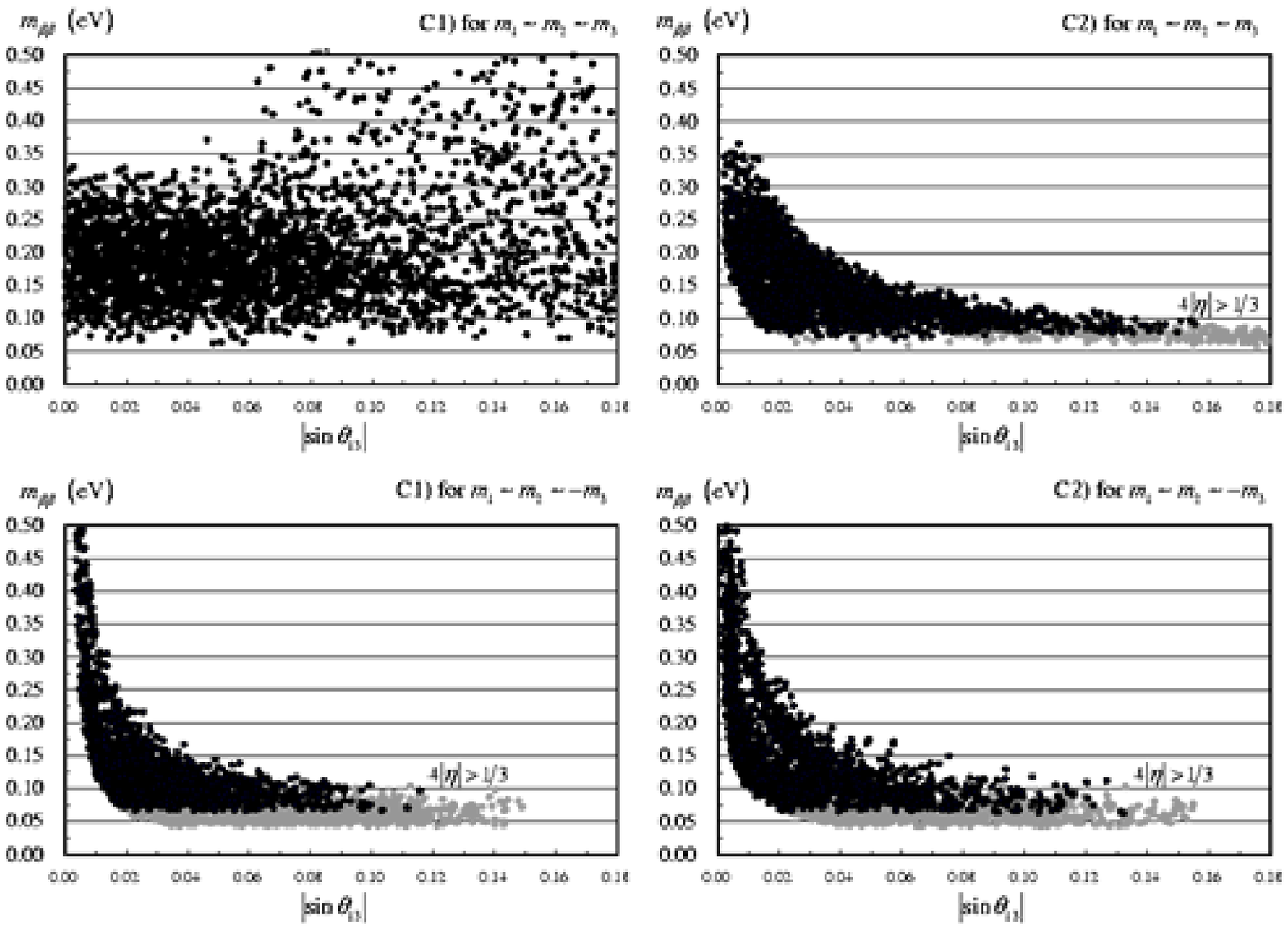}
\end{flushleft}
\vspace{-3mm}
\caption{The same as in FIG.\ref{Fig:degenerate-R} but for $m_{\beta\beta}$.}
\label{Fig:degenerate-ee}
\end{figure}

The largest size of the mass scales is $d_0$, which is approximately $m_{\beta\beta}(=M_{ee})$. Then, we find the following  
mass ordering:
\begin{eqnarray}
&&
m^2_{\beta\beta}:\Delta m^2_{atm}:\Delta m^2_\odot\sim 1:\vert\eta\vert:\left(\eta  - \frac{r^\prime \varepsilon ^2}{2}\right)\eta.
\label{Eq:Degenerate-13-1-mass-ordering}
\end{eqnarray}
Shown in FIG.\ref{Fig:degenerate-ee}-C1) for $m_1\sim m_2\sim m_3$ is the estimation of $m_{\beta\beta}$  as a function 
of $\vert\sin\theta_{13}\vert$.  We observe that $m_{\beta\beta}$ can take larger values as $\vert\sin\theta_{13}\vert$ gets larger.
This is because
\begin{itemize}
\item $\vert\eta\vert$ can be as small as possible if $\vert\sin\theta_{13}\vert={\mathcal O}(0.01)$,
\end{itemize}
where the size of $\Delta m_ \odot ^2/\Delta m_{atm}^2$ is taken care of by $\sin\theta_{13}$ via $\varepsilon^2$.  
It is in fact this value of $\vert\sin\theta_{13}\vert={\mathcal O}(0.01)$ that 
$m_{\beta\beta}$ begins to increase. The figure shows that $m_{\beta\beta}$ is in the range of 
\begin{eqnarray}
&&
0.1~{\rm eV}\mapleq m_{\beta\beta} \mapleq 0.5~{\rm eV}.
\label{Eq:Degenerate-13-1-ee}
\end{eqnarray}
This predicted values of $m_{\beta\beta}$ are inside 0.22 eV$\mapleq m_{\beta\beta}\mapleq$
1.6 eV obtained from the Heidelberg-Moscow experiment \cite{ee-exp}.  Our prediction 
is within reach of future planned experiments measuring $m_{\beta\beta}$ 
\cite{ee-exp-future}.  

\vspace{5mm}
\centerline{\label{subsec:3-4-1-2}{\small  \it 1-2. Category C2) with $s_{23}\sim -\sigma /\sqrt{2}$}}
\vspace{3mm}
The quasi degenerate mass pattern in this case has the mass matrix with 
$M_{\mu\tau}\rightarrow -M_{\mu\tau}$ in $M^{\rm C1)}_\nu$, and is given by
\begin{eqnarray}
&&
M^{\rm C2)}_\nu
=
d_0 \left( {\begin{array}{*{20}c}
   {1 - \eta } & {\left( {q\eta  + \varepsilon } \right)\eta } & { - \sigma \left
( {q\eta  - \varepsilon } \right)\eta }  \\
   {\left( {q\eta  + \varepsilon } \right)\eta } & {1 + r^\prime \varepsilon \eta }
 & { - \sigma \left( {1 - s\eta } \right)\eta }  \\
   { - \sigma \left( {q\eta  - \varepsilon } \right)\eta } & { - \sigma \left( {1 
- s\eta } \right)\eta } & {1 - r^\prime \varepsilon \eta }  \\
\end{array}} \right).
\label{Eq:Degenerate-12-1}
\end{eqnarray}
Neutrino masses are predicted to be:
\begin{eqnarray}
&&
m_1  \approx \left( {1 - \eta  - \frac{{\sqrt 2 \varepsilon \eta }}{{\sin 2\theta 
_{12} }}} \right)d_0,
\quad
m_2  \approx \left( {1 - \eta  + \frac{{\sqrt 2 \varepsilon \eta }}{{\sin 2\theta 
_{12} }}} \right)d_0,
\quad
m_3  \approx \left( {1 + \eta } \right)d_0,
\label{Eq:Degenerate-12-1-masses}
\end{eqnarray}
and
\begin{eqnarray}
&&
\Delta m_ \odot ^2  \approx \frac{{4\sqrt 2 \varepsilon \eta d_0^2 }}{{\sin 2
\theta _{12} }},
\quad
\Delta m_{atm}^2  \approx 4\vert\eta\vert d_0^2,
\label{Eq:Degenerate-12-1-squared}
\end{eqnarray}
and mixing angles are
\begin{eqnarray}
&&
\tan 2\theta _{12}  \approx \frac{{4\sqrt 2 \varepsilon }}{{s\eta  - r^{\prime 2} \varepsilon ^2 }},
\quad
\tan 2\theta _{13}  \approx   - \sqrt 2 \sigma q\eta, 
\quad
\cos 2\theta_{23}  \approx  - r^\prime \varepsilon.
\label{Eq:Degenerate-12-1-angles}
\end{eqnarray}

The prediction of $\tan 2\theta_{12}$ indicates that $\vert\eta\vert\sim \vert \varepsilon\vert$
 to have $\tan 2\theta_{12}={\mathcal{O}}(1)$. The requirement of 
 $\vert\eta\vert\sim \vert \varepsilon\vert$ gives
\begin{eqnarray}
&&
\cos 2\theta _{23} \approx \frac{\sigma r^\prime s\tan 2\theta_{12}}{4q}\sin\theta_{13},
\label{Eq:Degenerate-12-1-simplified-sin13}
\end{eqnarray}
leading to $\tan 2\theta_{12} \sim \cos 2\theta _{23}/\sin\theta _{13}$. 
The mass hierarchy is characterized 
by $\varepsilon$ instead of $\eta$ in C1), and can be cast into the form of
\begin{eqnarray}
&&
\frac{{\Delta m_ \odot ^2 }}{{\Delta m_{atm}^2 }} \approx  - \frac{{\sqrt 2 \kappa \cos 
2\theta _{{\rm{23}}} }}{{r^\prime \sin 2\theta _{12} }},
\label{Eq:Degenerate-12-1-simplified}
\end{eqnarray}
where $\kappa=\eta/\vert\eta\vert$, from which we find that  
$0.006\mapleq\vert\cos 2\theta _{23}\vert \mapleq 0.1$ from $\sqrt{2} \vert\cos 2\theta _{23}/r^\prime\vert\mapleq 0.045$ for 
$1/3\leq\vert r^\prime\vert \leq 3$ to yield 
$\Delta m_ \odot ^2/\Delta m_{atm}^2 \approx  \sqrt{2} \vert\cos 2\theta _{23}/r^\prime\vert=0.025-0.048$. 
From Eqs.(\ref{Eq:Degenerate-12-1-simplified-sin13}) and (\ref{Eq:Degenerate-12-1-simplified}), we observe that
\begin{eqnarray}
&&
\frac{{\Delta m_ \odot ^2 }}{{\Delta m_{atm}^2 }} \approx  - \frac{\sqrt 2 \sigma\kappa s }{4q \cos 2\theta_{12}}
\sin\theta_{13}.
\label{Eq:Degenerate-12-1-simplified-2}
\end{eqnarray}
The effective neutrino mass $m_{\beta\beta}$ is given by $(1 - \eta) d_0$. 
There is a possibility of having an enhanced $m_{\beta\beta}$ for $\eta\sim 0$, which satisfies
\begin{eqnarray}
&&
m^2_{\beta\beta}:\Delta m^2_{atm}:\Delta m^2_\odot\sim 1:\vert\eta\vert:\vert\eta\varepsilon\vert.
\label{Eq:Degenerate-12-1-mass-ordering}
\end{eqnarray}

The predictions of $\Delta m_ \odot ^2 /\Delta m_{atm}^2$, and $\cos 2\theta_{23}$
are plotted in FIG.\ref{Fig:degenerate-R}-C2) and FIG.\ref{Fig:degenerate-cos23}-C2), both  
for $m_1\sim m_2 \sim m_3$, where $\eta$ is restricted to satisfy $4\vert\eta\vert \leq 1/3$ to give 
$\vert m^2_1\vert \sim \Delta m^2_{atm}/4\vert\eta\vert \geq 3\Delta m^2_{atm}$. In all figures, plots are 
marked by the black dots satisfying $4\vert\eta\vert \leq 1/3$ and, for comparison, by the grey dots satisfying 
$4\vert\eta\vert > 1/3$.   In  FIG.\ref{Fig:degenerate-cos23}-C2) for $m_1\sim m_2\sim m_3$, $\cos 2\theta_{23}$ does not vanish 
and bounded as  $\vert\cos 2\theta_{23}\vert \mapleq 0.1$. This result is consistent with our expectation based on
\begin{itemize}
\item $\Delta m_ \odot ^2/\Delta m_{atm}^2 \approx  \sqrt{2} \vert\cos 2\theta _{23}/r^\prime\vert$ giving 
$0.006\mapleq\vert\cos 2\theta _{23}\vert \mapleq 0.1$.
\end{itemize}
From this figure, we find that
\begin{eqnarray}
&&
0.005 \mapleq\vert\cos 2\theta_{13}\vert \mapleq 0.1,
\label{Eq:Degenerate-12-1-cos23}
\end{eqnarray}
while $\vert\sin\theta_{13}\vert$ covers all the allowed values owing to the parameter $q$ in Eq.(\ref{Eq:Degenerate-12-1-angles}). 

Let us estimate $m_{\beta\beta}$ for $\sin^22\theta_{12}=0.8$.  The choice of 
$\varepsilon=0.02$ leads to $\Delta m_ \odot ^2/\Delta m_{atm}^2=3.2\times 10^{-2}$,
 which gives $\vert\eta\vert d^2_0=6.4\times 10^{-4}$ for $\Delta m_ \odot ^2 = 8\times 10^{-5}$
 eV$^2$, corresponding to $\Delta m_{atm}^2=2.56\times 10^{-3}$ eV$^2$.  This 
estimation gives $m_{\beta\beta} = 2.5\times 10^{-2}/\sqrt{\vert\eta\vert}$
 eV, leading to $m_{\beta\beta}= 0.18\sqrt{\vert\varepsilon/\eta\vert}$ eV.  
In FIG.\ref{Fig:degenerate-ee}-C2) for $m_1\sim m_2\sim m_3$, the larger $m_{\beta\beta}$ is realized by the smaller 
$\vert\sin\theta_{13}\vert$.  This feature is a result of 
\begin{itemize}
\item Eq.(\ref{Eq:Degenerate-12-1-mass-ordering}) supplemented by $\vert\eta\vert\sim \vert \varepsilon\vert$ giving $\vert\eta\vert \sim \vert\sin\theta_{13}\vert$,
\end{itemize}
which indicates that $\sin\theta_{13}\rightarrow 0$ as $\eta\rightarrow 0$.  The figure shows 
\begin{eqnarray}
&&
0.07~{\rm eV} \mapleq m_{\beta\beta}\mapleq 0.35 ~{\rm eV},
\label{Eq:Degenerate-12-1-ee}
\end{eqnarray}
for the black dots.

\subsubsection{\label{subsubsec:3-4-2}$m_1 \sim m_2 \sim -m_3$}
\centerline{\label{subsec:3-4-2-1}{\small  \it 2-1. Category C1) with $s_{23}\sim\sigma /\sqrt{2}$}}
The texture is given by
\begin{eqnarray}
&&
M^{\rm C1)}_\nu
=
d_0 \left( {\begin{array}{*{20}c}
   {1 + \eta } & {q\eta ^2  + \varepsilon } & { - \sigma \left( {q\eta ^2  - \varepsilon } \right)}  \\
   {q\eta ^2  + \varepsilon } & {\eta  + r^\prime \varepsilon } & { - \sigma }  \\
   { - \sigma \left( {q\eta ^2  - \varepsilon } \right)} & { - \sigma } & {\eta  - r^\prime \varepsilon }  \\
\end{array}} \right),
\label{Eq:Degenerate-dash-13-1}
\end{eqnarray}
which satisfies that $d - \sigma e - a = 0$ and
$b + \varepsilon\Delta b^\prime= {\mathcal O}(\eta^2,\varepsilon^2)$ to yield $\tan 2\theta _{12} = {\mathcal O}(1)$.
Neutrino masses are predicted to be:
\begin{eqnarray}
&&
m_1  \approx \left( {1 + \eta  + \frac{{\left( {2 + r^{\prime 2} } \right)\varepsilon ^2 }}{4} - \frac{{2q\eta ^2  + r^\prime \varepsilon ^2 }}{{\sqrt 2 \sin 2\theta _{12} }}} \right)d_0, 
\nonumber \\
&&
m_3  \approx \left( {1 + \eta  + \frac{{\left( {2  + r^{\prime 2} } \right)\varepsilon ^2 }}{4} + \frac{{2q\eta ^2  + r^\prime \varepsilon ^2 }}{{\sqrt 2 \sin 2\theta _{12} }}} \right)d_0, 
\nonumber \\
&&
m_3  \approx - \left( {1 - \eta  + \frac{{\left( {2  + r^{\prime 2} } \right)\varepsilon ^2 }}{2}} \right)d_0,
\label{Eq:Degenerate-dash-13-1-masses}
\end{eqnarray}
and
\begin{eqnarray}
&&
\Delta m_ \odot ^2  \approx \frac{2\sqrt 2 \left( 2q\eta ^2  + r^\prime\varepsilon ^2 \right)d_0^2 }{\sin 2\theta _{12}},
\quad
\Delta m_{atm}^2  \approx \left| 4\eta  - \frac{\left( 2  + r^{\prime 2}  \right)\varepsilon ^2}{2} \right|d_0^2,
\label{Eq:Degenerate-dash-13-1-squared}
\end{eqnarray}
and mixing angles are
\begin{eqnarray}
&&
\tan 2\theta _{12}  \approx \frac{2\sqrt 2 \left( 2q\eta ^2  + r^\prime \varepsilon ^2 \right)}{\left(2  + r^{\prime 2}\right)\varepsilon ^2 },
\quad
\tan 2\theta _{13}  \approx  - \sqrt 2 \sigma \varepsilon, 
\quad
\cos 2\theta_{23}  \approx  r^\prime\varepsilon.
\label{Eq:Degenerate-dash-13-1-angles}
\end{eqnarray}

The parameter $\varepsilon$ is common in $\cos 2\theta_{23}$ and $\sin\theta_{13}$ and yields
\begin{eqnarray}
\cos 2\theta _{23}  \approx  - \sqrt 2 \sigma r^\prime\sin \theta _{13}.
\label{Eq:Degenerate-dash-13-1-simple}
\end{eqnarray}
There is a constraint on $\eta$ to yield $\tan 2\theta _{12}={\mathcal{O}}(1)$, which requires that 
$\vert\eta\vert \mapleq \vert\varepsilon\vert$.
Another constraint arises from the requirement of $\Delta m_ \odot ^2 /\Delta m_{atm}^2={\mathcal O}(0.01)$.  
If $\vert\eta\vert \mapleq \varepsilon^2$, we obtain $\Delta m_{atm}^2\sim \varepsilon^2 d_0^2$, 
giving $\Delta m_ \odot ^2 /\Delta m_{atm}^2\mapgeq 1$.  Therefore, we need to have $\vert\eta\vert \gg \varepsilon^2$, 
suggesting that $\vert\eta\vert \mapgeq \vert\varepsilon\vert$.  The two constraints of $\vert\eta\vert \mapleq \vert\varepsilon\vert$ and 
$\vert\eta\vert \mapgeq \vert\varepsilon\vert$ are satisfied if $\vert\eta\vert \sim \vert\varepsilon\vert$.
The mass hierarchy is given by
\begin{eqnarray}
&& 
\frac{{\Delta m_ \odot ^2 }}{{\Delta m_{atm}^2 }}\approx \frac{1}{\sqrt 2\sin 2\theta _{12}}
\left( 2q\left| \eta  \right| + r^\prime \frac{\varepsilon^2}{\left| \eta  \right|}\right),
\label{Eq:Degenerate-dash-13-1-simplified}
\end{eqnarray}
from which we find that  
\begin{enumerate}
\item $\vert\eta\vert$ is constrained to be ${\mathcal O}(0.01)$ by $\Delta m_ \odot ^2 /\Delta m_{atm}^2$ if $qr^\prime > 0$,
\item $\vert\eta\vert$ is not constrained by $\Delta m_ \odot ^2 /\Delta m_{atm}^2$ but the size of
$2\left| \eta  \right| - (\vert r^\prime\vert \varepsilon^2/\vert q\eta \vert)$ is constrained if $qr^\prime < 0$.
\end{enumerate}
Larger values of $\vert\eta\vert$ ($\mapgeq 0.01$) are only possible for the texture with $qr^\prime < 0$.
This range of $\vert\eta\vert$ affects the size of the mass scale $d_0$, which is approximately $m_{\beta\beta}$. 
The following mass ordering is found in this texture:
\begin{eqnarray}
&&
m^2_{\beta\beta}:\Delta m^2_{atm}:\Delta m^2_\odot\sim 1:\vert\eta\vert:\eta ^2  + \frac{r^\prime}{2q}\varepsilon ^2.
\label{Eq:Degenerate-dash-13-1-mass-ordering}
\end{eqnarray}
The larger $m_{\beta\beta}$ is obtained for the smaller $\vert\eta\vert$.  Since $\vert\eta\vert$ can be as small as $0.01$, we expect that 
$m_{\beta\beta}\sim \sqrt{\Delta m^2_{atm}/\vert\eta\vert} \sim 0.5$ eV.  

The predictions $\Delta m^2_\odot/\Delta m^2_{atm}$ and $\cos 2\theta_{23}$
 are depicted in FIG.\ref{Fig:degenerate-R}-C1) and FIG.\ref{Fig:degenerate-cos23}
-C1) for $m_1\sim m_2\sim -m_3$.  In FIG.\ref{Fig:degenerate-cos23}-C1) for $m_1\sim m_2\sim -m_3$, there are almost straight lines 
originating from $\sin\theta_{13}\sim 0$. This appearance of these lines is due to 
\begin{itemize}
\item the proportionality of $\cos 2\theta_{23}$ to $\sin\theta_{13}$ of Eq.(\ref{Eq:Degenerate-dash-13-1-simple}).
\end{itemize}
From these figures for the black dots, we find that
\begin{eqnarray}
&&
\left| \cos 2\theta_{23}\right| \mapleq 0.15,
\quad
\vert\sin \theta_{13}\vert \mapleq 0.06.
\label{Eq:Degenerate-dash-13-1-cos23}
\end{eqnarray}
Shown in FIG.\ref{Fig:degenerate-ee}-C1) for $m_1\sim m_2\sim -m_3$ is the size of $m_{\beta\beta}$, which is in the range of 
\begin{eqnarray}
&&
0.07~{\rm eV}\mapleq m_{\beta\beta} \mapleq 0.50~{\rm eV}.
\label{Eq:Degenerate-dash-13-1-ee}
\end{eqnarray}
The figure shows that larger values of $m_{\beta\beta}$ are produced by smaller values of $\vert\sin\theta_{13}\vert$.  
This behavior can be understood because we know that 
\begin{itemize}
\item $m_{\beta\beta}$ gets larger as $\vert\eta\vert$ gets smaller from Eq.(\ref{Eq:Degenerate-dash-13-1-mass-ordering}),
\end{itemize}
where the smallness of $\vert\eta\vert$ is converted into that of $\vert\sin\theta_{13}\vert$ from $\vert\eta\vert\sim\vert\epsilon\vert$.
.

\vspace{5mm}
\centerline{\label{subsec:3-4-2-2}{\small  \it 2-2. Category C2) with $s_{23}\sim -\sigma /\sqrt{2}$}}
\vspace{3mm}
The quasi degenerate mass pattern is given by:
\begin{eqnarray}
&&
M^{\rm C2)}_\nu
=
d_0 \left( {\begin{array}{*{20}c}
   {1 + \eta } & {\left( {q + \varepsilon } \right)\eta } & { - \sigma \left( {q - \varepsilon } \right)\eta }  \\
   {\left( {q + \varepsilon } \right)\eta } & {\eta  + r^\prime \varepsilon } & \sigma   \\
   { - \sigma \left( {q - \varepsilon } \right)\eta } & \sigma  & {\eta  - r^\prime \varepsilon }  \\
\end{array}} \right),
\label{Eq:Degenerate-dash-12-1}
\end{eqnarray}
which satisfies that  $d + \sigma e - a = 0$ and 
$\varepsilon b^\prime +\Delta b = {\mathcal O}(\eta\varepsilon)$ to yield $\tan 2\theta _{12} = {\mathcal O}(1)$.
Neutrino masses are predicted to be:
\begin{eqnarray}
&&
m_1  \approx \left( 1 + \eta   - \frac{{3r^{\prime 2} \varepsilon ^2 }}{4} - \frac{{\sqrt 2 \left( {2 - qr^\prime} \right)\varepsilon \eta }}{{2\sin 2\theta _{12} }} \right)d_0,
\nonumber \\
&&
m_2  \approx \left( 1 + \eta   - \frac{{3r^{\prime 2} \varepsilon ^2 }}{4} + \frac{{\sqrt 2 \left( {2 - qr^\prime} \right)\varepsilon \eta }}{{2\sin 2\theta _{12} }} \right)d_0,
\nonumber \\
&&
m_3  \approx - \left( 1 - \eta   - \frac{3r^{\prime 2} \varepsilon ^2}{4}\right) d_0,
\label{Eq:Degenerate-dash-12-1-masses}
\end{eqnarray}
and
\begin{eqnarray}
&&
\Delta m_ \odot ^2  \approx 
{\frac{{2\sqrt 2 \left( {2 - qr^\prime} \right)\varepsilon \eta d_0^2 }}{{\sin 2\theta _{12} }}},
\quad
\Delta m_{atm}^2  \approx \left| {4\eta   + \frac{{3r^{\prime 2} \varepsilon ^2 }}{2}} \right|d_0^2, 
\label{Eq:Degenerate-dash-12-1-squared}
\end{eqnarray}
and mixing angles are
\begin{eqnarray}
&&
\tan 2\theta _{12}  \approx 
\frac{{2\sqrt 2 \left( {qr^\prime - 2} \right)\varepsilon \eta }}{{2q^2\eta ^2  + 3r^{\prime 2}\varepsilon ^2}},
\quad
\tan 2\theta _{13}  \approx  \sqrt 2 \sigma q\eta, 
\quad
\cos 2\theta_{23}  \approx  -\varepsilon r^\prime 
.
\label{Eq:Degenerate-dash-12-1-angles}
\end{eqnarray}

To obtain $\tan 2\theta_{12}={\mathcal O}(1)$ is possible if $\vert\eta\vert\sim\vert\varepsilon\vert$, which is 
translated into $\vert\cos 2\theta_{23}\vert\propto\vert\sin\theta_{13}\vert$. The mass hierarchy is given by
\begin{eqnarray}
&& \frac{{\Delta m_ \odot ^2 }}{{\Delta m_{atm}^2 }}\approx 
\frac{{\kappa \left( {qr^\prime - 2} \right)\cos 2\theta _{23} }}{{\sqrt 2 r^\prime\sin 2\theta _{12} }},
\label{Eq:Degenerate-dash-12-1-simplified}
\end{eqnarray}
where $\kappa$ stands for the sign of $\eta$. 
The effective neutrino mass $m_{\beta\beta}$ given by $(1 + \eta) d_0$ satisfies 
the same ordering as Eq.(\ref{Eq:Degenerate-12-1-mass-ordering}) for $m_1\sim m_2\sim m_3$ described by
\begin{eqnarray}
&&
m^2_{\beta\beta}:\Delta m^2_{atm}:\Delta m^2_\odot\sim 1:\vert\eta\vert:\vert\eta\varepsilon\vert.
\label{Eq:Degenerate-dash-12-1-mass-ordering}
\end{eqnarray}
This relation shows that large sizes of $m_{\beta\beta}$ favor small values of $\vert \eta\vert$.

The predictions of $\Delta m_ \odot ^2 /\Delta m_{atm}^2$, and $\cos 2\theta_{23}$ 
are plotted in FIG.\ref{Fig:degenerate-R}-C2) and FIG.\ref{Fig:degenerate-cos23}-C2), both for $m_1\sim m_2\sim -m_3$, 
as functions of $\vert\sin\theta_{13}\vert$.  The gross features are similar to those for C1).   We find that
\begin{eqnarray}
&&
0.005 \mapleq \vert\cos 2\theta_{23}\vert \mapleq 0.30,
\quad
\vert\sin\theta_{13}\vert \mapleq 0.14,
\label{Eq:Degenerate-dash-12-1-sin13}
\end{eqnarray}
for $4\vert\eta\vert \leq 1/3$.  
The effective neutrino mass $m_{\beta\beta}$ is plotted in FIG.\ref{Fig:degenerate-ee}-C2) for $m_1\sim m_2\sim -m_3$, 
where $m_{\beta\beta}$ gets larger as $\vert\sin \theta_{13}\vert$ gets smaller as indicated by 
$m_{\beta\beta} \sim\sqrt{\Delta m^2_{atm}/\vert\sin\theta_{13}\vert}$. 
We observe that
\begin{eqnarray}
&&
m_{\beta\beta}\mapgeq 0.07 ~{\rm eV},
\label{Eq:Degenerate-dash-12-1-ee}
\end{eqnarray}
which becomes as large as 0.5 eV for $\sin\theta_{13}\sim 0$.

\section{\label{sec:4}Summary and Discussions}
We have clarified the effects from the $\mu$-$\tau$ symmetry breaking in neutrino 
mass textures.  Possible forms of the textures are summarized in TABLE \ref{Tab:Lists},
 where the relations among $\cos 2\theta_{23}$, $\sin\theta_{13}$, and $\Delta m^2_\odot/\Delta m^2_{atm}$
 are listed.  These relations reflect the general property discussed in Ref.
\cite{Another-mu-tau}.  We have found the relations:
\begin{eqnarray}
&&
\vert\cos 2\theta_{23}\vert \sim \sin\theta_{23},
\quad
\vert\cos 2\theta_{23}\vert\sin\theta_{13}\sim \frac{\Delta m^2_\odot}{\Delta m^2_{atm}},
\quad
\vert\cos 2\theta_{23}\vert\sim \frac{\Delta m^2_\odot}{\Delta m^2_{atm}},
\label{Eq:new-relations-ee-summary}
\end{eqnarray}
depending on the type of textures, where the possible signs are ignored.  Two 
qualitatively different textures are classified as C1), and C2), respectively, 
according to the behavior of $\sin\theta_{13}\rightarrow 0$, and $\sin\theta_{12}\rightarrow 0$ 
in the $\mu$-$\tau$ symmetric limit.  It would be curious to have $\sin\theta_{12}=0$ in the symmetric limit because 
the textures indeed give $\sin^22\theta_{12}={\mathcal{O}}(1)$, which does not 
seem to vanish.  The reason is that, in the textures for C2), $\tan 2\theta_{12}$ 
is expressed as
\begin{eqnarray}
&&
\tan 2\theta _{12}  \approx \frac{{4\sqrt 2 \varepsilon }}{{2\left( {s-p} \right)
\eta  -  r^{\prime 2}\varepsilon ^2 }},
\label{Eq:Angle-12-summary-1}\\
&&
\tan 2\theta _{12}  \approx \frac{{4\sqrt 2 \varepsilon }}{{2p\eta  + r^{\prime 
2} \varepsilon ^2 }},
\label{Eq:Angle-12-summary-2}\\
&&
\tan 2\theta _{12}  \approx \frac{{4\sqrt 2 \varepsilon }}{{2s\eta  - r^{\prime 2} \varepsilon ^2 }},
\label{Eq:Angle-12-summary-3}\\
&&
\tan 2\theta _{12}  \approx \frac{{2\sqrt 2 \left( {qr^\prime-2} \right)\varepsilon \eta}}{{ 2q^2\eta ^2 + 3r^{\prime 2} \varepsilon ^2}}
\label{Eq:Angle-12-summary-4},
\end{eqnarray}
respectively, for the normal mass hierarchy, the inverted mass hierarchy, the 
quasi degenerate mass pattern II with $m_1\sim m_2\sim m_3$, and 
the quasi degenerate mass pattern II with $m_1\sim m_2\sim -m_3$.  It is thus obvious that $\tan 2\theta_{12}\rightarrow 0$
 as $\varepsilon\rightarrow 0$, restoring the $\mu$-$\tau$ symmetry.  From Eqs.(\ref{Eq:Angle-12-summary-1})
 -(\ref{Eq:Angle-12-summary-3}),  we have observed that $\tan 2\theta_{12}$ is roughly estimated to be
\begin{eqnarray}
&&
\tan 2\theta _{12}  \sim \frac{\cos 2\theta_{23}}{\sin\theta_{13}},
\label{Eq:Angle-12-summary-5}
\end{eqnarray}
which is specific to C2).

The differences in the same mass hierarchy 
can be found in Eq.(\ref{Eq:new-relations-ee-summary}) as in TABLE.\ref{Tab:Lists},
 where precise determination of $\sin\theta_{13}$, $\sin^22\theta_{23}$, and 
$\Delta m^2_\odot/\Delta m^2_{atm}$ may select one of the relations.  The differences 
in the prediction of $\cos 2\theta_{23}$ can also be read off from the figures, 
where each texture shows its specific pattern.  From the scattered plots, we may exclude some of 
textures with certain ranges of mixing angles.  When future experiments observe 
\begin{enumerate}
\item $\vert\sin\theta_{13}\vert= 0.16-0.18$ as large as the present upper bound, textures giving
\begin{itemize}
\item the normal mass hierarchy
\begin{itemize}
\item in C1) if $\vert\cos 2\theta_{23}\vert \mapleq 0.05$,
\item in C2) if $\vert\cos 2\theta_{23}\vert \mapleq 0.03$,
\end{itemize}
\item the inverted mass hierarchy
\begin{itemize}
\item in C1) with $m_1\sim m_2$ if $\vert\cos 2\theta_{23}\vert \mapleq 0.10$,
\item in C1) with $m_1\sim -m_2$  for $\eta\neq 0$ if $\vert\cos 2\theta_{23}\vert\mapgeq 0.10$ and for $\eta\neq 0$, 
\item in C2) with $m_1\sim m_2$ if $\vert\cos 2\theta_{23}\vert \mapgeq 0.07$,
\end{itemize}
\item the quasi degenerate mass pattern I exclusively in C1)
\begin{itemize}
\item with the normal mass ordering if $\vert\cos 2\theta_{23}\vert \mapleq 0.05$,
\item with the inverted mass ordering if $\vert\cos 2\theta_{23}\vert \mapleq 0.10$,
\end{itemize}
\item the quasi degenerate mass pattern II
\begin{itemize}
\item in C1) with $m_1\sim m_2\sim m_3$ if $\vert\cos 2\theta_{23}\vert \mapleq 0.07$,
\item in C2) with $m_1\sim m_2\sim m_3$ if $\vert\cos 2\theta_{23}\vert \mapgeq 0.10$,
\item in C1) and C2) with $m_1\sim m_2\sim -m_3$,
\end{itemize}
\end{itemize}
are excluded;
\item $\vert\sin\theta_{13}\vert\mapleq 0.02$ as an almost vanishing $\vert\sin\theta_{13}\vert$, textures giving
\begin{itemize}
\item the normal mass hierarchy
\begin{itemize}
\item in C1) if $\vert\cos 2\theta_{23}\vert \mapgeq 0.13$.
\item in C2) if $\vert\cos 2\theta_{23}\vert \mapgeq 0.05$,
\end{itemize}
\item the inverted mass hierarchy
\begin{itemize}
\item in C1) with $m_1\sim m_2$ if $\vert\cos 2\theta_{23}\vert\mapgeq 0.10$, 
\item in C2) with $m_1\sim -m_2$ for $\vert\eta\vert \leq 0.001$, 
\end{itemize}
\item the quasi degenerate mass pattern I exclusively in C1)
\begin{itemize}
\item with the normal mass ordering for 
$\vert\eta\vert \leq 0.01$ if $\vert\cos 2\theta_{23}\vert \mapleq 0.05$ and $\vert\cos 2\theta_{23}\vert \mapgeq 0.20$,
\item with the inverted mass ordering for 
$\vert\eta\vert \leq 0.01$ if $\vert\cos 2\theta_{23}\vert \mapleq 0.07$,
\end{itemize}
\item the quasi degenerate mass pattern II
\begin{itemize}
\item in C1) and C2) with $m_1\sim m_2\sim \pm m_3$ if $\vert\cos 2\theta_{23}\vert \mapgeq 0.10$,
\end{itemize}
\end{itemize}
are excluded.
\end{enumerate}
Similarly, if $m_{\beta\beta}$ is observed to be $\sim 0.35$ eV, textures giving the quasi degenerate mass pattern II 
\begin{itemize}
\item in C1) with $m_1\sim m_2\sim m_3$ if $\vert\sin\theta_{13}\vert\mapleq 0.04$,
\item in C2) with $m_1\sim m_2\sim m_3$ and in C1) and C2) with $m_1\sim m_2\sim -m_3$ if $\vert\sin\theta_{13}\vert \mapgeq 0.02$,
\end{itemize}
are excluded.

Most of textures contain two small parameters $\varepsilon$, and $\eta$,  The 
smallness of $\varepsilon$ is obviously ascribed to the tiny $\mu$-$\tau$ symmetry 
breaking.  However, the reason to have the small $\eta$ is not clear. There is a possibility 
that textures for the normal mass hierarchy may be protected by the 
electron number conservation, whose tiny violation can take care of the smallness 
of $\eta$ \cite{eNumber-example,eNumber}. As stated in Eq.(\ref{Eq:Normal-13-1-simplified-ee}), 
the flavor neutrino masses naturally satisfy that 
$M_{ee}:M_{ei}:M_{ij}\sim\eta^2:\eta:1$ for  $i,j$=$\mu,\tau$.  We have also shown that 
the smallness of $\varepsilon$ is always linked to that of $\cos 2\theta_{23}$.
Another interesting possibility arises from the texture where the tiny quantities are just supplied by 
the $\mu$-$\tau$ symmetry breaking parameter $\varepsilon$.  It is described by the specific form
\begin{eqnarray}
&&
M_\nu
=
d_0 \left( {\begin{array}{*{20}c}
   { - 2} & {q + \varepsilon } & { - \sigma \left( {q - \varepsilon } \right)}  \\
   {q + \varepsilon } &  {1 - r + r^\prime\varepsilon } & { - \sigma \left( {1 + 
r} \right)}  \\
   { - \sigma \left( {q - \varepsilon } \right)} & { - \sigma \left( {1 + r} 
\right)} & {1 - r - r^\prime\varepsilon }  \\
\end{array}} \right),
\label{Eq:Normal-13-2-p=0}
\end{eqnarray}
for the quasi degenerate mass pattern I with $r\neq 0$.  If $r=0$, this texture 
gives the inverted mass hierarchy.  These two textures are solely available 
for C1) with $m_1 \sim -m_2$, and does not contain the additional small parameter $\eta$.
  Furthermore, we have demonstrated that these two textures can provide
\begin{eqnarray}
&&
\frac{\Delta m^2_\odot}{\Delta m^2_{atm}}\sim \sin^2\theta_{13},
\label{Eq:MassSquared-summary}
\end{eqnarray}
where $\sin^2\theta_{13}$ is, respectively, estimated to be:
\begin{eqnarray}
&&
0.02\mapleq \vert\sin\theta_{13}\vert\mapleq 0.10,~
0.14\mapleq \vert\sin\theta_{13}\vert\mapleq 0.18,~{\rm and}~
0.04\mapleq \vert\sin\theta_{13}\vert\mapleq 0.10,
\label{Eq:sin13-inverted-summary}
\end{eqnarray}
for the inverted mass hierarchy, the quasi degenerate mass pattern I with the normal mass ordering, and 
the quasi degenerate mass pattern I with the inverted mass ordering.
It should be noted here that we have to retain all the terms of ${\mathcal{O}}(\sin^2\theta_{13})$,
 which consist of any second order combinations of $\sin\theta_{13}(\approx \varepsilon)$,
 and $\cos 2\theta_{23}(\approx 2\Delta)$. 

The similar predictions are obtained in the textures for C2) which predict 
$\Delta m^2_\odot\propto \eta\varepsilon$ (or $\Delta m^2_\odot\propto \varepsilon$),
 $\sin\theta_{13}\propto \eta$ and $\tan 2\theta_{12} \propto \varepsilon/\eta$. From 
the collaboration between $\eta$ and $\varepsilon$, we have found that
\begin{eqnarray}
&&
\frac{\Delta m^2_\odot}{\Delta m^2_{atm}}\left( \propto \eta\varepsilon\right) \sim \tan 2\theta_{12}\sin^2\theta_{13},
\quad
\vert\sin\theta_{13}\vert \mapgeq 0.06,
\label{Eq:MassSquared-2-summary}
\end{eqnarray}
for the normal mass hierarchy, and
\begin{eqnarray}
&&
\frac{\Delta m^2_\odot}{\Delta m^2_{atm}}\left( \propto \varepsilon\right) \sim \tan 2\theta_{12}\sin\theta_{13},
\quad
0.002\mapleq \vert\sin\theta_{13}\vert\mapleq 0.03,
\label{Eq:MassSquared-3-summary}
\end{eqnarray}
for the inverted mass hierarchy.  These are the relations discussed in Ref.\cite{AtmDeviation,Theta31AndMass}.

We expect the following scenarios based on the symmetry argument to emerge: Underlying dynamics creating neutrino 
masses may possess
\begin{itemize}
\item the approximate $\mu$-$\tau$ symmetry to induce the textures Eq.(\ref{Eq:Inverted-13-2}) 
and Eq.(\ref{Eq:Normal-13-2}) in the case of C1), respectively, for the quasi degenerate mass pattern I, and the inverted mass hierarchy, or
\item the approximate $\mu$-$\tau$ symmetry as well as the approximate electron 
number conservation to induce the texture Eq.(\ref{Eq:Normal-13-1}) in the case 
of C1), or Eq.(\ref{Eq:Normal-12-1}) in the case of C2), both for the normal mass hierarchy.
\end{itemize}
To see the corresponding lagrangians, 
${\mathcal{L}}^{C1)}_{\rm inverted}$ for Eq.(\ref{Eq:Inverted-13-2}), 
${\mathcal{L}}^{C1)}_{\rm normal}$ for Eq.(\ref{Eq:Normal-13-2}), 
${\mathcal{L}}^{C1-e)}_{\rm normal}$ for Eq.(\ref{Eq:Normal-13-1}),
 and 
${\mathcal{L}}^{C2-e)}_{\rm normal}$ for Eq.(\ref{Eq:Normal-12-1}), 
it is convenient 
to use $\nu _ \pm   = (\nu _\mu   \pm ( - \sigma )\nu _\tau )/\sqrt{2}$, where $\nu_{+(-)}$ is an even (odd) 
state with respect to the $\mu$-$\tau$ exchange.  In terms of $\nu_\pm$ the lagrangians are described by
\begin{eqnarray}
 - {\mathcal{L}}^{C1)}_{\rm inverted} &=& 
\frac{d_0}{2}\left[ 2\left(\nu _ +  \nu _ +  - \nu _e \nu _e
\right)   + \sqrt 2 q\left( {\nu _e \nu _ +   + \nu _ +  \nu _e } \right) \right]
  + {\mathcal{L}}_b,
\nonumber\\
 - {\mathcal{L}}^{C1)}_{\rm normal} &=& 
\frac{d_0}{2}\left[ 2\left(\nu _ +  \nu _ +   - r\nu _ -  \nu _ - - \nu _e \nu _e
\right)   + \sqrt 2 q\left( {\nu _e \nu _ +   + \nu _ +  \nu _e } \right) \right]
  + {\mathcal{L}}_b,
\nonumber\\
 - {\mathcal{L}}^{C1-e)}_{\rm normal} &=& 
\frac{d_0}{2}\left[ 2\nu _ -  \nu _ - +\eta\left( s \left(\nu_+\nu_+-\nu_-\nu_-
\right) + \sqrt 2 \left( {\nu _e \nu _ +   + \nu _ +  \nu _e }\right) \right)
+
p^\prime\eta^2\nu _e \nu _e\right]
+
{\mathcal{L}}_b,
\nonumber\\
 - {\mathcal{L}}^{C2-e)}_{\rm normal} &=& 
\frac{d_0}{2}\left[ 2\nu _ +  \nu _ + +\eta\left( s \left(\nu_-\nu_- - \nu_+\nu_+
\right) + \sqrt 2 \left( {\nu _e \nu _ +   + \nu _ +  \nu _e }\right)\right)
+
p^\prime\eta^2\nu _e \nu _e\right]
+
{\mathcal{L}}_b,
\label{Eq:lagrangian}
\end{eqnarray}
for $p^\prime$ from $p = p^\prime\eta $, where ${\mathcal{L}}_b$ is given by
\begin{eqnarray}
{\mathcal{L}}_b &=& \frac{\varepsilon d_0}{2} \left[ \sqrt 2 \left( {\nu _e \nu 
_ -   + \nu _ -  \nu _e } \right)
+ r^\prime\left( {\nu _ +  \nu _ -   + \nu _ -  \nu _ +  } \right) \right],
\label{Eq:lagrangian_break}
\end{eqnarray}
as the $\mu$-$\tau$ symmetry breaking lagrangian.  In ${\mathcal{L}}^{C1)}_{\rm normal,inverted}$, the relative 
strength of $\nu_e\nu_e$ over $\nu_+\nu_+$ should be exactly $-1$. Similarly, 
in ${\mathcal{L}}^{C1)}_{\rm inverted}$ and ${\mathcal{L}}^{C2-e)}_{\rm normal}$ 
(or ${\mathcal{L}}^{C1-e)}_{\rm normal}$), the $\mu$-$\tau$ symmetric term of $\nu_+\nu_+$ (or $\nu_-\nu_-$) that 
remains at $\eta=0$ is absent.  The appearance of such specific forms of the textures 
may call for additional underlying symmetries.  For instance, the absence of $\nu_+\nu_+$ (or $\nu_-\nu_-$)
may be achieved by introducing a discrete symmetry that gives the difference between 
$\nu_+\nu_+$, and $\nu_-\nu_-$.  

It is generally expected in the quasi degenerate mass pattern II that the effective 
neutrino mass $m_{\beta\beta}$ is larger than ${\mathcal{O}}(\sqrt{\Delta m^2_{atm}})$.
The predicted $m_{\beta\beta}$ satisfies the mass ordering of
\begin{eqnarray}
&&
m^2_{\beta\beta} : \Delta m^2_{atm} : \Delta m^2_\odot\sim 1:\vert\eta\vert:\eta^2,
\label{Eq:m-ee-summary-1}
\end{eqnarray}
for C1) with $m_1\sim m_2\sim m_3$,
\begin{eqnarray}
&&
m^2_{\beta\beta}:\Delta m^2_{atm}:\Delta m^2_\odot\sim 1:\vert\eta\vert:\eta ^2  + \frac{qr^\prime}{2}\varepsilon ^2,
\label{Eq:m-ee-summary-2}
\end{eqnarray}
for C1) with $m_1\sim m_2\sim -m_3$, and
\begin{eqnarray}
&&
m^2_{\beta\beta} : \Delta m^2_{atm} : \Delta m^2_\odot\sim 1:\vert\eta\vert:\vert\eta
\varepsilon\vert,
\label{Eq:m-ee-summary-3}
\end{eqnarray}
for C2) with $m_1\sim m_2\sim \pm m_3$. The mass scale of $m_{\beta\beta}$ is, therefore, given by 
$\sqrt{\Delta m^2_{atm}/\vert\eta\vert}$.  The three figures, FIG.\ref{Fig:degenerate-ee} for C2) with 
$m_1\sim m_2\sim m_3$ and for C1) and C2) with $m_1\sim m_2\sim -m_3$, show the same shape characterized by 
the rapid rise in $m_{\beta\beta}$ around $\vert\eta\vert(\sim \vert\sin\theta_{13}\vert) = 0$ as indicated by 
Eqs.(\ref{Eq:m-ee-summary-2}) and (\ref{Eq:m-ee-summary-3}). This behavior generally disappears 
for C1) because $\eta$ is not directly related to $\sin\theta_{13}(\sim\varepsilon)$.  The texture 
giving $m_1\sim m_2\sim m_3$ shows no rise at $\sin\theta_{13}=0$.  However, the texture giving  
$m_1\sim m_2\sim -m_3$ also shows this rapid rise.  It is because $\eta$ is related to $\sin\theta_{13}$ 
by Eq.(\ref{Eq:Degenerate-dash-13-1-simplified}).  It should be noted that, in textures with $m_1\sim m_2\sim -m_3$,
the approximate $L_\mu - L_\tau$ conservation is respected.  The results of our calculations 
turn out to give similar behaviors of $\cos 2\theta_{23}$ and $m_{\beta\beta}$ versus $\sin\theta_{13}$ to 
those predicted in Ref.\cite{LeLmuPlot}, which has discussed the neutrino oscillation phenomenology 
based on the approximate $L_\mu - L_\tau$ conservation.  Namely, $\vert\cos 2\theta_{23}\vert$ is 
proportional to $\vert\sin\theta_{13}\vert$ and $m_{\beta\beta}$ rapidly increases 
as $\vert\sin\theta_{13}\vert\rightarrow 0$.  

The flavor neutrino masses receive radiative corrections if the neutrino masses are 
created at relatively higher scale such as $10^{10}$ TeV.  It has been argued 
that neutrinos in the normal mass hierarchy do not receive large radiative effects,
 and the form of the texture restores, at low energies, its original one \cite{Normal}.
  However, other two mass patterns are likely influenced by these radiative 
corrections \cite{InvertedDegenerate}.  So, the proposed textures may not be 
respected by observed neutrinos if they are the textures determined at high 
energies.  Conversely, we should discuss how the textures realized at low energies 
are evolved into those realized at high energies, which may exhibit some good 
ordering.  Any underlying dynamics that respects the approximate $\mu$-$\tau$ symmetry 
should induce, at low energies, one of our textures 
because the symmetry structure is not drastically disturbed.

\vspace{3mm}
\noindent
\centerline{\small \bf ACKNOWLEGMENTS}

The authors are grateful to I. Aizawa, and T. Kitabayashi for useful discussions.
  The work of M.Y. is supported by the Grants-in-Aid for Scientific Research on 
Priority Areas (No 13135219) from the Ministry of Education, Culture, Sports, 
Science, and Technology, Japan.

\appendix
\section{\label{sec:Appendix}Estimation of Neutrino Masses and Mixing Angles}
In this Appendix, we show our estimation of neutrino masses and mixing angles discussed 
in Ref.\cite{Another-mu-tau}. This estimation is based on the formula in Ref.\cite{FormulaNoCP}, 
where $M_\nu$ is parameterized by
\begin{eqnarray}
&& M_\nu = \left( {\begin{array}{*{20}c}
	M_{ee} & M_{e\mu} & M_{e\tau}  \\
	M_{e\mu} & M_{\mu\mu} & M_{\mu\tau}  \\
	M_{e\tau} & M_{\mu\tau} & M_{\tau\tau}  \\
\end{array}} \right).
\label{Eq:NuMatrixEntries}
\end{eqnarray}
There is a general formula for deriving mixing angles and masses given by
\begin{eqnarray}
&&
\tan 2\theta _{12} = \frac{2X}{\lambda _2  - \lambda _1},
\label{Eq:ExactMixingAngle12}\\
&&
\tan 2\theta _{13}  = \frac{2 Y}{\lambda _3 - M_{ee}},
\label{Eq:ExactMixingAngle13}\\
&&
\left( M_{\tau\tau} - M_{\mu\mu}\right)\sin 2\theta_{23}  - 2 M_{\mu\tau}\cos 2\theta_{23}= 2s_{13} X,
\label{Eq:ExactMixingAngle23}
\end{eqnarray}
and
\begin{eqnarray}
&&
m_1  = c_{12}^2 \lambda_1  + s_{12}^2 \lambda _2  - 2c_{12} s_{12} X,
\quad
m_2 = s_{12}^2 \lambda_1  + c_{12}^2 \lambda _2  + 2c_{12} s_{12} X,
\nonumber\\
&&
m_3  = c_{13}^2 \lambda _3  + 2c_{13} s_{13}Y + s_{13}^2 M_{ee},
\label{Eq:ExactMasses}
\end{eqnarray}
where
\begin{eqnarray}
&&
X = \frac{c_{23} M_{e\mu} - s_{23} M_{e\tau}}{c_{13}},
\quad
Y = s_{23} M_{e\mu} + c_{23} M_{e\tau},
\nonumber\\
&&
\lambda_1  = c_{13}^2 M_{ee} - 2c_{13} s_{13} Y + s_{13}^2 \lambda_3,
\quad
\lambda_2  = c_{23}^2 M_{\mu\mu} + s_{23}^2 M_{\tau\tau} - 2s_{23} c_{23} M_{\mu\tau},
\nonumber\\
&&
\lambda_3  = s_{23}^2 M_{\mu\mu} + c_{23}^2 M_{\tau\tau} + 2s_{23} c_{23} M_{\mu\tau}.
\label{Eq:Parameters}
\end{eqnarray}
By inserting the mass matrix Eq.(\ref{Eq:M_13-break}) into these equations, we can calculate neutrino masses and mixing angles for any types of textures.
By using Eqs.(\ref{Eq:ExactMixingAngle12}) and (\ref{Eq:Parameters}), we derive
\begin{eqnarray}
&&
\Delta m_ \odot ^2  = \frac{\lambda^2_2  - \lambda^2_1 }{\cos 2\theta _{12}},
\label{Eq:SolarMass2-1}
\end{eqnarray}
which can be transformed into
\begin{eqnarray}
&&
\Delta m_ \odot ^2  = \frac{2 (m_1  + m_2 )X}{\sin 2\theta _{12}},
\label{Eq:SolarMass2-2}
\end{eqnarray}
instead of Eq.(\ref{Eq:SolarMass2-1}) if $\theta_{12}\neq 0$.

For C1), masses and mixing angles are given by 
\begin{eqnarray}
&&
m_1  \approx \frac{{a + d - \sigma e - \left( {d + \sigma e - a} \right)t_{13}^2  +2 \left(\sigma e\Delta  +\varepsilon d^\prime \right)\Delta }}{2} - \frac{{X}}{{\sin 2\theta _{12} }},
\nonumber \\
&&
m_2  \approx \frac{{a + d - \sigma e - \left( {d + \sigma e - a} \right)t_{13}^2  +2 \left(\sigma e\Delta  +\varepsilon d^\prime \right)\Delta }}{2} + \frac{{X}}{{\sin 2\theta _{12} }},
\nonumber \\
&&
m_3  \approx d + \sigma e + \left( {d + \sigma e - a} \right)t_{13}^2  - 2\left( \sigma e\Delta  +  \varepsilon d^\prime \right)\Delta,
\label{Eq:Masses-13}
\end{eqnarray}
and
\begin{eqnarray}
&&
\tan 2\theta _{12}  \approx \frac{{2X}}{d - \sigma e - a + \left( {d + \sigma e - a} \right)t_{13}^2  +2 \left( \sigma e\Delta  +  \varepsilon d^\prime\right)\Delta },
\nonumber \\
&&
\tan 2\theta _{13}  \approx \frac{{2Y}}{d + \sigma e - a - 2\left( \sigma e\Delta  + \varepsilon d^\prime \right)\Delta },
\nonumber \\
&&
\cos 2\theta_{23} \approx 2\Delta,
\quad
\sin 2\theta _{23} \approx \sigma,
\label{Eq:Mixing-13}
\end{eqnarray}
with
\begin{eqnarray}
&&
X \approx \sqrt 2 \left( {b + \varepsilon b^\prime \Delta } \right),
\quad
Y \approx \sqrt 2 \sigma \left( {\varepsilon b^\prime  - b\Delta } \right),
\quad
\Delta  \approx  - \frac{{\sigma \varepsilon d^\prime + \sqrt 2 s_{13} b }}{{2e}},
\label{Eq:X-Y-Delta-13}
\end{eqnarray}
where we see that $\sin\theta_{13}\rightarrow 0$ as $\varepsilon\rightarrow 0$.  For C2), masses and mixing angles are given by 
\begin{eqnarray}
&&
m_1  \approx \frac{a + d + \sigma e - \left( {d - \sigma e - a} \right)t_{13}^2  - 2\left( \sigma e\Delta  -\varepsilon d^\prime \right)\Delta }{2} - \frac{X}{{\sin 2\theta _{12} }},
\nonumber \\
&&
m_2  \approx \frac{a + d + \sigma e - \left( {d - \sigma e - a} \right)t_{13}^2  - 2\left( \sigma e\Delta -\varepsilon d^\prime \right)\Delta }{2} + \frac{X}{{\sin 2\theta _{12} }},
\nonumber \\
&&
m_3  \approx d - \sigma e + \left( {d - \sigma e - a} \right)t_{13}^2  +2 \left( \sigma e\Delta  - \varepsilon d^\prime \right)\Delta ,
\label{Eq:Masses-12}
\end{eqnarray}
and
\begin{eqnarray}
&&
\tan 2\theta _{12}  \approx \frac{{2X}}{d + \sigma e - a + \left( {d - \sigma e - a} \right)t_{13}^2  - 2\left( \sigma e\Delta  -  \varepsilon d^\prime  \right)\Delta },
\nonumber \\
&&
\tan 2\theta _{13}  \approx \frac{{2Y}}{d - \sigma e - a +2 \left(\sigma e\Delta -\varepsilon d^\prime \right)\Delta },
\nonumber \\
&&
\cos 2\theta_{23} \approx 2\Delta,
\quad
\sin 2\theta _{23} \approx -\sigma,
\label{Eq:Mixing-12}
\end{eqnarray}
with
\begin{eqnarray}
&&
X \approx \sqrt 2 \left( {\varepsilon b^\prime  + b\Delta } \right),
\quad
Y \approx  - \sqrt 2 \sigma \left( {b - \varepsilon b^\prime \Delta } \right),
\quad
\Delta  \approx \frac{{\sigma d^\prime  - \sqrt 2 s_{13} b^\prime }}{{2e + \sqrt 2 s_{13} b}}\varepsilon,
\label{Eq:X-Y-Delta-12}
\end{eqnarray}
where we see that $\sin\theta_{12}\rightarrow 0$ as $\varepsilon\rightarrow 0$. It should be stressed again that the smallness of $\sin^2\theta_{13}$ is not guaranteed by the $\mu$-$\tau$ symmetry because $Y$ is mainly proportional to $b$.  To obtain its smallness needs an additional requirement.

From Eqs.(\ref{Eq:X-Y-Delta-13}) and (\ref{Eq:X-Y-Delta-12}), we can obtain that
\begin{eqnarray}
&&
s_{13}  \approx \frac{{2 eb^\prime  + \sigma bd^\prime }}{{\sqrt 2 \left[ {\sigma e\left( {d + \sigma e - a} \right) - b^2 } \right]}}\varepsilon,
\quad
\cos 2\theta _{23}  \approx  - \frac{{ \left( {d + \sigma e - a} \right)d^\prime } + 2 bb^\prime}{{\sigma e\left( {d + \sigma e - a} \right) - b^2 }}\varepsilon,
\label{Eq:s13Delta-13}
\end{eqnarray}
for $d + \sigma e - a\neq 0$, from Eq.(\ref{Eq:Mixing-13}) for C1), and
\begin{eqnarray}
&&
s_{13}  \approx  - \frac{2\sqrt 2  eb-\varepsilon^2b^\prime d^\prime}{2\sigma e\left(d - \sigma e - a\right)},
\quad
\cos 2\theta _{23}  \approx \frac{\left( d - \sigma e - a \right)d^\prime  + 2bb^\prime }{\sigma e\left( {d - \sigma e - a} \right) - b^2}\varepsilon,
\label{Eq:s12Delta-13}
\end{eqnarray}
for $d - \sigma e - a\neq 0$, from Eq.(\ref{Eq:Mixing-12}) for C2).  Since $b\sim 0$ in C2), $\cos 2\theta_{23}$ is further reduced to 
\begin{eqnarray}
&&
\cos 2\theta _{23}  \approx \frac{\sigma d^\prime}{e}\varepsilon.
\label{Eq:s12Delta-13-cos23}
\end{eqnarray}


\begin{table}[!htbp]
    \caption{\label{Tab:Lists}The textures consistent with the observed results, where the 
    overall scale $d_0$ is omitted, $\sin\theta_{13}=0$ ($\sin\theta_{12}=0$) corresponds to 
    textures with $s_{23}\approx \sigma/\sqrt{2}$ ($s_{23}\approx -\sigma/\sqrt{2}$) for 
    $\sigma=\pm 1$, and $R=\Delta m_ \odot ^2 /\Delta m_{atm}^2 $. In the quasi degenerate 
    mass pattern II, the estimate of the effective neutrino mass $m_{\beta\beta}$ is shown.  
    In some textures, the constraint on $\eta$ arises to yield $\tan 2\theta_{12}={\mathcal {O}}(1)$ 
    as shown in the each texture's last line if it is required.}
    \begin{center}
    \begin{tabular}{|c|c|}
\hline
C1) $\sin\theta_{13}=0$& C2) $\sin\theta_{12}=0$\\
\hline 
\multicolumn{2}{|c|}{Normal Mass Hierarchy}\\
\hline
$
\left( {\begin{array}{*{20}c}
   {p\eta } & {\eta  + \varepsilon } & { - \sigma \left( {\eta  - \varepsilon } \right)}  \\
   {\eta  + \varepsilon } & {1 + r^\prime \varepsilon } & {\sigma \left( {1 - s\eta } \right)}  \\
   { - \sigma \left( {\eta  - \varepsilon } \right)} & {\sigma \left( {1 - s\eta } \right)} & {1 - r^\prime \varepsilon }  \\
\end{array}} \right)
$
&
$
\left( {\begin{array}{*{20}c}
   {p\eta } & {\eta  + \varepsilon } & { - \sigma \left( {\eta - \varepsilon } \right)}  \\
   {\eta  + \varepsilon } & {1 + r^\prime \varepsilon } & { -\sigma \left( {1 -s\eta } \right)}  \\
   { - \sigma \left( {\eta  - \varepsilon } \right)} & { -\sigma \left( {1 - s\eta } \right)} & {1 - r^\prime \varepsilon }  \\
\end{array}} \right)
$
\\
$
\cos 2\theta _{23}  \approx  -\sqrt 2 \sigma r^\prime \sin \theta _{13}
$ 
&
$
\cos 2\theta _{23}  \approx \frac{{\sigma  \left( {s - p} \right)r^\prime\tan 2\theta _{12} }}{2}\sin \theta _{13} 
$
\\
$R \approx \frac{s+p}{\sqrt{2}\sin 2\theta_{12}}\eta^2$
&
$R \approx \frac{s^2-p^2}{2\cos 2\theta_{12}}\sin^2\theta _{13}$
\\
($\ast$) $\vert\eta\vert\gg\varepsilon^2$
&
($\ast$) $\vert\eta\vert\sim\vert\varepsilon\vert$
\\
\hline
\multicolumn{2}{|c|}{Inverted Mass Hierarchy}\\
\hline
$
\left( {\begin{array}{*{20}c}
   2 - p\eta & {\eta  + \varepsilon } & { - \sigma \left( {\eta  - \varepsilon } \right)}  \\
   {\eta  + \varepsilon } & {1 + r^\prime \varepsilon } & { - \sigma\left( {1 -s\eta } \right) }  \\
   { - \sigma \left( {\eta  - \varepsilon } \right)} & { - \sigma\left( {1 -s\eta } \right) } & {1 - r^\prime \varepsilon }  \\
\end{array}} \right)
$
&
$
\left( {\begin{array}{*{20}c}
   {2 - p\eta } & {\eta  + \varepsilon } & { - \sigma \left( {\eta  - \varepsilon } \right)}  \\
   {\eta  + \varepsilon } & {1 + r^\prime \varepsilon } & \sigma\left( {1 -s\eta } \right)   \\
   { - \sigma \left( {\eta  - \varepsilon } \right)} & \sigma\left( {1 -s\eta } \right)  & {1 - r^\prime \varepsilon }  \\
\end{array}} \right)
$
\\
$
\cos 2\theta _{23}  \approx  - \sqrt 2 \sigma r^\prime \sin \theta _{13} 
$
&
$
\cos 2\theta _{23}  \approx \frac{{\sigma \left( {p-s} \right)r^\prime \tan 2\theta _{12} }}{2}\sin \theta _{13} 
$
\\
$
R\approx \frac{{2\sqrt 2 }}{{\sin 2\theta _{12} }}\left(\eta  + r^\prime \sin^2\theta_{13} \right)
$
&
$
R\approx {\frac{{\sqrt 2 \sigma \left( {p-s} \right)}}{{\cos 2\theta _{12} }}\sin \theta _{13} }
$
\\
($\ast$) $m_1\sim m_2$
&
($\ast$) $\vert\eta\vert\sim\vert\varepsilon\vert$, $m_1\sim m_2$
\\
\hline
$
\left( {\begin{array}{*{20}c}
   { - (2 - \eta)} & {q + \varepsilon } & { - \sigma \left( {q - \varepsilon } \right) }  \\
   {q + \varepsilon } & {1 + r^\prime \varepsilon } & { - \sigma }  \\
   { - \sigma \left( {q - \varepsilon } \right)d_0 } & { - \sigma } & {1 - r^\prime \varepsilon }  \\
\end{array}} \right)
$
&
\\
$
\cos 2\theta _{23}  \approx  \frac{{2\sqrt 2 \sigma \left( {q + r^\prime} \right)}}{{2 - qr^\prime }}\sin \theta _{13} 
$
&
$\cdots$
\\
$R\sim \sin^2\theta_{13}$ for $\eta$=0
&
\\
($\ast$) $m_1\sim -m_2$
&
\\
\hline 
\multicolumn{2}{|c|}{Quasi Degenerate Mass Pattern I ($\vert m_{1,2,3}\vert \sim \sqrt{\Delta m^2_{atm}}$)}\\
\hline
$
\left( {\begin{array}{*{20}c}
   { - (2 - \eta)} & {q + \varepsilon } & { - \sigma \left( {q - \varepsilon } \right)}  \\
   {q + \varepsilon } & 1 - r + r^\prime\varepsilon & { - \sigma \left( {1 + r} \right)}  \\
   { - \sigma \left( {q - \varepsilon } \right)} & { - \sigma \left( {1 + r} \right)} & 1 - r - r^\prime\varepsilon   \\
\end{array}} \right)
$
&
\\
$
\cos 2\theta_{23}\approx \frac{{2\sqrt 2 \sigma \left[ {q + r^\prime \left( {1 - r} \right)} \right]}}{{2 + 2r - qr^\prime}}\sin \theta _{13} 
$
&
$\cdots$
\\
$R\sim \sin^2\theta_{13}$ for $\eta$=0
&
\\
($\ast$) $\vert r\cos 2\theta_{12}\vert \sim 1$, $m_1\sim -m_2$
&
\\
\hline
\multicolumn{2}{|c|}{Quasi Degenerate Mass Pattern II ($m^2_{1,2,3} \gg \Delta m^2_{atm}$ and $\kappa = \eta/\vert\eta\vert$) 
}\\
\hline
$
\left( {\begin{array}{*{20}c}
   {1 - \eta } & {\left( {q\eta  + \varepsilon } \right)\eta } & { - \sigma \left( {q\eta  - \varepsilon } \right)\eta }  \\
   {\left( {q\eta  + \varepsilon } \right)\eta } & {1 + r^\prime \varepsilon \eta } & {\sigma \left( {1 - s\eta } \right)\eta }  \\
   { - \sigma \left( {q\eta  - \varepsilon } \right)\eta } & {\sigma \left( {1 - s\eta } \right)\eta } & {1 - r^\prime \varepsilon \eta }  \\
\end{array}} \right)
$
&
$
\left( {\begin{array}{*{20}c}
   {1 - \eta } & {\left( {q\eta  + \varepsilon } \right)\eta } & { - \sigma \left( {q\eta  - \varepsilon } \right)\eta }  \\
   {\left( {q\eta  + \varepsilon } \right)\eta } & {1 + r^\prime \varepsilon \eta } & { - \sigma \left( {1 - s\eta } \right)\eta }  \\
   { - \sigma \left( {q\eta  - \varepsilon } \right)\eta } & { - \sigma \left( {1 - s\eta } \right)\eta } & {1 - r^\prime \varepsilon \eta }  \\
\end{array}} \right)
$
\\
$
0.07\mapleq m_{\beta\beta}\mapleq 0.50~{\rm eV}
$
&
$
0.07\leq m_{\beta\beta}\mapleq 0.35~{\rm eV}
$
\\
$
\cos 2\theta _{23}  \approx  - \sqrt 2 \sigma r^\prime\sin \theta _{13} 
$
&
$
\cos 2\theta _{{\rm{23}}}  \approx  \frac{{\sigma r^\prime s\tan 2\theta _{12} }}{{4q}}\sin \theta _{13} 
$
\\
$
 R \approx \frac{\sqrt 2 \kappa}{\sin 2\theta _{12} } \left( q\eta  - r'\sin ^2 \theta _{13}\right)
$
&
$
 R \approx - \frac{{\sqrt 2 \sigma \kappa s}}{{4 q\cos 2\theta _{12} }}{\rm{sin}}\theta _{13} 
$
\\
($\ast$) $m_1\sim m_2\sim m_3$

&
($\ast$) $\vert\eta\vert\sim\vert\varepsilon\vert$, $m_1\sim m_2\sim m_3$
\\
\hline
$
\left( {\begin{array}{*{20}c}
   {1 + \eta } & {q\eta ^2  + \varepsilon } & { - \sigma \left( {q\eta ^2  - \varepsilon } \right)}  \\
   {q\eta ^2  + \varepsilon } & {\eta  + r^\prime \varepsilon } & { - \sigma }  \\
   { - \sigma \left( {q\eta ^2  - \varepsilon } \right)} & { - \sigma } & {\eta  - r^\prime \varepsilon }  \\
\end{array}} \right)
$
&
$
\left( {\begin{array}{*{20}c}
   {1 + \eta } & {\left( {q + \varepsilon } \right)\eta } & { - \sigma \left( {q - \varepsilon } \right)\eta }  \\
   {\left( {q + \varepsilon } \right)\eta } & {\eta  + r^\prime \varepsilon } & \sigma   \\
   { - \sigma \left( {q - \varepsilon } \right)\eta } & \sigma  & {\eta  - r^\prime \varepsilon }  \\
\end{array}} \right)
$
\\
$
0.07\leq m_{\beta\beta}\mapleq 0.50~{\rm eV}
$
&
$
0.07\leq m_{\beta\beta}\mapleq 0.50~{\rm eV}
$
\\
$
\cos 2\theta _{23}  \approx  - \sqrt 2 \sigma r^\prime\sin \theta _{13} 
$
&
$
\cdots$
\\
$
R\approx \frac{{\sqrt 2 }}{{\sin 2\theta _{12} }}\left( {q\left| \eta  \right| + \frac{{r^\prime }}{{\left| \eta  \right|}}\sin ^2 \theta _{13}} \right)
$
&
$
R\approx \frac{{\kappa \left( {qr^\prime - 2} \right)}}{{\sqrt 2 r^\prime\sin 2\theta _{12} }}\cos 2\theta _{23} 
$
\\
($\ast$) $m_1\sim m_2\sim -m_3$

&
($\ast$) $\vert\eta\vert\sim\vert\varepsilon\vert$, $m_1\sim m_2\sim -m_3$
\\
\hline
    \end{tabular}
    \end{center}
\end{table}
\newpage
\noindent
\vspace{-3mm}
\noindent
\noindent
\noindent
\noindent
\noindent
\noindent
\noindent
\noindent
\noindent
\noindent
\noindent
\noindent
\noindent
\noindent

\begin{thebibliography}{}
\bibitem{PMNS} 
	B. Pontecorvo, \Journal{\JETPUSSR}{7}{172}{1958} [\Journal{\ZETP}{34}{247}{1958}];
	Z. Maki, M. Nakagawa and S. Sakata, \Journal{\PTP}{28}{870}{1962}. 

\bibitem{SK}
	Y. Fukuda {\it et al.}, [Super-Kamiokande Collaboration], \Journal{\PRL}{81}{1562}{1998}; \Journal{\PRL}{82}{2430}{1999};
	T. Kajita for the collaboration, \Journal{\NPSUPPL}{77}{123}{1999}.
	See also,
	T. Kajita and Y. Totsuka, \Journal{\RMP}{73}{85}{2001};
	A. Suzuki, ``Evidence for Neutrino Mass", Talk given at {\it Neutrino 2006: The XXII International Conference on Neutrino Physics and Astrophysics}, 
	Santa Fe, New Mexico, USA (June 13-19, 2006).

\bibitem{Experiments}
	Y. Suzuki, ``Accelerator and Atmospheric Neutrinos", Talk given at {\it XXII International Symposium on Lepton-Photon Interactions at High Energy}, Uppsala, Sweden (June 30-July 5, 2005);
	A. Poon, ``Solar and Reactor Neutrinos", Talk given at {\it XXII International Symposium on Lepton-Photon Interactions at High Energy}, Uppsala, Sweden (June 30-July 5, 2005);

\bibitem{OldSolor}
    J.N. Bahcall, W.A. Fowler, I. Iben and R.L. Sears, \Journal{\APJ}{137}{344}{1963};
    J. Bahcall, \Journal{\PRL}{12}{300}{1964};
    R. Davis, Jr., \Journal{\PRL}{12}{303}{1964};
    R. Davis, Jr., D.S. Harmer and K.C. Hoffman, \Journal{\PRL}{20}{1205}{1968}; 
    J.N. Bahcall, N.A. Bahcall and G. Shaviv, \Journal{\PRL}{20}{1209}{1968};
    J.N. Bahcall and R. Davis, Jr., \Journal{\SCI}{191}{264}{1976}.

\bibitem{Sun}
	Y. Fukuda {\it et al.}, [Super-Kamiokande Collaboration], \Journal{\PRL}{81}{1158}{1998}; [\Journal{\Erratum}{81}{4297}{1998}];
	B.T. Clevel {\it et al.}, \Journal{\APJ}{496}{505}{1998};
	W. Hampel {\it et al.}, [GNO Collaboration],  \Journal{\PLB}{447}{1271}{1999};
	Q.A. Ahmed. {\it et al.}, [SNO Collaboration], \Journal{\PRL}{87}{071301}{2001}; \Journal{\PRL}{89}{011301}{2002}.

\bibitem{K2K}
	S. H. Ahn, {\it et al.}, [K2K Collaboration], \Journal{\PLB}{511}{178}{2001}; \Journal{\PRL}{90}{041801}{2003}.

\bibitem{Reactor} 
	M. Apollonio, {\it et al.}, [CHOOZ Collaboration], \Journal{\EPJ}{27}{331}{2003};
	K. Eguchi, {\it et al.}, [KamLAND collaboration], \Journal{\PRL}{90}{021802}{2003};
	K. Inoue, [KamLAND collaboration], \Journal{\NJP}{6}{147}{2004}.

\bibitem{Seesaw} 
	T. Yanagida, in {\it Proceedings of the Workshop on Unified Theories and 
	Baryon Number in the Universe} edited by A. Sawada and A. Sugamoto 
	(KEK Report No.79-18, Tsukuba, 1979), p.95; \Journal{\PTP}{64}{1103}{1980};  
	M. Gell-Mann, P. Ramond and R. Slansky, in {\it Supergravity} edited by P. van Nieuwenhuizen and D.Z. Freedmann (North-Holland, Amsterdam 1979), p.315; 
	R.N. Mohapatra and G. Senjanovi\'{c}, \Journal{\PRL}{44}{912}{1980}. See also,
	P. Minkowski, \Journal{\PLBOLD}{B67}{421}{1977}.

\bibitem{type2seesaw}
	R.N. Mohapatra and G. Senjanovi\'{c}, \Journal{\PRD}{23}{165}{1981};
	C. Wetterich, \Journal{\NPB}{187}{343}{1981}.
	See also, J. Schechter and J.W.F. Valle, \Journal{\PRD}{22}{2227}{1980}.

\bibitem{Zee}
	A. Zee, \Journal{\PLBOLD}{93B}{389}{1980}; \Journal{\PLBOLD}{161B}{141}{1985};
	L. Wolfenstein, \Journal{\NPB}{175}{93}{1980};
	S. T. Petcov, \Journal{\PLBOLD}{115B}{401}{1982}.

\bibitem{Babu}
	A. Zee, \Journal{\NPBOLD}{264B}{99}{1986}; 
	K. S. Babu, \Journal{\PLB}{203}{132}{1988}; 
	D. Chang, W-Y. Keung and P.B. Pal, \Journal{\PRL}{61}{2420}{1988}; 
	J. Schechter and J.W.F. Valle, \Journal{\PLB}{286}{321}{1992}.

\bibitem{NuData} 
	G.L. Fogli, E. Lisi, A. Marrone, A. Palazzo, \Journal{\PPNP}{57}{742}{2006}.
See also,
	S. Goswami, A. Bandyopadhyay and S. Choubey, \Journal{\NPSUPPL}{143}{121}{2005};
	G. Altarelli, \Journal{\NPSUPPL}{143}{470}{2005};
	A. Bandyopadhyay, \Journal{\PLB}{608}{115}{2005}.

\bibitem{PositiveSolor}
O. Mena and S. Parke, \Journal{\PRD}{69}{117301}{2004}.

\bibitem{NeutrinoSummary}
See for example, 
	R. N. Mohapatra and A. Y. Smirnov, ``Neutrino Mass and New Physics", [arXive: hep-ph/0603118] (to be published 
	in the Annual Review of Nuclear and Particle Science Vol. 56 (2006)).
See also,
	R. Peccei, ``Physics Beyond the Standard Model - What we know, don't know, and what it means for neutrinos",  
	Talk given at {\it Neutrino 2006: The XXII International Conference on Neutrino Physics and Astrophysics}, 
	Santa Fe, New Mexico, USA (June 13-19, 2006);
	R.N. Mohapatra, ``Models of Neutrino Mass", 
	Talk given at {\it Neutrino 2006: The XXII International Conference on Neutrino Physics and Astrophysics}, 
	Santa Fe, New Mexico, USA (June 13-19, 2006).

\bibitem{Nishiura}
	T. Fukuyama and H. Nishiura, in {\it Proceedings of International Workshop on Masses and Mixings of Quarks and Leptons} edited by Y. Koide (World Scientific, Singapore, 1997), p.252; ``Mass Matrix of Majorana Neutrinos", [arXive:hep-ph/9702253];
	Y. Koide, H. Nishiura, K. Matsuda, T. Kikuchi and T. Fukuyama, \Journal{\PRD}{66}{093006}{2002};
	Y. Koide, \Journal{\PRD}{69}{093001}{2004};
	K. Matsuda and H. Nishiura, \Journal{\PRD}{69}{117302}{2004}; \Journal{\PRD}{71}{073001}{2005}; \Journal{\PRD}{72}{033011}{2005}; \Journal{\PRD}{73}{013008}{2006}. 

\bibitem{mu-tau}
	R.N. Mohaptra and S. Nussinov, \Journal{\PRD}{60}{013002}{1999};
	C.S. Lam, \Journal{\PLB}{507}{214}{2001}; \Journal{\PRD}{71}{093001}{2005};
	Z.Z. Xing, \Journal{\PRD}{61}{057301}{2000}; \Journal{\PRD}{64}{093013}{2001};
	``Neutrino Telescopes as a Probe of Broken $\mu$-$\tau$ Symmetry", [arXive: hep-ph/0605219];
	``Nearly Tri-bimaximal Neutrino Mixing and CP Violation from $\mu$-$\tau$ Symmetry Breaking", [arXive: hep-ph/0607091];
	E. Ma and M. Raidal, \Journal{\PRL}{87}{011802}{2001}; [\Journal{\Erratum}{87}{159901}{2001}];
	A. Datta and P.J. O'Donnell, \Journal{\PRD}{72}{113002}{2005};
	Y.H. Ahn, S.K. Kang, C.S. Kim and J. Lee, \Journal{\PRD}{73}{093005}{2006}.
	
\bibitem{mu-tau0}
	T. Kitabayashi and M. Yasu\`{e}, \Journal{\PLB}{524}{308}{2002}; 
	\Journal{\IJMP}{17}{2519}{2002}; \Journal{\PRD}{67}{015006}{2003};
	I. Aizawa, M. Ishiguro, T. Kitabayashi and M. Yasu\`{e}, \Journal{\PRD}{70}{015011}{2004};
	I. Aizawa, T. Kitabayashi and M. Yasu\`{e}, \Journal{\PRD}{71}{075011}{2005}.

\bibitem{mu-tau1}
	W. Grimus and L. Lavoura,  \Journal{\JHEP}{0107}{045}{2001}; \Journal{\EPJ}{28}{123}{2003}; \Journal{\PLB}{572}{189}{2003}; \Journal{\PLB}{579}{113}{2004}; \Journal{\JPG}{30}{1073}{2004}; \Journal{\JHEP}{0508}{013}{2005};
	W. Grimus, A.S. Joshipura, S. Kaneko, L. Lavoura and M. Tanimoto, \Journal{\JHEP}{0407}{078}{2004}; 
	W. Grimus, A.S. Joshipura, S. Kaneko, L. Lavoura, H. Sawanaka and M. Tanimoto, \Journal{\NPB}{713}{151}{2005}; 
	W. Grimus, S. Kaneko, L. Lavoura, H. Sawanaka and M. Tanimoto, \Journal{\JHEP}{0601}{110}{2006}. 

\bibitem{mu-tau2}
	R.N. Mohapatra,  \Journal{\JHEP}{0410}{027}{2004};
	R.N. Mohapatra and S. Nasri,  \Journal{\PRD}{71}{033001}{2005};
	R.N. Mohapatra, S. Nasri and Hai-Bo Yu, \Journal{\PLB}{615}{231}{2005}; \Journal{\PRD}{72}{033007}{2005};
	\Journal{\PLB}{639}{318}{2006}.

\bibitem{eNumber-example}
	T. Kitabayashi and M. Yasue, \Journal{\PLB}{524}{308}{2002} in Ref.\cite{mu-tau0};
	I. Aizawa, M. Ishiguro, T. Kitabayashi and M. Yasu\`{e}, in Ref.\cite{mu-tau0}.

\bibitem{eNumber}
	M. Frigerio and A.Yu. Smirnov, \Journal{\NPB}{640}{233}{2002};
	R.N. Mohapatra and W. Rodejohann, \Journal{\PRD}{72}{053001}{2005}.

\bibitem{Another-mu-tau}
	K. Fuki and M. Yasu\`{e}, \Journal{\PRD}{73}{055014}{2006}.

\bibitem{Another-mu-tau-pre}
	W. Grimus, A.S. Joshipura, S. Kaneko, L. Lavoura, H. Sawanaka and M. Tanimoto, in Ref.\cite{mu-tau1}.

\bibitem{AtmDeviation}
	A. de Gouv\`{e}a, \Journal{\PRD}{69}{093007}{2004}.

\bibitem{Theta31AndMass}
	R.N. Mohapatra, in Ref.\cite{mu-tau2};
	W. Grimus, A.S. Joshipura, S. Kaneko, L. Lavoura, H. Sawanaka and M. Tanimoto, 
in Ref.\cite{mu-tau1};
	R.N. Mohapatra and W. Rodejohann, in Reg.\cite{eNumber};
	F. Plentinger and W. Rodejohann, \Journal{\PLB}{625}{264}{2005}.
See also, A. Joshipura, ``Summary of Model Predictions for $U_{e3}$", Talk given 
at {\it the 5th Workshop on "Neutrino Oscillations and their Origin (NOON2004)},
 Tokyo, Japan (February 11-15, 2004), [arXive:hep-ph/0411154].

\bibitem{PlusMinusNu}
	Riazuddin, \Journal{\JHEP}{0310}{009}{2003}.

\bibitem{type2seesawinSO10}
	For a review, see for example,
	R. N. Mohapatra and A. Y. Smirnov, in \cite{NeutrinoSummary}.

\bibitem{leptonic-mu-tau}
	See for example, 
	W. Grimus and L. Lavoura,  \Journal{\JHEP}{0107}{045}{2001}, \Journal{\EPJ}{28}{123}{2003}, and \Journal{\JPG}{30}{73}{2004} in Ref.\cite{mu-tau1}; 
	E. Ma and G. Rajasekaran, \Journal{\PRD}{68}{071302}{2003};
	R.N. Mohapatra, in Ref.\cite{mu-tau2};
	A. Joshipura, ``Universal 2-3 symmetry", [arXive: hep-ph/0512252];
	K. Matsuda and H. Nishiura, in Ref.\cite{Nishiura};	
	W. Grimus, A.S. Joshipura, S. Kaneko, L. Lavoura, H. Sawanaka and M. Tanimoto, in Ref.\cite{mu-tau1}.

\bibitem{recent_mu-tau-breaking}
For recent studies, see for example,
	R.N. Mohapatra, S. Nasri and Hai-Bo Yu, \Journal{\PLB}{636}{114}{2006}; 
	N. Haba and W. Rodejohann, \Journal{\PRD}{74}{017701}{2006}.

\bibitem{NewMassTexture}
	I. Aizawa and M. Yasu\`{e}, \Journal{\PRD}{73}{015002}{2006}.

\bibitem{QDGone}
See for example,
	B. Adhikary, ``Soft Breaking of $L_\mu$-$L_\tau$ Symmetry: Light Neutrino Spectrum and Leptogenesis", [arXive: hep-ph/0604009].
	
\bibitem{TheoryMass-ee}
See for example,
	S. Pascoli and S.T. Petcov, \Journal{\NPSUPPL}{138}{233}{2005};
	S. Pascoli, S.T. Petcov and T. Schwetz, \Journal{\NPB}{734}{24}{2005};
	M. Hirsch, E. Ma, J.W.F. Valle and A.V. del Moral, \Journal{\PRD}{72}{091301}{2005}; [\Journal{\Erratum}{72}{119904}{2005}];
	M. Hirsch, ``Phenomonology of Double Beta Decay", 
	Talk given at {\it Neutrino 2006: The XXII International Conference on Neutrino Physics and Astrophysics}, 
	Santa Fe, New Mexico, USA (June 13-19, 2006).

\bibitem{AotherAnalysis}
	A. Merle and W. Rodejohann, \Journal{\PRD}{73}{073012}{2006}.

\bibitem{LeLmuPlot}
	S. Choubey and W. Rodejohann, \Journal{\EPJ}{40}{259}{2005}.

\bibitem{LmueMinusLtau}
	B. Adhikary, in Ref.\cite{QDGone};
	T. Ota and W. Rodejohann, \Journal{\PLB}{639}{322}{2006}.
See also,
	X. G. He et al., \Journal{\PRD}{43}{22}{1991};  \Journal{\PRD}{44}{2118}{1991}; 
	E. Ma, D. P. Roy and S. Roy, \Journal{\PLB}{525}{101}{2002}.

\bibitem{AbsoluteMass}
See for example, 
	C. Giunti, \Journal{\ACTAB}{36}{3215}{2005}.

\bibitem{ee-exp}
	H.V. Klapdor-Kleingrothaus {\it et al.}, \Journal{\EPJA}{12}{147}{2001}.

\bibitem{ee-exp-future}
See for example, 
	S. Elliott, ``Intro to experimental program"
	Talk given at {\it Neutrino 2006: The XXII International Conference on Neutrino Physics and Astrophysics}, 
	Santa Fe, New Mexico, USA (June 13-19, 2006).

\bibitem{Normal}
See for example, 
	S. Antusch, J. Kersten, M. Lindner and M. Ratz, \Journal{\NPB}{674}{401}{2003};
	S. Antusch, J. Kersten, M. Lindner, M. Ratz and M.A. Schmidt, \Journal{\JHEP}{0503}{024}{2005}.

\bibitem{InvertedDegenerate}
For recent studies, see for example, 
	R.N. Mohapatra, M.K. Parida and G. Rajasekaran, \Journal{\PRD}{71}{057301}{2005}; 
	J.W. Mei, \Journal{\PRD}{71}{073012}{2005}; 
	S. Luo, J.W. Mei and Z.Z. Xing, \Journal{\PRD}{72}{053014}{2005};
	S. Luo and Z.Z. Xing, \Journal{\PLB}{632}{341}{2006}.

\bibitem{FormulaNoCP}
	I. Aizawa, T. Kitabayashi and M. Yasu\`{e}, in Ref.\cite{mu-tau0}.
\end{thebibliography}
\end{document}